\documentclass[twocolumn]{aastex701}

\usepackage{pifont}% http://ctan.org/pkg/pifont
\usepackage[version=3]{mhchem}
\usepackage{gensymb}
\usepackage{amssymb}% http://ctan.org/pkg/amssymb
\usepackage{pifont}% http://ctan.org/pkg/pifont
\usepackage{threeparttablex}
\usepackage{natbib}

\begin{document}

\title{The chemical diversity of giant-planet nurseries as revealed by ALMA}

\author[0000-0003-2014-2121]{Alice S. Booth} 
\altaffiliation{Clay Postdoctoral Fellow}
\affiliation{Center for Astrophysics \textbar\, Harvard \& Smithsonian, 60 Garden St., Cambridge, MA 02138, USA}
\email[show]{alice.booth@cfa.harvard.edu}

\author[0000-0002-0150-0125]{Jenny Calahan}
\affiliation{Center for Astrophysics \textbar\, Harvard \& Smithsonian, 60 Garden St., Cambridge, MA 02138, USA}
\email{jenny.calahan@cfa.harvard.edu}

\author[0000-0002-7935-7445]{Milou Temmink}
\affiliation{Leiden Observatory, Leiden University, 2300 RA Leiden, the Netherlands}
\email{temmink@strw.leidenuniv.nl}

\author[0000-0002-7212-2416]{Lisa Wölfer}
\affiliation{Department of Earth, Atmospheric, and Planetary Sciences, Massachusetts Institute of Technology, Cambridge, MA 02139, USA}
\email{lwoelfer@mit.edu}

\author[0000-0002-6042-1486]{Jamila Pegues}
\altaffiliation{STScI Postdoctoral Fellow}
\affiliation{Space Telescope Science Institute, Baltimore, MD 21218, USA}
\email{jamila.pegues@stsci.edu}

\author[0000-0003-1413-1776]{Charles J.\ Law}
\altaffiliation{NASA Hubble Fellowship Program Sagan Fellow}
\affiliation{Department of Astronomy, University of Virginia, Charlottesville, VA 22904, USA}
\email{cjl8rd@virginia.edu}

\author[0009-0006-1929-3896]{Lucy Evans}
\affiliation{School of Physics and Astronomy, University of Leeds, Leeds LS2 9JT, UK}
\email{L.E.Evans@leeds.ac.uk}

\author[0000-0003-3674-7512]{Margot Leemker}
\affiliation{Dipartimento di Fisica, Università degli Studi di Milano, Via Celoria 16, 20133 Milano, Italy}
\email{margot.leemker@unimi.it}

\author[0000-0003-2493-912X]{Shota Notsu}
\affiliation{Department of Earth and Planetary Science, Graduate School of Science, The University of Tokyo, 7-3-1 Hongo, Bunkyo-ku, Tokyo 113-0033, Japan}
\affiliation{Star and Planet Formation Laboratory, Pioneering Research Institute, RIKEN, 2-1 Hirosawa, Wako, Saitama 351-0198, Japan}
\email{shota.notsu@eps.s.u-tokyo.ac.jp}

\author[0000-0001-8798-1347]{Karin Öberg}
\affiliation{Center for Astrophysics \textbar\, Harvard \& Smithsonian, 60 Garden St., Cambridge, MA 02138, USA}
\email{koberg@cfa.harvard.edu}

\author[0000-0001-6078-786X]{Catherine Walsh}
\affiliation{School of Physics and Astronomy, University of Leeds, Leeds LS2 9JT, UK}
\email{c.walsh1@leeds.ac.uk}

\author[0000-0001-7591-1907]{Ewine F. van Dishoeck}
\affiliation{Leiden Observatory, Leiden University, 2300 RA Leiden, the Netherlands}
\affiliation{Max-Planck-Institut für Extraterrestrische Physik, Giessenbachstrasse 1, 85748 Garching, Germany}
\email{ewine@strw.leidenuniv.nl}

\begin{abstract}
With the giant exoplanet occurrence rate peaking around stars of 1.5–2 solar masses, there is strong motivation to characterize the disks that set their formation conditions. Observations with the Atacama Large Millimeter/submillimeter Array (ALMA) allow us to investigate both the availability of different molecules in disks and infer the radial distribution of elemental abundances, enabling us to make connections to exoplanet systems. Here we present a survey of six transition disks around young F-, A-, and B-type stars using ALMA. We find \ce{^{13}C^{18}O}, CS, SO, and \ce{H_2CO} in all six systems, as well as ten additional molecules in a subset of disks, including detections of \ce{H_2S}, \ce{^{33}SO}, and \ce{CH_3OCH_3}. Using these data, and literature data where available, we construct the first comprehensive picture of Herbig disk chemistry. We find clear correlations between molecular tracers of C/O$>$1 environments (e.g., CS, \ce{C_2H}) and disk mass, as traced by \ce{C^{18}O} line flux. In contrast, tracers of C/O$<$1 environments (e.g., SO, \ce{CH_3OH}) do not show significant correlations with disk mass. Interestingly, these molecules are relatively brighter in lower-mass disks, with their presence primarily linked to disks with central cavities and spirals. Finally, we show that the observed chemical diversity seen across Herbig disks leads to varying C/O regimes at the orbital radii of candidate proto-planets identified within these disks. When comparing these inferred disk C/O ratios with those measured for directly imaged exoplanets, we find a notable overlap and show that low C/O gas is common on 10's of au scales in Herbig disks. 

\end{abstract}

% \keywords{}

\section{Introduction}

Understanding the chemical composition of protoplanetary disks is crucial for interpreting exoplanet compositions and assessing the potential habitability of extrasolar systems \citep[e.g.;][]{2022ApJ...937...36P,2023ASPC..534.1031K}. Exoplanet population demographics from both radial velocity and direct imaging surveys show that gas-giant planet occurrence rates peak around 1.5–2.0~$M_{\odot}$ stars \citep{2010PASP..122..905J,2015A&A...574A.116R,2019AJ....158...13N,2021A&A...651A..72V,2022A&A...661A..63W,2025A&A...693A..54S}, and there is a notable population of wide-orbit giants with well-characterized atmospheres around F- and A-type stars in debris disk systems \citep[e.g.,][]{2023AJ....165....4W,2023A&A...673A..98B,2024A&A...682A..48L}. Given this growing wealth of exoplanet data, we need to systematically study the disks around young intermediate-mass ($\approx$1.5–3~$M_{\odot}$) stars - often referred to collectively as Herbig disks.

When comparing elemental abundances measured in the gas-phase for disks to those for exoplanet atmospheres, \citet{2024ApJ...969L..21B} highlight a potential dichotomy where wide-separation exoplanets have C/O$<$1.0 \citep[e.g.;][]{2024A&A...687A.298N} and spatially-resolved disk observations of Herbig systems report C/O~$\geq$1 but, this comparison only considers three Herbig systems (HD~163296 and MWC~480 \citealt{2021ApJS..257....7B}; and AB~Aurigae \citealt{2020A&A...642A..32R}). 
More generally, the presence and abundances of complex organic molecules (COMs) in disks trace the prebiotic potential of the chemical reservoir \citep{2014A&A...563A..33W}. Compared to T-Tauri disks, the warmer temperatures that shift snowlines outward and/or the higher masses of Herbig Ae disks provide a window into the COM reservoir available during planet formation \citep{2015Natur.520..198O, 2021NatAs...5..684B}. Constraining the availability of these molecules is crucial for considering the potential habitability of terrestrial planets or moons forming within disks.

% These molecules are difficult to detect due in part to their low abundances \citep{2016ApJ...823L..10W}

With the Atacama Large Millimeter/submillimeter Array (ALMA), we have now built up a picture of the dust and gas masses of the local population of disks around young ($\sim$1–10~Myr) intermediate-mass stars \citep{2022A&A...658A.112S,2024A&A...682A.149S,2025A&A...693A..49S,2025A&A...693A.286S}. These disks are typically more massive in both dust and gas than their T-Tauri counterparts and, due to their more luminous host stars, they are also warmer \citep[e.g.;][]{2021ApJ...908....8C,2021ApJS..257...17C,2021ApJS..257....4L}. The latter means that snowlines lie radially farther from the host stars, and some systems have no CO snowlines \citep[e.g.,][]{2018A&A...616A..19A,2021ApJS..257....5Z,2023A&A...678A.146B}. Meaning, in some cases, there are no significant reservoirs of CO ice and subsequently less CO depletion from the gas-phase into the ice-phase \citep[e.g.;][]{2016A&A...592A..83K,2023A&A...678A.146B,2025A&A...698A.231P,2025ApJ...984L..18T} and, therefore less potential for the in-situ formation of COMs via CO-ice mediated grain chemistry \citep{2018A&A...618A.182B, 2018A&A...616A..19A}. 

When looking for observational trends in disk chemistry with host star/system properties, several surveys have targeted small samples of Herbig disks \citep{2021ApJS..257....1O, 2022MNRAS.510.1148S, 2023ApJ...948...57P, 2024AJ....167..164B, 2024AJ....167..165B}. These works support a picture in which there are similarities in the observable chemistry of Herbig and T-Tauri disks but, some notable differences. Specifically, observations of transition disks around young, intermediate-mass stars have provided the first evidence of Class II disk gas with C/O~$<$~1, via the detections of SO and \ce{SO_2} \citep{2010A&A...524A..19F, abaur_chem, 2021A&A...651L...6B}, NO \citep{2023arXiv230300768L, 2024AJ....167..164B}, and thermally sublimating COMs, which trace warm ($>$100 K) low C/O gas within the water snowline \citep{2021A&A...651L...5V, 2022A&A...659A..29B, 2024ApJ...974...83Y, 2025ApJ...982...62E, 2025arXiv250414023B}. These low C/O ratios are traced on scales of tens of au with, in some cases, higher C/O ratios traced on larger scales  \citep[e.g., HD~100546][]{2024A&A...687A.299L}. These results therefore potentially alleviate some of the tension between disk and exoplanet compositions highlighted by \citet{2024ApJ...969L..21B}. Additionally, the presence of COMs has revealed that both the inheritance of interstellar organic-rich ices and local disk processes contribute to shaping the disk organic reservoir \citep{2021NatAs...5..684B, 2024ApJ...974...83Y}.

In this work, we aim to extend our understanding of disk chemistry in this class of disk with an ALMA survey of six nearby, gas-rich transition disks around young B-, A-, and F-type stars. Our observations target molecules such as \ce{^{13}C^{18}O}, SO, CS, and \ce{CH_3OH}, as well as other organic species and in some cases, increase the detections statistics of these species by over 50\%. These molecules allow us to measure the disk gas mass, C/O ratio, and potential to facilitate pre-biotic chemistry. In Section 2, we describe the sample; in Section 3, we detail the observations; and in Section 4, we present the results. In Section 5, we incorporate archival data where possible to build a more unified picture of Herbig disk chemistry, drawing connections to bulk system properties, disk substructures, and the directly imaged exoplanet population. 

\section{Targets}

\begin{table*}[t!]
\centering
\caption{Properties of star-disk systems observed in this survey.}
\begin{tabular}{ccccccccccc}
\hline
Target     & Distance  & $\mathrm{M_{*}}$ & Sp. Type & log$\mathrm{_{10}}$($\mathrm{L_{*}}$) & log$\mathrm{_{10}}$($\mathrm{M_{acc}}$) & $\mathrm{M_{dust}}$  & $\mathrm{V_{sys}}$ & P.A.   & Inc. & $\mathrm{R_{cav}}$ \\ 
&  [pc] & [$\mathrm{M_{\odot}}$] & & [$\mathrm{L_{\odot}}$] & [$\mathrm{M_{\odot}}~yr^{-1}$] & [$\mathrm{M_{\earth}}$] & [km~s$^{-1}$]& [\degree] & [$\degree$] & [au] \\ \hline \hline
HD 97048   & 184  & 2.80  &  B8    & 1.81  & -6.36 &  155.9$\pm$16.0 &  4.6  & –176     & 41 & 63  \\
CQ Tau     & 149  & 1.50 & F2-F3  & 0.82  & -7.18 &  44.2$\pm$4.8   &  6.2  & -50    &  35 & 50 \\
HD 135344B & 134  & 1.46 & F5-F8  &  0.71 & -6.94 &  35.2$\pm$3.8   &  7.1  &  -62   & 10  & 52  \\
HD 169142  & 114  & 1.55 & A9-F0  & 0.76  & -7.35 &  22.9$\pm$2.4   &  6.9  & -75    & 13 & 26  \\
MWC 758    & 155  & 1.67 & A9-F0  & 0.94  & -7.11 &  18.8$\pm$2.0   &  6.0  & -60    & 21  & 62  \\
HD 100453  & 104  & 1.60 & A9-F0  & 0.79  & -7.32 &  17.5$\pm$1.8   &  5.2  & –145    & 35 & 30  \\ \hline
\end{tabular}
\tablecomments{The stellar properties, distances and mass accretion rates are taken from \citet{2021A&A...650A.182G}, the disk dust masses from \citet{2022A&A...658A.112S}. The disk geometries and source velocities are from the survey paper by \citet{2024A&A...682A.149S} where the P.A. is defined as the angle clockwise from north to the blue-shifted side of the disk and, the cavity sizes are from \citet{2020ApJ...892..111F}.}
\label{table1}
\end{table*}

Our six targets were selected from the \citet{2024A&A...682A.149S} sample of Herbig disks with detections of \ce{C^{18}O} gas. We note that within 200~pc, there are only 18 out of a total 33 systems reported by \citet{2018A&A...620A.128V} with \ce{C^{18}O} gas detections. All of our targets are transition disks, which is the case for most of this class of disk where from the sample with \ce{C^{18}O} detections 13/18 have resolved dust cavities detected with ALMA \citep{2022A&A...658A.112S}. Transition disks were also targeted due to the observed molecular richness of systems like HD~100546 and Oph~IRS~48 \citep[e.g.;][]{2024AJ....167..164B,2024AJ....167..165B} in order to see if this is common for this class of disk. The basic properties of our target stars and their disks are listed in Table~\ref{table1}, with values collated from \citet{2021A&A...650A.182G}, \citet{2022A&A...658A.112S}, and \citet{2024A&A...682A.149S}. 
Given the relatively small number of nearby Herbig disks, our sample provides significant new data and complements previous detailed chemical studies of the full disks MWC~480 and HD~163296 (e.g.; \citealt{2021ApJS..257....1O}) and the transition disks AB~Aurigae, HD~100546, HD~142527, and Oph-IRS~48 \citep{2020A&A...642A..32R, Temmink2023, 2024AJ....167..164B, 2024AJ....167..165B}—the latter of which is not included in the \citet{2024A&A...682A.149S} sample. Notably, our sample includes mostly later-type F stars, with the exception of the B-type system HD~97048, thereby complementing studies listed above which predominantly focused on A-type stars.
There is also limited overlap when comparing to the surveys from \citet{2022MNRAS.510.1148S} with ALMA and \citet{2023ApJ...948...57P} with the SMA
where HD~100453 is covered in the former and MWC~758 in the latter.  
In the remainder of this section, we summarize the known dust and molecular gas properties of our target systems prior to our observations.

% Herbigs with C18O and no other line data <200pc: AK Sco, HD 104237, V892 Tau, HD 9672 (49 ceti, so thats a debris disk)
% 13CO only: HD 141569, 

\textit{HD~97048:} This is the most luminous star in our sample and is host to a radially-extended and ringed protoplanetary disk. The millimeter dust emission and near-infrared scattered light observations show multiple dust rings with the millimeter rings located at $\approx$50, 150 and 300~au and a clearly flared disk structure is detected at shorter wavelengths \citep{2006Sci...314..621L, 2016ApJ...831..200W, 2016A&A...595A.112G, 2017A&A...597A..32V,2024A&A...685A..52G}. 
The gas disk traced by \ce{^{12}CO} extends beyond 700~au with \ce{^{13}CO} and \ce{C^{18}O} also detected extending beyond 600~au and 500~au respectively \citep{2022ApJ...932..114L, 2024A&A...682A.149S}. In addition, there are detections of \ce{HCO^+}, \ce{H^{13}CO^+} and \ce{HC^{15}N} where the latter, rarer isotopologues, appear to be ringed \citep{2017A&A...597A..32V,2019A&A...629A..75B}. Most notably, there is a few Jupiter-mass planet candidate located within the dust gap at $\approx$130~au that has been identified by non-Keplerian \ce{^{13}CO} gas kinematics \citep{2019NatAs...3.1109P}. 

\textit{CQ~Tau:} This hosts a transition disk with a single ring peaking at $\approx$50~au in the millimeter dust, with some azimuthal brightness asymmetries likely linked to the spiral arms seen in the near-infrared data and CO line kinematics \citep{2019MNRAS.486.4638U,2020AJ....159..118U,2021A&A...648A..19W,2025ApJ...984L...9C}.
The gas disk, traced in \ce{^{12}CO}, \ce{^{13}CO}, and \ce{C^{18}O}, is more radially extended than the dust and gas is present within the dust cavity, though depleted relative to the ring. One compelling explanation for these spiral structures and the gas depletion within the dust cavity is an unseen companion on an inclined and eccentric orbit located within the dust cavity \citep{2022MNRAS.515.6109H}. 
Most recently, ringed SO emission was detected originating from the dust cavity edge  \citep{2025arXiv250616481Z}. 

\textit{HD~135344B:} This is also known as SAO~206462 and is part of a visual binary system where HD~135344A is an A-type star located $\approx$20.8" to the north-north-east of HD~135344B \citep{2006MNRAS.367..737B}. The millimeter disk of HD~135344B has been well observed with ALMA at high angular resolution ($\approx$0.1") over multiple wavelengths. These studies report a ring of millimeter dust peaking at $\approx$50~au, with a crescent-shaped potential vortex beyond this ring at $\approx$80~au \citep{2016ApJ...832..178V,2018A&A...619A.161C,2025ApJ...984L..22W}. These dust traps intersect with spiral arms seen in scattered light data which are also traced in CO gas \citep{2016A&A...595A.113S,2021MNRAS.507.3789C}. 
The gas disk appears to be azimuthally symmetric when traced in \ce{^{13}CO} and \ce{C^{18}O}, with a $\approx$30~au gas cavity compared to a $\approx$40~au dust cavity \citep{2016A&A...585A..58V}. Recent observations with VLT/SPHERE report the detection of a giant planet within this cavity on an $\approx$16~au orbit \citep{2025arXiv250706206S}. 

\textit{HD~169142:} This is a nearly face-on system with multiple rings of dust traced in the millimeter continuum and near-infrared scattered light, a proposed few Jupiter-mass proto-planet orbiting within a gap at $\approx$40~au, and a deep central gas cavity \citep{2017A&A...600A..72F,2019ApJ...881..159M,2019AJ....158...15P,2022A&A...663A..23L,2023MNRAS.522L..51H}. HD~169142 has a particularly molecule-rich protoplanetary disk with little CO freeze-out inferred from weak detection of \ce{N_2H^+} relative to the \ce{C^{18}O} line flux \citep{2023A&A...678A.146B}. From the reported detections of $\approx$14 molecules and isotopologues, the most relevant to this study are SO, CS, and \ce{CH_3OH} \citep{2023ApJ...952L..19L,2023A&A...678A.146B}. In addition, the unique detection of SiS in this system—which traces a \ce{^{12}CO} kinematic excess within the proposed planet-carved gap—makes this an intriguing system to study disk chemistry \citep{2022MNRAS.517.5942G,2023ApJ...952L..19L}.

\textit{MWC~758:} This is also known as HD~36112 and is another disk with clear spirals traced in near-infrared and CO emission lines \citep{2015A&A...578L...6B,2018ApJ...853..162B} potentially linked to a warped disk \citep{2025arXiv250711669W}. The underlying millimeter dust ring is clumpy, with two prominent azimuthal asymmetries \citep{2018ApJ...853..162B,2018ApJ...860..124D}. There are two proposed giant protoplanets in this system: MWC~758b, located interior to the dust ring at 20~au, and MWC~758c, located at 100~au at the end of a spiral arm \citep{2018A&A...611A..74R,2023NatAs...7.1208W}. In terms of other gas tracers, \citet{2023ApJ...948...57P} report detections of \ce{HCO^+}, CS, and tentatively \ce{DCO^+} with the Sub-Millimeter Array and \citet{2025arXiv250616481Z} detect SO with ALMA. 

\textit{HD~100453:} is a binary system where the detected disk is associated with HD~100453~A (hereafter, HD 100453). The companion, HD~100453~B, is a young M4 star located 120~au from HD~100543 and has been shown to drive two spiral arms traced both in scattered light and CO gas kinematics \citep{2009ApJ...697..557C, 2015ApJ...813L...2W,2018ApJ...854..130W,2020MNRAS.491.1335R,2023AJ....165..225F}. The HD~100543 millimeter dust disk is a single ring peaking at $\approx$30~au \citep{2020MNRAS.491.1335R}
Beyond CO isotopologues, \citet{2022MNRAS.510.1148S} report the detections of \ce{HCO^+}, \ce{HCN}, CN and \ce{CS} and a non-detection of \ce{C_2H} in this disk from $\approx$1" resolution ALMA observations. \citet{2025arXiv250414023B} recently confirmed the warm and oxygen-rich nature of this disk with detections of \ce{CH_3OH}, \ce{CH_3OH} isotopologues and \ce{CH_3OCHO} which are all likely the products of thermal ice sublimation of organic-rich ices within the \ce{H_2O} snowline. 

\section{Methods} \label{sec:observations} 

\subsection{ALMA observations}

This paper presents ALMA observations from the Cycle 11 program 2023.1.00252.S (PI A. Booth). These are Band 7 observations in the 290-315~GHz range with an angular resolution ranging from $\approx$0\farcs3 to 0\farcs6. The breakdown of the observing time, baselines, maximum recoverable scale (MRS), and the date of observations for each execution are listed in Table~\ref{obstable}. 
The maximum recoverable scale of 4-6" across the survey is well-suited to encompass the full CO gas disk of all our targets, except HD~97048, where the \ce{C^{18}O} gas disk extends beyond 2.5" radially \citep{2024A&A...682A.149S}. A subset of the line data for the HD~100453 disk from this program has been presented in \citet{2025arXiv250414023B} which focused on the \ce{CH_3OH}, \ce{CH_3OH} isotopologues and other COMs emissions. 

The data consist of two spectral settings (A and B) with a total of 25 spectral windows, all with a channel width of 488.281~kHz ($\approx$0.55~kms$^{-1}$ at the lowest frequencies). The full list of central sky frequencies and bandwidth of the spectral windows is listed in Table~\ref{spwtable}. We note that the observations were incomplete, with CQ~Tau and MWC~758 only being observed in one of the two settings (B) at 38\% of the requested sensitivity and, the HD~97048 data in settings A and B only reached 78\% and 33\% of the requested sensitivity goal, respectively. We therefore, supplement our data with CS J=7-6 observations of CQ~Tau and MWC~758 from the exoALMA Large Program \citep[2021.1.01123.L;][]{2025ApJ...984L...6T} where we take the CS low-resolution (0\farcs3) image cubes from the exoALMA webpage\footnote{\url{https://www.exoalma.com/data}} with full details on their data reduction and imaging procedure found in \citet{2025ApJ...984L...7L}. To complement our array of line data, we also show maps of higher angular resolution continuum data for all of our targets which are taken from various other ALMA programs \citep{2018ApJ...860..124D,2018A&A...619A.161C, 2019NatAs...3.1109P,2019MNRAS.486.4638U,2020MNRAS.491.1335R,2019AJ....158...15P}.

\subsection{Line identification and imaging}

The data for each setting were calibrated and self-calibrated within the ALMA data reduction pipeline by ALMA staff using CASA version 6.5.5 \citep{2007ASPC..376..127M,2022PASP..134k4501C,2023PASP..135g4501H}. The identification of bright lines was initially performed by eye from the product data cubes. The measurement sets were then continuum subtracted using CASA task \texttt{uvcontsub} with a fit order of 1 and channels where bright lines were present were excluded. 
The data were imaged using \texttt{tCLEAN} at a briggs robust value of +2.0 to maximize signal-to-noise with a channel width of 0.55~km~s$^{-1}$. 
This was performed with an analytical Keplerian mask\footnote{\url{https://github.com/richteague/keplerian_mask}} using the parameters listed in Table~\ref{table1} and down to a threshold of 3~$\times$ the rms measured in the line free channels of the dirty image. 
The exoALMA CS channel maps are convolved to the same beam size as the corresponding SO observations using CASA task \texttt{imsmooth}. 
We then generated Keplerian masked integrated intensity maps using the python package \texttt{BetterMoments}\footnote{\url{https://github.com/richteague/bettermoments}}. The resulting channel maps and integrated intensity maps are hosted on Zenodo (DOI:10.5281/zenodo.17653342) to enable others to utilize these images for analysis.

The disk integrated fluxes are extracted from Keplerian masks which contain the full radial extent of the molecular disk as traced in \ce{H_2CO} $J=4_{1,3}-3_{1,2}$ (the most extended molecule in these data). For the more radially extended disks - HD~97048 and HD~169142 - we also use a smaller 1.0" mask for species with more compact emission structures, e.g. SO and \ce{CH_3OH}. If a molecule is undetected, we give the 3$\sigma$ upper limit on the flux, where $\sigma$ is propagated from the rms in the channel maps and the number of pixels included in the mask \citep[e.g.,][]{Carney2019}. 
Matched filtering was also performed using an array of Keplerian masks tuned to each of the disks  to check for weak line detections across the full spectral coverage of the data \citep{Loomis2018}. 
We define a detection of a line to have a disk integrated flux of $>$3~$\sigma$ and a matched filter peak $>$3$\sigma$ and define a tentative detection where only one of these two conditions are met.
In addition, lines where significant $>$3~$\sigma$ signal is seen in the integrated intensity maps but the previous conditions are not met are considered tentative detections. We quantify the significance of the signal in the integrated intensity maps showing contours at the 3, 5, and 10$\sigma$ levels, where $\sigma$ is the rms noise propagated from the channel maps, accounting for the total number of channels collapsed to generate the map. The resulting image properties are listed in Table~\ref{table_images} and the properties of the transitions listed are collated in Table~\ref{molecules}. 

%These upper limits can be significantly higher than 3$\times$ the channel map rms because of width in velocity space, i.e. the number of channels, over which emission is expected particularly for inclined disks. 

\begin{ThreePartTable}
\begin{TableNotes}
\item{$^{*}$per 0.55~$\mathrm{kms^{-1}}$ channel for all lines aside from CS in CQ Tau and MWC~758 which have 0.2~$\mathrm{kms^{-1}}$ channels.}
\item{\textbf{$^{**}$}Where if line flux is below the 3$\sigma$ level we report the 3$\sigma$ value as an upper-limit. Note these errors do no include the 10\% absolute flux calibration of ALMA, see ALMA Technical Handbook: \url{https://almascience.eso.org/documents-and-tools/cycle12/alma-technical-handbook}.}
\item{$^{***}$Where \ding{51} denotes a line detection, \textbf{?} a tentative detection and \ding{55} a non-detection.}
\end{TableNotes}
\centering
\begin{longtable*}{ccccccc}
\caption{Image properties and disk integrated fluxes across the survey} \\ \hline
Molecule & Transition & Source & Beam & RMS$^{*}$ & Integrated Flux $\pm$ Error$^{**}$ & Detected$^{***}$ \\
         &            &      & [$\farcs \times \farcs$ ($^{\circ}$)] & [mJy $\mathrm{beam^{-1}}$] & {[}mJy km s$^{-1}${]}   &    \\ \hline \hline  

\ce{^{13}C^{18}O}   & $J=3-2$  
& HD 97048 & 0.64$\times$0.39 (4.0) & 3.0 & 1669.0$\pm$23.0 & \ding{51}\\
& & HD 135344B & 0.87$\times$0.57 (-86.0) & 1.9 & 306.0$\pm$15.0 & \ding{51}\\
& & HD 169142 & 0.54$\times$0.39 (-86.0) & 1.5 & 574.0$\pm$10.0 & \ding{51}\\
& & HD 100453 & 0.67$\times$0.63 (-69.0) & 1.2 & 20.0$\pm$6.0 & \ding{51}\\
& & CQ Tau & 0.77$\times$0.56 (-39.0) & 2.4 & 91.0$\pm$15.0 & \ding{51}\\
& & MWC 758 & 0.84$\times$0.54 (-44.0) & 2.4 & 144.0$\pm$14.0 & \ding{51}\\
 \hline 

CS & $J=6-5$  
& HD 97048 & 0.62$\times$0.45 (15.0) & 1.6 & 1652.0$\pm$12.0& \ding{51}\\
& & HD 135344B & 0.59$\times$0.43 (90.0) & 1.2 & 508.0$\pm$13.0 & \ding{51}\\
&  & HD 169142 & 0.57$\times$0.45 (-81.0) & 1.0 & 2389.0$\pm$6.0 & \ding{51}\\
& & HD 100453 & 0.56$\times$0.48 (57.0) & 1.2 & 61.0$\pm$7.0 & \ding{51}\\
& 7-6 & CQ Tau &  0.8$\times$0.58 (-39.0) & 5.2  &   454$\pm$6.0 & \ding{51}\\
&  & MWC 758 & 0.88$\times$0.56 (-43.0)  &  6.7 &  460$\pm$6.0 & \ding{51}\\
\hline 

SO  & $7(7)-6(6)$
 & HD 97048 & 0.67$\times$0.41 (4.0) & 2.6 & 70.0$\pm$20.0& \ding{51}\\
& & HD 135344B & 0.89$\times$0.6 (-87.0) & 1.7 & 106.0$\pm$13.0 & \ding{51}\\
& & HD 169142 & 0.57$\times$0.41 (-86.0) & 1.2 & 69.0$\pm$8.0 & \ding{51}\\
& & HD 100453 & 0.68$\times$0.64 (-65.0) & 1.1 & 258.0$\pm$5.0 & \ding{51}\\
& & CQ Tau & 0.8$\times$0.58 (-39.0) & 2.7 & 173.0$\pm$17.0 & \ding{51}\\
\hline 

\ce{^{33}SO}  & $J=8(7)-7(6)$
 & HD 97048 & 0.67$\times$0.41 (4.0) & 2.3 & $<51.0$ & \ding{55}\\
& & HD 135344B & 0.89$\times$0.6 (-87.0) & 1.4 & $<$31.0 & \ding{55}\\
& & HD 169142 & 0.57$\times$0.41 (-86.0) & 0.9 & $<$18.0& \ding{55} \\
& & HD 100453 & 0.68$\times$0.64 (-65.0) & 0.9 & 43.0$\pm$4.0  & \ding{51}\\
\hline 

\ce{SO_2} & $J=17_{1,17}-16_{0,16}$
 & HD 97048 & 0.64$\times$0.39 (5.0) & 2.9 & $<$70.0  & \ding{55}\\
& & HD 135344B & 0.84$\times$0.63 (-89.0) & 1.4 & $<$31.0 & \ding{55}\\
& & HD 169142 & 0.54$\times$0.39 (-86.0) & 1.4 & $<$28.0  & \ding{55}\\
& & HD 100453 & 0.66$\times$0.63 (-67.0) & 1.3 & 124.0$\pm$6.0 & \ding{51}\\
& & CQ Tau & 0.77$\times$0.56 (-38.0) & 2.7 & 76.0$\pm$18.0 & \ding{51}\\
& & MWC 758 & 0.84$\times$0.54 (-43.0) & 2.7 & $<$47.0  & \ding{55}\\
\hline 

\ce{H_2CS} &  $J=9_{1,9}-8_{1,8}$  
& HD 97048 & 0.6$\times$0.43 (16.0) & 1.6 & 240.0$\pm$12.0& \ding{51}\\
& & HD 135344B & 0.58$\times$0.42 (90.0) & 1.2 & $<$40.0  & \ding{51}\\
& & HD 169142 & 0.56$\times$0.45 (-79.0) & 1.0 & 140.0$\pm$7.0 & \ding{51}\\
& & HD 100453 & 0.49$\times$0.45 (66.0) & 1.2 & $<$22.0 & \ding{55}\\
& $J=9_{1,8}-8_{1,7}$  & CQ Tau & 0.77$\times$0.56 (-38.0) & 2.9 & $<$56.0 & \ding{55}\\
& & MWC 758 & 0.84$\times$0.54 (-43.0) & 2.7 & $<$47.0 & \ding{55}\\
\hline 

\ce{H_2S} & $J=3_{3,0}-3_{2,1}$
 & HD 97048 & 0.67$\times$0.41 (4.0) & 2.6 & $<$60.0  & \ding{55}\\
& & HD 135344B & 0.88$\times$0.66 (90.0) & 1.2 & $<$25.0  & \ding{55}\\
& & HD 169142 & 0.57$\times$0.41 (-86.0) & 1.2 & 26.0$\pm$8.0 & \ding{51}\\
& & HD 100453 & 0.69$\times$0.65 (-64.0) & 1.1 & $<$14.0  & \ding{55}\\
& & CQ Tau & 0.81$\times$0.58 (-38.0) & 2.7 & $<$51.0  & \ding{55}\\
& & MWC 758 & 0.88$\times$0.56 (-43.0) & 2.9 & $<$48.0  & \ding{55}\\
\hline 

\ce{H_2CO} & $J=4_{1,3}-3_{1,2}$
 & HD 97048 & 0.67$\times$0.41 (5.0) & 2.7 & 1248.0$\pm$27.0 & \ding{51}\\
& & HD 135344B & 0.88$\times$0.66 (-89.0) & 1.2 & 453.0$\pm$8.0 & \ding{51}\\
& & HD 169142 & 0.56$\times$0.41 (-86.0) & 1.2 & 1738.0$\pm$10.0 & \ding{51}\\
& & HD 100453 & 0.69$\times$0.64 (-65.0) & 1.0 & 500.0$\pm$4.0 & \ding{51}\\
& & CQ Tau & 0.8$\times$0.57 (-37.0) & 2.5 & 693.0$\pm$16.0 & \ding{51}\\
& & MWC 758 & 0.87$\times$0.53 (-46.0) & 2.6 & 434.0$\pm$15.0 & \ding{51}\\
\hline 

\ce{CH_3OH} & $J=6_{1,5}-6_{0,6}$  
& HD 97048 & 0.65$\times$0.39 (5.0) & 2.6 & $<$62.0 & \ding{55}\\
& & HD 135344B & 0.84$\times$0.63 (-89.0) & 1.2 & $<$27.0 &\textbf{?}\\
& & HD 169142 & 0.54$\times$0.4 (-86.0) & 1.2 & 96.0$\pm$8.0 & \ding{51}\\
& & HD 100453 & 0.67$\times$0.63 (-64.0) & 1.1 & 132.0$\pm$5.0 & \ding{51}\\
& & CQ Tau & 0.77$\times$0.56 (-37.0) & 2.6 & 54.0$\pm$17.0 & \ding{51}\\
& & MWC 758 & 0.85$\times$0.54 (-42.0) & 2.5 & $<$43.0& \ding{55}\\
\hline 

\ce{HC_3N} & $J=32-31$  
& HD 97048 & 0.62$\times$0.45 (16.0) & 1.6 & 110.0$\pm$12.0 & \ding{51}\\
& & HD 135344B & 0.6$\times$0.44 (-89.0) & 3.3 & $<$106.0 & \ding{55}\\
& & HD 169142 & 0.58$\times$0.46 (-82.0) & 1.0 & 66.0$\pm$6.0 & \ding{51}\\
& & HD 100453 & 0.52$\times$0.46 (68.0) & 1.3 & $<$22.0 & \ding{55}\\
& $J=33-32$ & CQ Tau & 0.8$\times$0.57 (-37.0) & 2.5 & $<$47.0 & \ding{55}\\
& & MWC 758 & 0.87$\times$0.53 (-46.0) & 2.6 & $<$44.0 & \ding{55}\\
\hline 

\ce{c-C_3H_2} & $J=6_{3,3}-5_{4,2}$
& HD 97048 & 0.65$\times$0.39 (5.0) & 2.1 & 94.0$\pm$22.0 & \ding{51}\\
& & HD 135344B & 0.84$\times$0.63 (-90.0) & 0.9 & $<$21.0  & \ding{55}\\
& & HD 169142 & 0.54$\times$0.4 (-86.0) & 1.0 & 51.0$\pm$7.0 & \ding{51}\\
& & HD 100453 & 0.66$\times$0.63 (-67.0) & 1.0 & $<$13.0 & \ding{55}\\
& & CQ Tau & 0.77$\times$0.56 (-37.0) & 2.7 & $<$53.0 & \ding{55}\\
& & MWC 758 & 0.85$\times$0.54 (-43.0) & 2.7 & $<$46.0 & \ding{55}\\
\hline 

\ce{CH_3CN} & $J=17_0-16_0$
& HD 97048 & 0.65$\times$0.39 (5.0) & 2.6 & 59.0$\pm$21.0 & \textbf{?}\\
& & HD 135344B & 0.84$\times$0.63 (-90.0) & 1.2 & $<$27.0 & \ding{55}\\
& & HD 169142 & 0.54$\times$0.4 (-86.0) & 1.3 & $<$27.0 & \ding{55}\\
& & HD 100453 & 0.66$\times$0.63 (-67.0) & 1.2 & 43.0$\pm$6.0 & \ding{51}\\
& & CQ Tau & 0.77$\times$0.56 (-37.0) & 2.6 & $<$50.0 & \ding{55}\\
& & MWC 758 & 0.85$\times$0.54 (-43.0) & 2.6 & $<$44.0 & \ding{55}\\
\hline 

\ce{CH_3OCHO} &   $J=27-26$ blend   
& HD 97048 &  0.62$\times$0.45 (16.0) & 1.6 & $<$35.0 & \ding{55} \\
& & HD 135344B & 0.6$\times$0.44 (-89.0) & 2.6 & $<$84.0 & \ding{55} \\
& & HD 169142 & 0.58$\times$0.46 (-82.0) & 1.0 & $<$18.0 & \ding{55} \\
& & HD 100453 & 0.52$\times$0.46 (68.0) &1.24 & 38$\pm$5.0  & \ding{51} \\ \hline 

\ce{CH_3OCH_3} & $J=16_{1,16}-15_{0,15}$ blend     
& HD 97048  & 0.63$\times$0.45 (16.0) & 1.6 & $<$36.0  & \textbf{?}\\
& & HD 135344B &   0.6$\times$0.43 (90.0) & 1.2 & $<$38.0 & \ding{55}\\
& & HD 169142 &  0.58$\times$0.45 (-82.0) & 1.0 & $<$18.0 & \ding{55}\\
& & HD 100453 & 0.51$\times$0.46 (66.0) & 1.2 & $<$22.0& \ding{55}\\ \hline 

HDO & $J=6_{2,5}-5_{3,2}$
& HD 97048 & 0.64$\times$0.39 (5.0) & 2.9 & $<$70.0 & \ding{55}\\
& & HD 135344B & 0.84$\times$0.63 (-89.0) & 1.4 & $<$31.0 & \ding{55}\\
& & HD 169142 & 0.54$\times$0.39 (-86.0) & 1.4 & $<$28.0 & \ding{55}\\
& & HD 100453 & 0.66$\times$0.63 (-67.0) & 1.3 & $<$17.0 & \ding{55}\\
& & CQ Tau & 0.77$\times$0.56 (-38.0) & 2.8 & $<$54.0& \ding{55}\\
& & MWC 758 & 0.84$\times$0.54 (-43.0) & 2.9 & $<$50.0 & \ding{55}\\
\hline 

SiO & $J=7-6$
& HD 97048 & 0.6$\times$0.43 (15.0) & 1.6 & $<$37.0 & \ding{55}\\
& & HD 135344B & 0.58$\times$0.42 (-90.0) & 1.2 & $<$39.0 & \ding{55}\\
& & HD 169142 & 0.56$\times$0.44 (-78.0) & 1.0 & $<$19.0 & \ding{55}\\
& & HD 100453 & 0.49$\times$0.45 (68.0) & 1.2 & $<$21.0 & \ding{55}\\
\hline 

\insertTableNotes

\end{longtable*}
\label{table_images}
\end{ThreePartTable}

\section{Results} \label{sec:results} 
\subsection{Disk integrated line fluxes}

\begin{figure*}
\centering
\includegraphics[width=0.95\hsize]{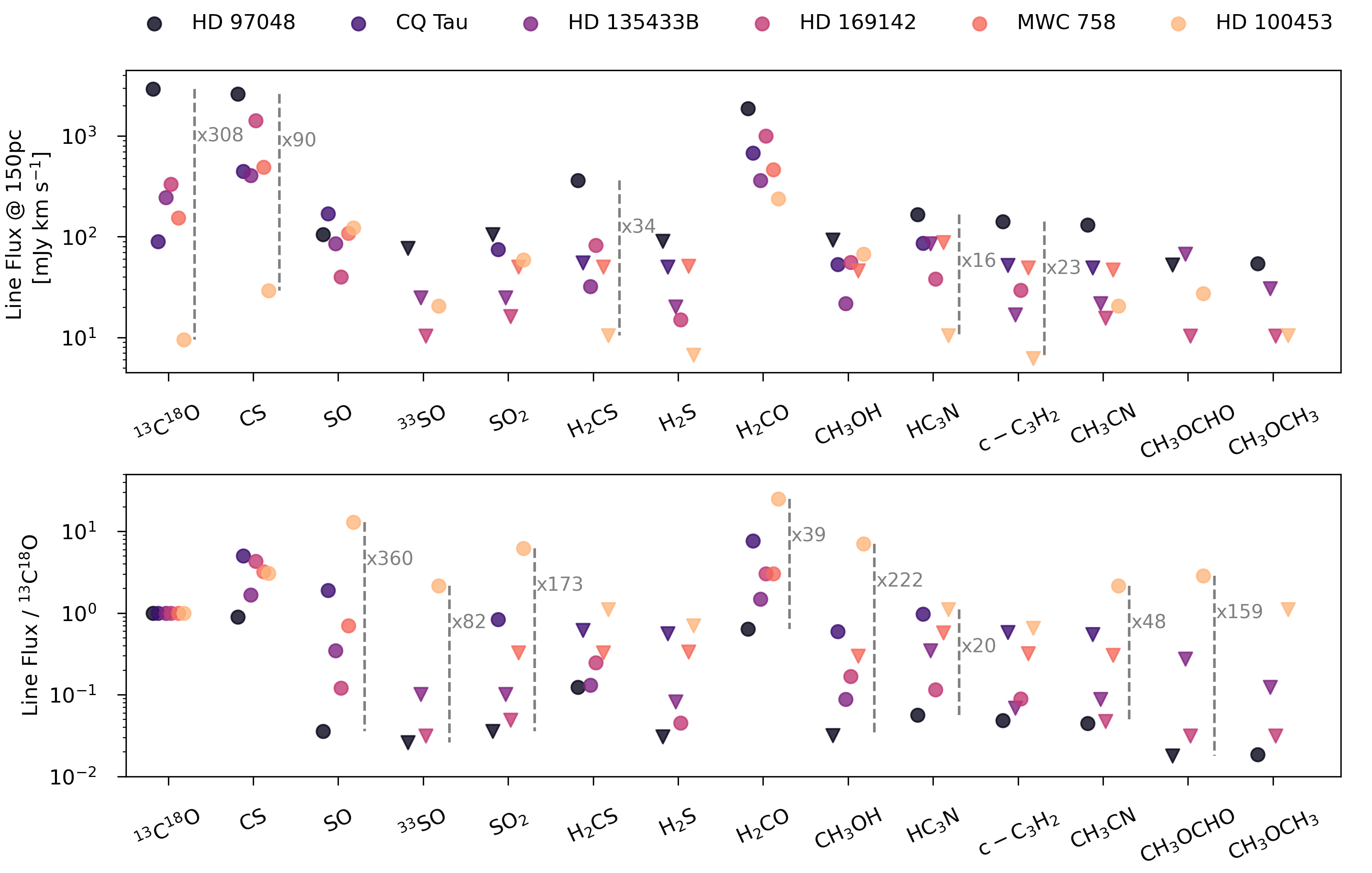}
\caption{Top: Disk-integrated fluxes for the molecules detected in at least one disk across the sample. Bottom: Disk-integrated fluxes relative to the \ce{^{13}C^{18}O} J=3-2 line flux for each disk. The disks listed in order of decreasing dust mass as listed in Table~\ref{table1}.
Both detections and tentative detections are shown as circles whereas the triangles are 3~$\sigma$ upper limits - as listed in Table~\ref{table_images}.
The dashed lines and numbers note the factor difference between the highest flux detection and lowest detection or upper-limit where this difference is greater than one order of magnitude. For most of the lines, the $\pm$1$\sigma$ error bars are smaller than the plot markers.}
\label{fig:fluxes}
\end{figure*}

To make a first comparison of the observable chemistry in our target disks we compare the disk-integrated line fluxes for the molecules detected in at least one disk as listed in Table~\ref{table_images}. These are shown in Figure \ref{fig:fluxes} where the line fluxes are normalized to a common distance of 150~pc and the disks are listed in order of decreasing dust mass (as listed in Table~\ref{table1}). For species with multiple lines targeted, we show the brightest transition detected in the survey. For the CS in the CQ~Tau and MWC~758 disks, we convert the $J=7–6$ line fluxes to $J=6–5$ assuming a line ratio consistent with optically thin emission in LTE with an excitation temperature of 25~K (as determined for CS in the HD~163296 disk by \citealt{2025ApJ...985...84L}). Similarly, for \ce{H_2CS} and \ce{HC_3N}, the brighter lines are in setting A, so those fluxes are converted for CQ~Tau and MWC~758 assuming the same temperatures for CS and \ce{H_2CS} and, a slightly warmer temperature of 40~K for \ce{HC_3N} (consistent with that seen for the Herbig disks HD~163296 and MWC~480 in \citealt{2021ApJS..257....9I}). Additionally, the \ce{^{33}SO} lines are located at the edge of a spectral window and are unfortunately missed in the CQ~Tau and MWC~758 data sets.

From the comparisons in Figure~\ref{fig:fluxes}, there is a clear spread in the disk-integrated fluxes across many of the detected molecules. 
If we are in the optically thin regime these fluxes can be used as a proxy for the column density of these species, but if some of the lines are optically thick then this does not hold. In the optically thin case we still require an assumed gas temperature to map to a number of molecules which for a temperature range of 20-100~K will change the column density of molecules by up to a factor of a few. We assume the gas is optically thin for discussions that follow but are cognizant that this simplification may not be the case for every molecule or disk.
The most significant spread is for \ce{^{13}C^{18}O} and CS, which span $\approx$2 orders of magnitude each where the former should trace the bulk CO gas mass of the disks. Interestingly the organics and carbon chains, \ce{H_2CS}, \ce{HC_3N} and \ce{c-C_3H_2} show at least 1 order of magnitude spread whereas the SO and \ce{CH_3OH} detections are much more clustered as well as the \ce{H_2CO}. This clustering may indicate that the emitting column of low C/O tracers SO and \ce{CH_3OH} is similar across the sample whereas the presence of the high C/O tracers, \ce{HC_3N} and \ce{c-C_3H_2}, depend more on the total mass and size of the disk. 

Some of the variations from disk to disk may be expected given the different disk masses of each system and in particular, the dust masses from \citet{2022A&A...658A.112S} listed in Table~\ref{table1} span almost an order of magnitude. 
There will also be variations due to the excitation temperature of the disk gas but without additional transitions we cannot test by measuring rotational temperatures for most molecules in the sample. To investigate the disk mass hypothesis, we also show in Figure \ref{fig:fluxes} the line fluxes normalized to the \ce{^{13}C^{18}O} line flux, which should be a good proxy for the optically thin CO gas mass in each disk \citep[e.g.;][]{2017NatAs...1E.130Z}. For the line ratios, the spread reduces significantly for CS as well as \ce{H_2CS}, \ce{HC_3N} and \ce{c-C_3H_2} — to less than one order of magnitude — but for the rest of the species, in most cases, the differences become more disparate. Many of the oxygen-bearing molecules show at least an order of magnitude spread with SO, \ce{SO_2} and \ce{CH_3OH} showing the most variation. These differences were also shown in \citet{2024AJ....167..165B} when comparing the HD~100546 and IRS~48 transition disks. Even though both of these disks have detections of SO isotopologues and \ce{CH_3OH}, the latter disk is $>50$ times brighter in terms of relative line fluxes of these species. 
 
From these comparisons it is clear that if the same reservoir of oxygen-rich gas relative to CO was present across all of the disks, species like \ce{CH_3OH} and \ce{SO_2} would be detected across the sample. These results therefore point towards the presence of species like SO, \ce{SO_2} and \ce{CH_3OH} (and \ce{CH_3OCHO}), being independent, or at least not strongly positively correlated to the total reservoir of gas. This makes sense given that \ce{^{13}C^{18}O} will probe the bulk disk gas whereas these low C/O tracers are only expected to be in the gas-phase on smaller physical scales. It is also important to note that optically thick dust may obscure emission on scales of a few to 10's of au \citep[e.g.;][]{2021A&A...646A...3L}. Interestingly, HD~100453 which has the lowest \ce{^{13}C^{18}O} line flux is notably relatively brighter in all of the other molecules (where detected) when compared to the other disks. Given the bright COMs emissions and warm $>100$~K gas in this disk \citep{2025arXiv250414023B} the low relative \ce{^{13}C^{18}O} J=3-2 line flux may also reflect a higher average CO gas temperature rather than just a lower total disk gas mass.
In contrast, HD~97048, which has the brightest \ce{^{13}C^{18}O} emission, appears relatively weaker line fluxes of other molecules. For some molecules, the non-detections are due to a lack of sensitivity, in particular, the detection of \ce{H_2S} in HD~169142 is below the relative line flux upper-limits of the other five disks and the same can be said for some of the non-detections of \ce{HC_3N}, \ce{c-C_3H_2} and \ce{CH_3OCHO} in some of the sources.

\subsection{Emission maps across the sample}

\begin{figure*}
     \centering
     \vspace{0.75cm}
    \includegraphics[width=\linewidth]{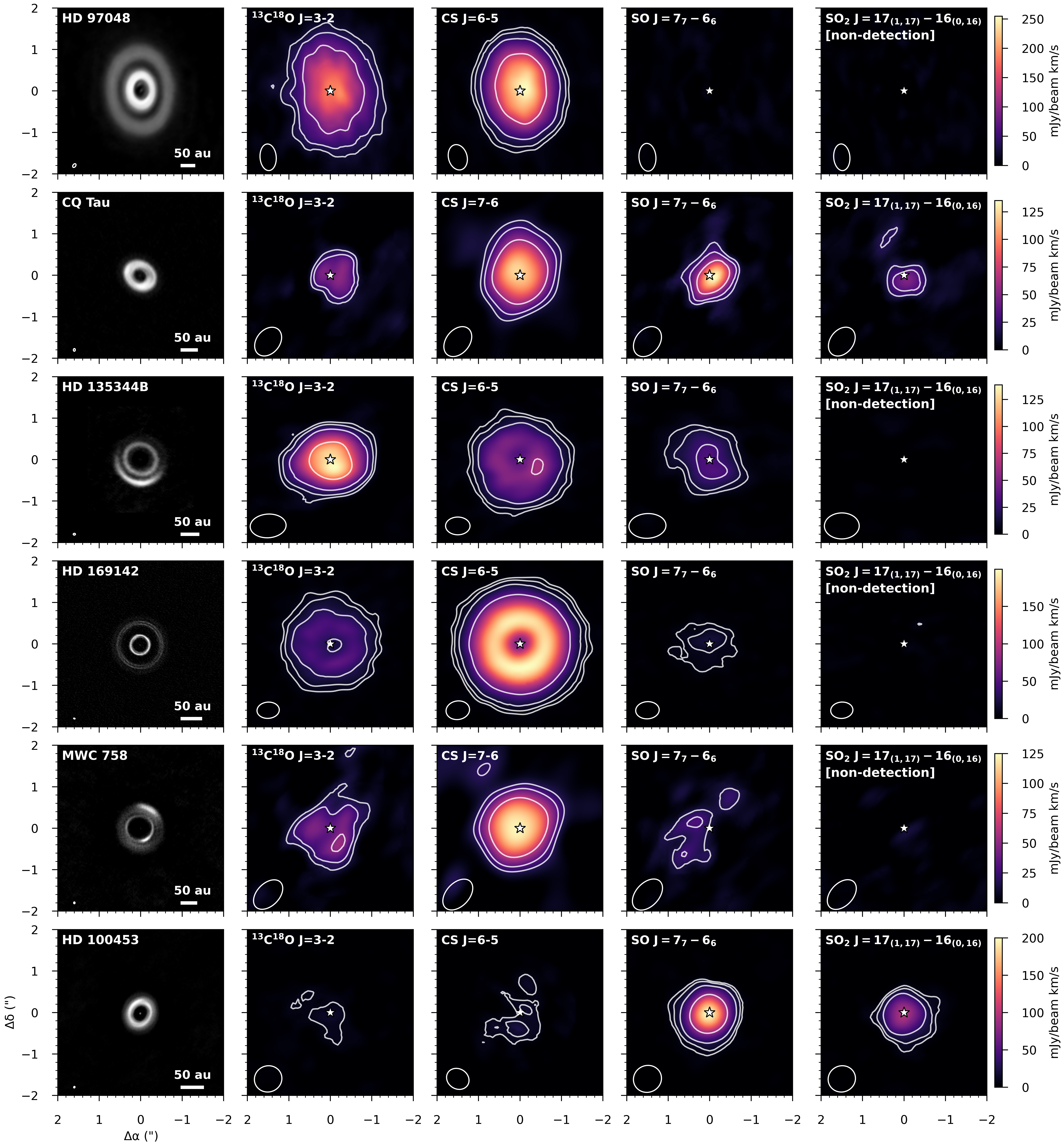}
    \caption{Integrated intensity maps of the dust continuum emission (taken from \citealt{2018ApJ...860..124D, 2018A&A...619A.161C, 2019NatAs...3.1109P, 2019MNRAS.486.4638U, 2020MNRAS.491.1335R, 2019AJ....158...15P}) and molecular line emissions of \ce{^{13}C^{18}O}, CS, SO, and \ce{SO_2} are presented. Each row of line maps shares a common color bar scale and the contours indicate the 3, 5, 10, and 30~$\sigma$ levels of the integrated line emission. A scale bar in the leftmost column represents 50~au and the position of the central star is marked with a star symbol. Non-detections and tentative detections are noted. Note that the CS J=7-6 data taken from \citep[][]{2025ApJ...984L...6T}.} 
    \label{maps1}
\end{figure*}

\begin{figure}
    \centering
    \includegraphics[width=\linewidth]{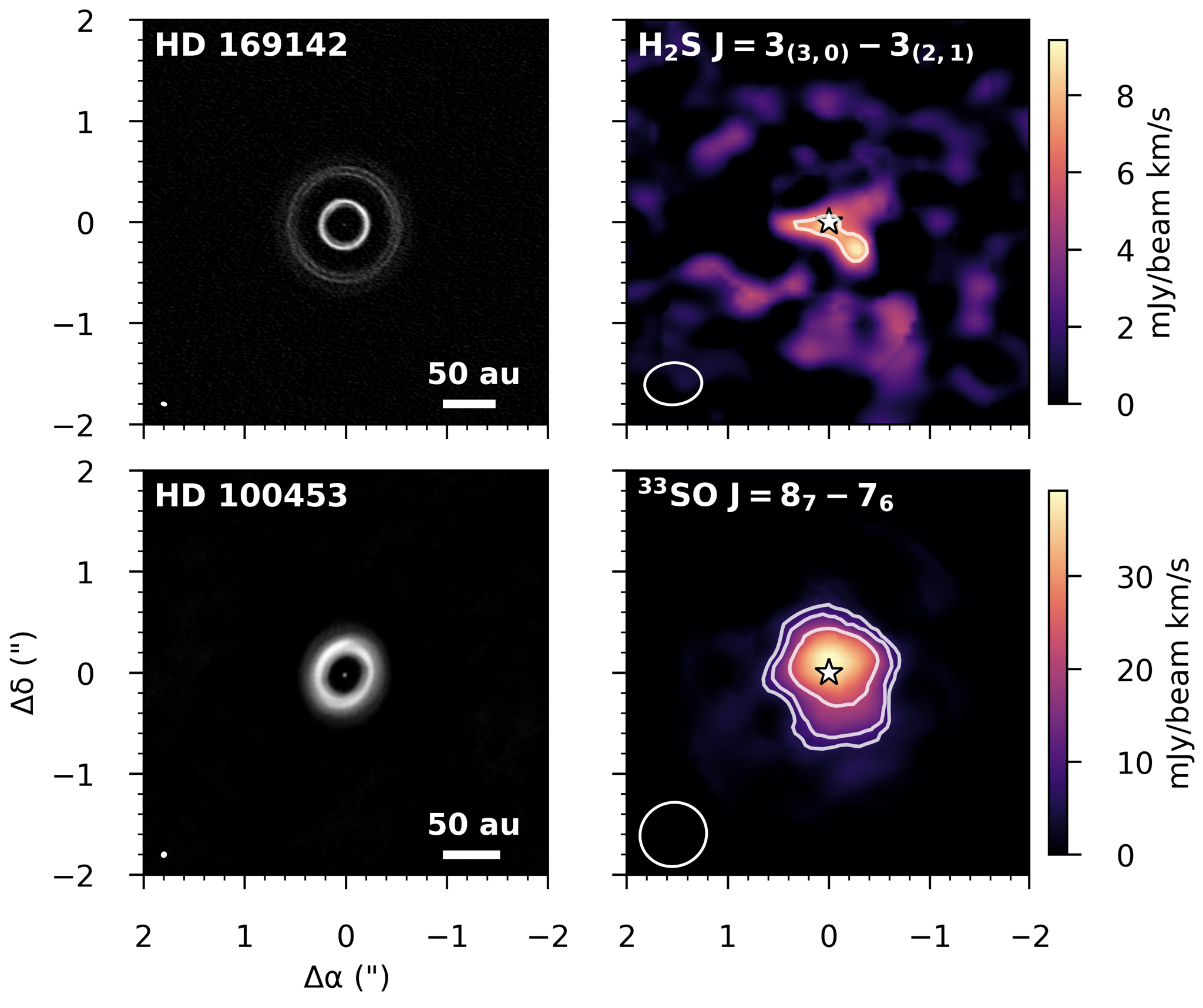}
    \caption{Same as for Figure~\ref{maps1} for \ce{H_2S} in HD~169142 and \ce{^{33}SO} in HD~100453.}
    \label{fig:h2s}
\end{figure}

\begin{figure*}
     \centering
     \vspace{0.5cm}
    \includegraphics[width=\linewidth]{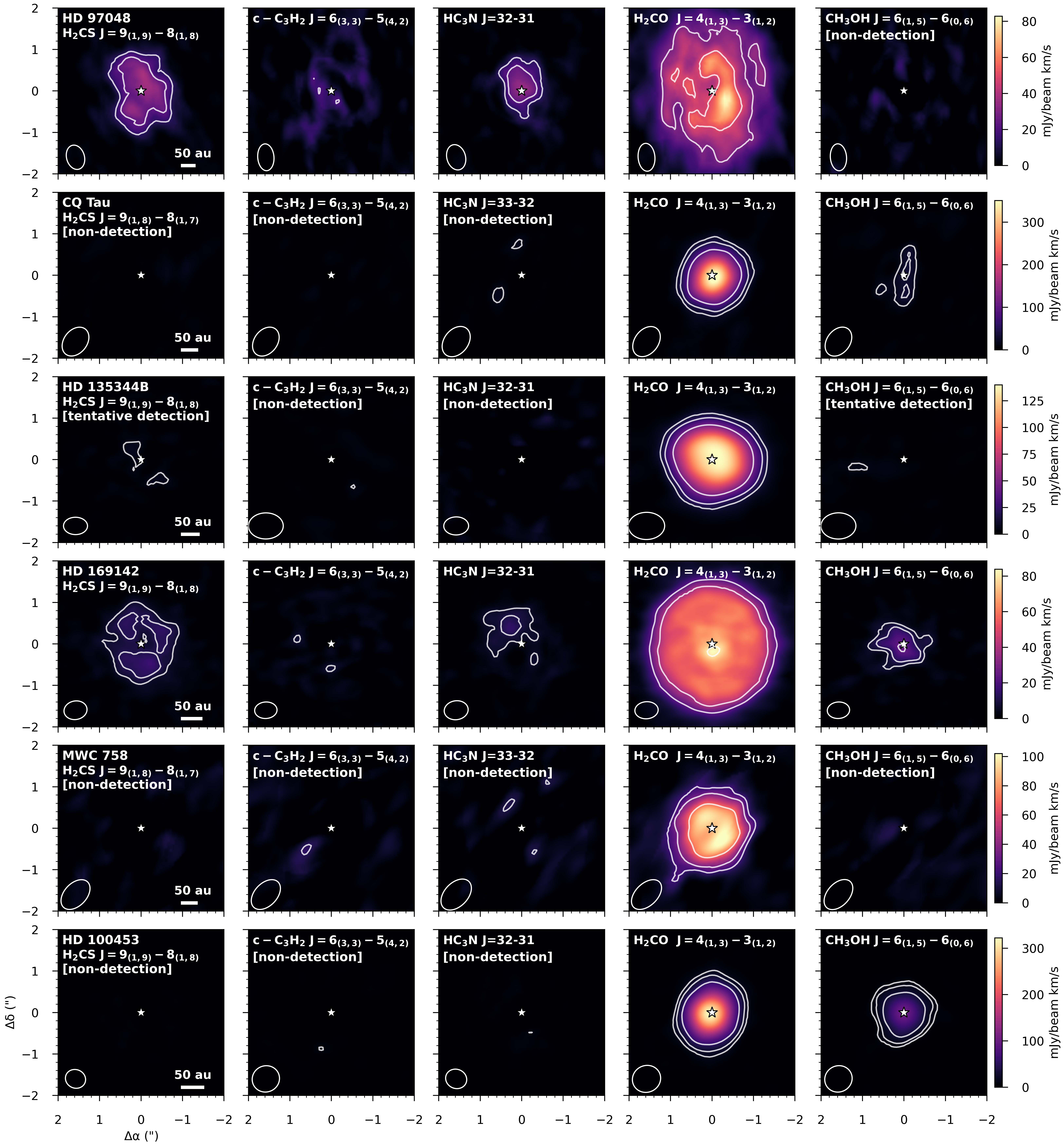}
    \caption{Same as for Figure~\ref{maps1} for \ce{H_2CS}, \ce{H_2CO}, \ce{c-C_3H_2}, \ce{HC_3N} and \ce{CH_3OH}. Note the transitions in each column are not all the same as where possible the brighter detected line was chosen.}      
    \label{fig:maps2}
\end{figure*}

% Here, we highlight the main features in the Keplerian-masked integrated intensity maps of the molecules detected in the survey. 

Figure~\ref{maps1} presents the moment maps of the simple molecules: \ce{^{13}C^{18}O}, CS, SO, and \ce{SO_2}. The color maps are all set to the same intensity scale for each disk to emphasize that in some sources, SO is the brightest species, whereas in others it is CS or \ce{^{13}C^{18}O}. In 4 out of 6 disks, \ce{^{13}C^{18}O} exhibits a ring-like morphology, likely highlighting gas depletion within the central dust cavities. In the remaining two disks, HD~97048 and HD~100453, the absence of a clear ring may be due to the large beam size compared to the dust cavity size and/or in the case of HD~100453 may be due to the relatively low signal-to-noise of the detection. The emission from SO (and \ce{SO_2}, detected in the two SO-brightest disks) is generally more compact and centrally peaked, whereas CS is more radially extended and sometimes resolves into a ring. When both CS and SO are clearly detected, they are most often spatially offset —as seen most clearly in HD~169142—though in other disks, radial and azimuthal offsets are less distinct given the resolution of these data. Although detected in the disk integrated flux, clear SO emission is not present in the HD~97048 map suggesting weak extended emission rather than compact emission associated with the cavity wall, but this will be investigated further in Section~4.3. Additional S-species detections are shown in Figure~\ref{fig:h2s} where compact $>$3~$\sigma$ emission from \ce{H_2S} is detected in HD~169142 while \ce{^{33}SO} is only detected in HD~100453. The disks with the brightest SO and \ce{SO_2} emissions therefore do not show \ce{H_2S}, which is interesting given that \ce{H_2S} is more abundant than both SO and \ce{SO_2} in warm proto-stellar gas and cometary ices \citep[e.g.;][]{2018MNRAS.476.4949D}. 

\begin{figure*}
     \centering
     \vspace{0.5cm}
    \includegraphics[width=\linewidth]{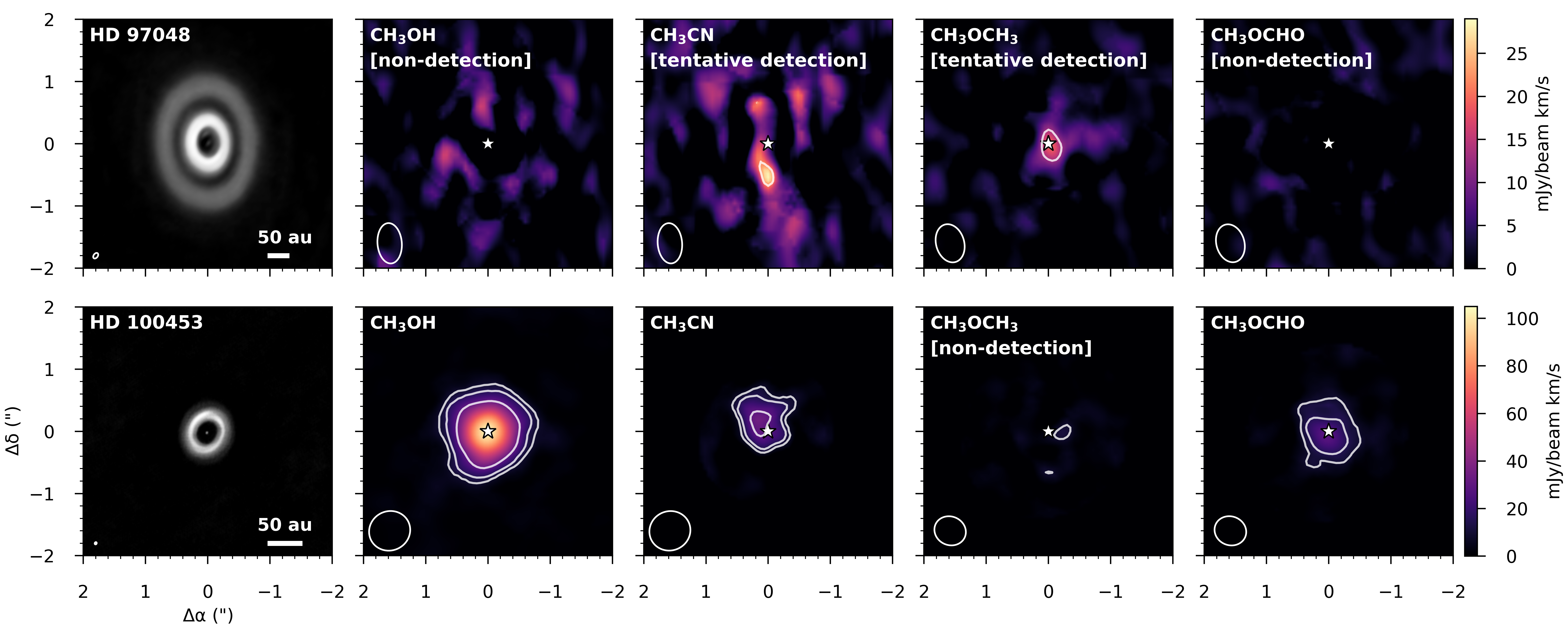}
    \caption{Same as for Figure~\ref{maps1} for \ce{CH_3OH}, \ce{CH_3CN}, \ce{CH_3OCH_3} and \ce{CH_3OCHO} in HD~97048 and HD~100453.} 
    \label{fig:maps4}
\end{figure*}

Figure~\ref{fig:maps2} presents the moment maps of the organics, hydrocarbons and some of the complex organic molecules targeted in the survey: \ce{H_2CS}, \ce{H_2CO}, \ce{c-C_3H_2}, \ce{HC_3N} and \ce{CH_3OH}.
In all cases, \ce{H_2CO} is the brightest and most radially extended of these species and in some disks there are sub-structures. In HD~97048, the \ce{H_2CO} emission appears ringed whereas in HD~169142 the radial emission structure appears relatively smooth. In comparison, there are hints of azimuthal asymmetries in \ce{H_2CO} in HD~135344~B and a central cavity in the MWC~758 disk.  
Similar to CS, where detected \ce{H_2CS} also appears ring-like and \ce{HC_3N} detections have a similar radial extent to the \ce{H_2CS} lines but appear azimuthally asymmetric.
There is weak and extended \ce{c-C_3H_2} emission in HD~169142 that is significant in the disk-integrated flux but not the images (note that \citealt{2023A&A...678A.146B} report a detection of extended and ringed \ce{c-C_3H_2} in the HD~169142 disk). Overall, the three \ce{CH_3OH} detections are relatively compact, similar to SO, and are much smaller in extent compared to the mm-disk and the \ce{H_2CO} gas. 
We also note a tentative detection of \ce{CH_3OH} in HD~135344~B discussed in Section 4.3. Among the larger COMs, as shown in Figure~\ref{fig:maps4}, only compact emission from \ce{CH_3CN} is confidently detected in HD~100453, with a tentative off-source peak in HD~97048. For \ce{CH_3OCH_3} and \ce{CH_3OCHO}, there are only on-source detections in HD~97048 and HD~100453, respectively. The significance of these detections will be discussed in Section 4.3.

\begin{figure*}[t!]
    \centering
    \includegraphics[width=\linewidth]{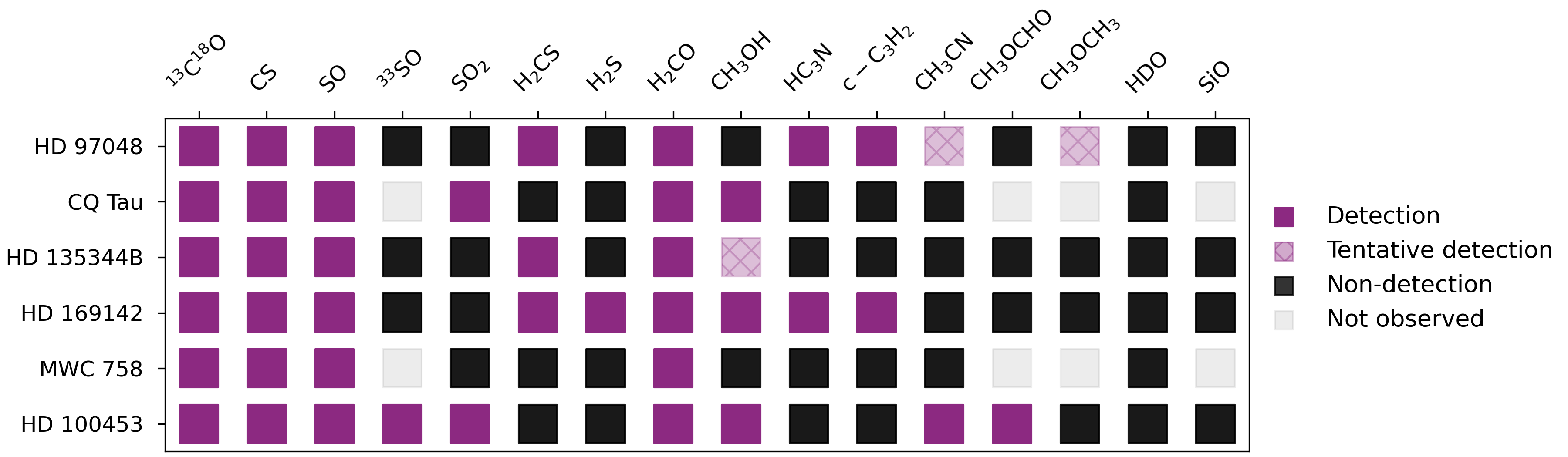}
    \caption{Summary of molecules detected, tentatively detected and not detected across the sample. Note that not observed refers to the molecules only present in setting A where observations of CQ~Tau and MWC~758 were not taken and the CS detections in these two disks are from \citet{2025ApJ...984L...6T}.}
    \label{fig:detections}
\end{figure*}

\subsection{Molecule detection summary}

In Figure~\ref{fig:detections}, we visualize our molecule detections in each disk. A thorough discussion on presence of weak lines and tentative detections in the data is provided in Appendix~B where we use both matched filtering \citep{Loomis2018} and the spectral stacking tool \textit{GoFish} \citep{GoFish}. We find that \ce{^{13}C^{18}O}, CS (when including archival data), \ce{H_2CO} and SO are detected in all six disks. The SO in HD~97048 appears to be distinct to the other systems and exhibits weak extended emission likely co-spatial with CS - as seen in the outer ring of the HD~100546 disk \citep{2024AJ....167..164B} - rather than more compact emission associated with the central dust cavity. Generally, the presence of \ce{SO_2} and \ce{H_2CS} in a disk appears to be anti-correlated and \ce{^{33}SO} is only present in the disk with the brightest SO. \ce{H_2S} is only detected in HD~169142 adding to the very few detections of \ce{H_2S} in Class II disks \citep{2018A&A...616L...5P,2022A&A...665A..61R}.  The carbon chains \ce{HC_3N} and \ce{c-C_3H_2} co-occur in disks with \ce{H_2CS}. Among the larger oxygen-bearing organics, \ce{CH_3OH} is present in 4/6 disks and potentially 5/6 when including a tentative detection. Although exhibiting SO emission, \ce{CH_3OH} is not detected in MWC~758 but this may just be due to the sensitivity limitations of current data. Interestingly, \ce{CH_3CN} is detected in both the most oxygen-rich (i.e., brightest SO and \ce{SO_2}) and carbon-rich (i.e., brightest CS and \ce{c-C_3H_2}) disks in the sample: HD~100453 and HD~97048, respectively. While for the larger COMs, \ce{CH_3OCHO} is detected only in HD~100453 and \ce{CH_3OCH_3} is tentatively detected only in HD~97048. Finally, we also report non-detections of SiO and HDO where observed. 

\section{Discussion} \label{sec:discussion} 

\subsection{Linking line fluxes to bulk system properties}

In order to try and understand the observed chemical diversity across our sample we compare the disk-integrated line fluxes to an array of bulk system properties that could influence the observable disk chemistry. We consider the host star mass, stellar luminosity, mass accretion rate, the disk-integrated 1.3 mm and \ce{C^{18}O} J=2–1 fluxes, and the size of the millimeter dust disk. All of these values are taken from \citet{2021A&A...650A.182G}, \citet{2022A&A...658A.112S}, \citet{2024A&A...682A.149S}  For this comparison, we use the following lines: $\mathrm{^{13}C^{18}O}$ $J=3-2$, CS $J=6-5$, SO ${J=7_7-6_6}$, $\mathrm{H_2CO}$ ${J=4_{(1,3)}-3_{(1,2)}}$, $\mathrm{CH_3OH}$ ${J=6_{(1,5)}-6_{(0,6)}}$, $\mathrm{HC_3N}$ J=32–31, $\mathrm{CH_3CN}$ ${J=17_{0}-16_{0}}$, and \ce{C_2H} $\mathrm{N=2–1}, J=\frac{9}{2}-\frac{7}{2}$. Although the latter is not included in our survey, there is coverage of \ce{C_2H} lines in 2 out of the 6 sources in our sample. We also include flux measurements from other Herbig disks in the literature where available and these are listed in Table~\ref{table:literature_data}. Where identical transitions have not been targeted, we convert fluxes to match our selected transitions by assuming line ratios consistent with optically thin emission and a specified excitation temperature as detailed in Appendix~B. The total sample spans a range of host star masses from approximately 1.5 to 2.8~$\mathrm{M_{\odot}}$, and more than an order-of-magnitude range in stellar luminosity, mass accretion rate, and disk-integrated 1.3 mm and \ce{C^{18}O} $J=2-1$ fluxes. We also report the Pearson correlation coefficient (r) and p-value, calculated using the \texttt{pearsonr} function from \texttt{scipy.stats} \citep{2020SciPy-NMeth}, to test for significant linear correlations between the variables where, for this analysis, we only consider detections and exclude upper limits.

\begin{figure*}[t!]
\centering
    \includegraphics[width=0.98\linewidth]{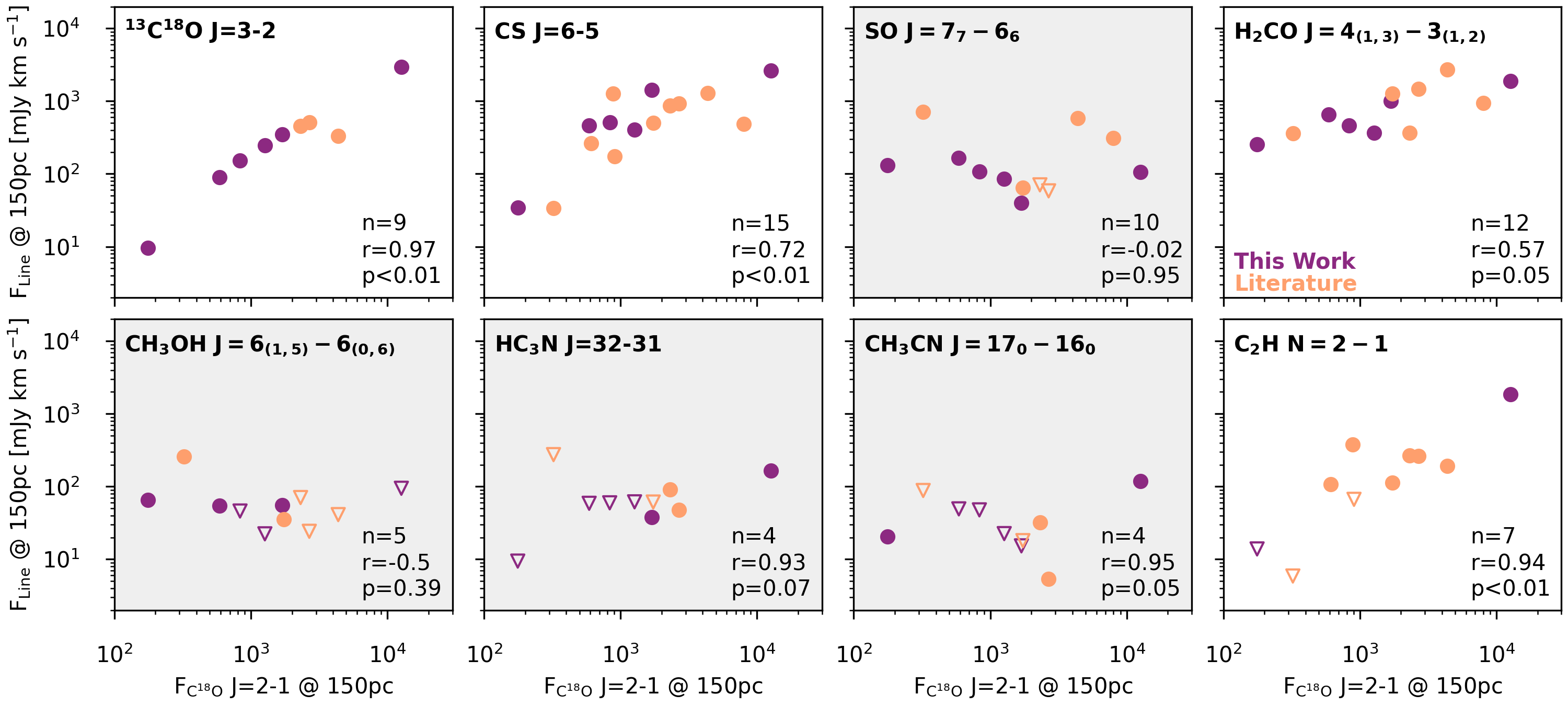}
\caption{Disk integrated line fluxes normalized to 150~pc and compared to the \ce{C^{18}O} J=2-1 line. The colored points are the disks within this survey and the translucent points are other Herbig disks from the literature (as described in Section 4.2). The triangles denote 3~$\sigma$ upper-limits. Numbers in the bottom right of each panel show the resulting Pearson correlation coefficient (r) and p-value for each set of variables.}
    \label{fig:scatter_c18o}
\end{figure*}

Here we discuss the main takeaways when comparing to the gas mass reservoir in the outer disk as this resulted in some of the strongest correlations. In Figure~\ref{fig:scatter_c18o} we show the relationship between different molecules and the \ce{C^{18}O} J=2–1 line flux, all scaled to 150~pc, where \ce{^{13}C^{18}O} shows the clearest linear correlation with the \ce{C^{18}O}. This result is expected as \ce{^{13}C^{18}O}, like \ce{C^{18}O}, traces the bulk CO gas mass of the disk. There is one slight outlier to the linear trend and this is the HD~142527 disk where, likely the large central gas cavity results in a lower \ce{^{13}C^{18}O} line flux relative to \ce{C^{18}O} \citep{Temmink2023}. Looking at the ratio of \ce{C^{18}O} to \ce{^{13}C^{18}O} (after scaling the $J=3-2$ flux to $J=2-1$ assuming 30~K) we find a range across the sample from $\approx$13 to 56 with HD~100453 having the highest and HD~97048 the lowest. This ratio does assume the same emitting area for the two isotopologues which is likely not correct, but 
these values are all lower than the ISM isotope ratio of $\approx$69 \citep{1999RPPh...62..143W} indicating that the \ce{C^{18}O} $J=2-1$ line is at least partially optically thick in all of these disks. 

For the simple molecules known to trace high C/O ratios we find that both CS and \ce{C_2H} correlate positively with the gas mass in the outer disk as traced by \ce{C^{18}O}. In the images these species trace a similar radial extent to the \ce{^{13}C^{18}O} emission and therefore will track the bulk gas reservoir in the disk to some degree, as is broadly consistent with predictions from disk chemical models \citep{2010ApJ...722.1607W,2018A&A...616A..19A,2018A&A...609A..93C}.    
Interestingly, there are hints at correlations between \ce{CH_3CN} and \ce{HC_3N} with  \ce{C^{18}O}, where these species, like \ce{C_2H}, have been shown to trace high C/O gas \citep{2023NatAs...7...49C}. But, as there are only four disks with \ce{CH_3CN} and \ce{HC_3N} detections we do not have the numbers to say anything concrete. The molecules SO and \ce{CH_3OH} which trace low C/O ratios do not show any significant correlation to the disk \ce{C^{18}O} line flux. Given the compact nature of the emission from SO and \ce{CH_3OH} in most disks, the presence and abundance of these species may depend more on the thermal structure of the disk on 10's of au scales, i.e., the emitting area of the low C/O gas, rather than the total mass reservoir. Additionally, \ce{H_2CO} does  show a potentially significant correlation (p=0.05) with \ce{C^{18}O} flux and the variation in fluxes for this species is maybe due to the mix of possible formation pathways for \ce{H_2CO} in disks. We would expect both gas-phase formation of \ce{H_2CO} in disks in addition to the thermal and non-thermal desorption of \ce{H_2CO} from grain surfaces \citep{2015ApJ...809L..25L,2020ApJ...890..142P,2021ApJ...906..111T, 2025arXiv250204957E}.
The lack of CO ice in some Herbig disks, a precursor for \ce{H_2CO} formation on the grains, may also play a significant role in the observable chemistry as discussed in detail in \citet{2023ApJ...948...57P}.

\begin{figure*}
\centering
    \includegraphics[width=0.85\linewidth]{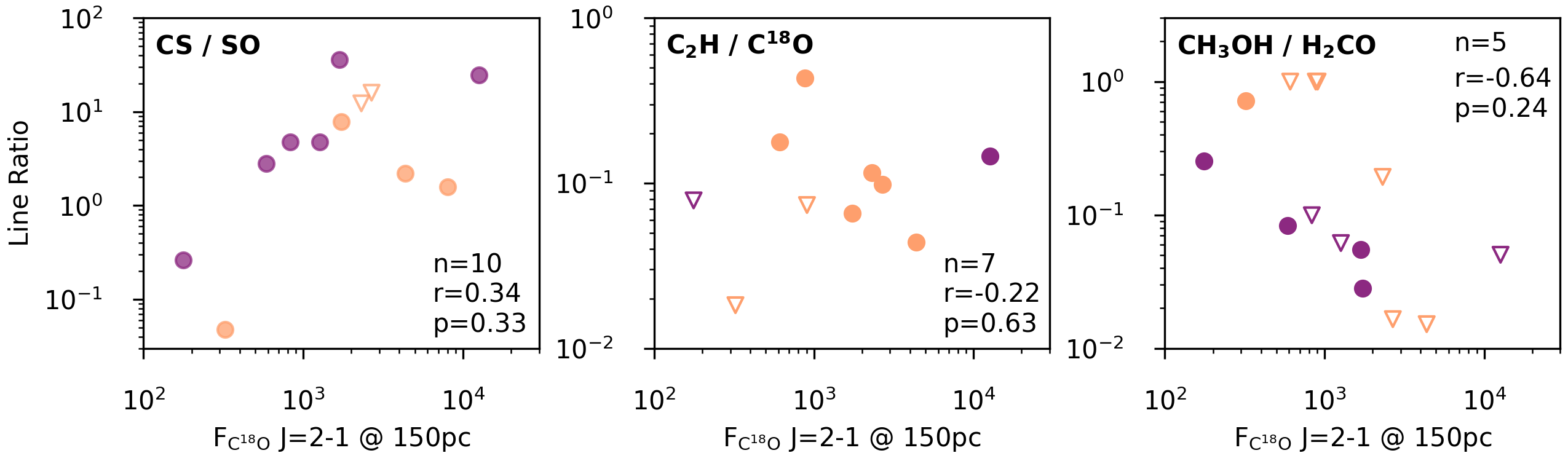}
\caption{Disk-integrated line flux ratios compared to the \ce{C^{18}O} J=2-1 line. The purple points are the disks within this survey and the orange points are from the literature (as described and listed in Appendix~B and Table~\ref{table:literature_data}). The triangles denote 3~$\sigma$ upper-limits. Numbers in the bottom right of each panel show the number of detections and the resulting Pearson correlation coefficient (r) and p-value for each set of variables where n notes the number of data points considered in the correlation test.}
    \label{fig:scatter_c18o_ratios}
\end{figure*}

We take these comparisons one step further and compute the line flux ratios of CS/SO, \ce{C_2H}/\ce{C^{18}O} and \ce{CH_3OH}/\ce{H_2CO}. These are shown in Figure~\ref{fig:scatter_c18o_ratios} plotted against the \ce{C^{18}O} line fluxes. Generally, lower mass disks have lower inferred bulk C/O ratios from the disk integrated CS/SO line flux ratio, although this result is not statistically significant. Since disk size and \ce{C^{18}O} line flux are correlated, this can be interpreted as smaller disks lacking cool ($<$50 K) gas in the outer disk where high C/O ratios are expected, therefore, the warm, low C/O gas dominates the observed emission. The \ce{C_2H}/\ce{C^{18}O} ratio is scattered, as seen by \citet{2020ApJ...890..142P}, but the lowest mass disks with bright emission from low C/O tracers - Oph-IRS~48 and HD~100435 - do show non-detections of \ce{C_2H}.
This is similar to the result from \citet{2021A&A...653L...9V} where 
bright \ce{C_2H} emission is linked to the mass of the dust reservoir beyond the CO snowline. Finally, there is an anti-correlation between \ce{CH_3OH}/\ce{H_2CO} and \ce{C^{18}O} where smaller disks are also more \ce{CH_3OH}-rich, where \ce{CH_3OH} will again, trace the warm ($>$100~K) gas reservoir.   

The complete results of this comparison across all six source properties are shown in Appendix Figure~\ref{fig:scatter_all} including the for \ce{C^{18}O} $J=2-1$ line which correlates with all other bulk properties considered. The \ce{^{13}C^{18}O} correlates with all properties but the the disk mm-flux. The simple molecules that trace high C/O gas show a mix of correlations where CS trends positively with all but the mass accretion rate and \ce{C_2H} all but the mm-flux and dust disk size. In comparison, \ce{H_2CO} trends particularly well with the mm-flux in addition to the host star mass, mass accretion rate and disk size. The low C/O tracers do not correlate with any bulk system property which is particularly notable for SO given we have 10 data points in the comparison. Additionally in Appendix~D we discuss and compare the line fluxes of \ce{CS}, \ce{H_2CO} and \ce{C_2H} from the Herbig population to the T-Tauri population as collated by \citep{2023ApJ...948...57P} and shown in Figure~\ref{fig:ttauris}. This is only a qualitative comparison, but shows that these are not two distinct chemical regimes and there is likely some chemical continuum across spectral types at least for these species.

\subsection{Linking low C/O tracers to disk sub-structures}

The variation we see in disk line fluxes for low C/O tracers across Herbig disks is not dependent on the properties of the host star, the disk mass or the disk size. The disk density and temperature structure and evolution is expected to impact the observable disk chemistry. Here we discuss any connections between the disk physical structure, i.e., central cavities, rings and spirals, and the detection or non-detection of SO and other co-occurring tracers of low C/O gas such as \ce{SO_2} and \ce{CH_3OH} which, are thought in be primary present within the water snowline.

SO emission has now been detected in 10 Herbig transition disks and seems to trace different chemical regimes. In all but HD~97048, AB~Aurigae and HD~142527, this molecule is found to trace gas inside or at the edge of the disk's central cavity \citep[][this work, Temmink et al. (in prep.)]{2021A&A...651L...6B,Booth23a,2023ApJ...952L..19L,2025arXiv250616481Z}. In the specific case of AB~Aurigae, the SO emission can be linked to late in-fall onto the disk both driving gravitationally instability and generating a local ``hot-spot" of SO where this in-falling material meets the disk \citep{2025ApJ...981L..30S}. Additionally, the ring of SO emission is inferred to be cold $\approx$20~K \citep{Dutrey24} and may be similar to the outer ring in the HD~100546 system \citep{2024AJ....167..164B}. For HD~97048, the lack of SO at the cavity edge is puzzling. As listed in Table~\ref{table1}, the dust cavity in HD~97048 is 63~au and when compared to the other disks in the sample, and taking into consideration the stellar luminosities, warm SO emission would be expected to be present in the HD~97048 disk - unlike the case for the HD~142527 disk where the dust trap edge is cool \citep{Temmink2023}. But, the lack of \ce{CH_3OH} in HD~97048 may indicate the temperatures here are not sufficient to liberate \ce{H_2O} from the dust grains which would be required to drive the gas-phase formation of SO; via the barrier-less reaction of S + OH \citep[e.g.,][]{2018A&A...617A..28S}. Additionally, or alternatively, we may need to consider the photo-dissociation timescales of these species given the high luminosity ($\approx$65$L_{\odot}$) of HD~97048 but disk specific modeling is required to test this hypothesis. 

Aside from HD~97048, all of the other Herbig transition disks with SO are also host to spiral arms in the near-infrared scattered light. This link between SO and dynamically perturbed disks is explored in detail by \citet[][see discussion and references therein]{2025arXiv250616481Z} who postulate that the local heating in the disk due to dynamical perturbations could drive the formation of SO in the disk either from the thermal desorption of S-rich ices and/or gas-phase chemistry. Higher angular resolution observations are needed to see if any SO emission can clearly be spatially linked to spirals but the lack of SO in HD~97048, a disk with rings in the scattered light \citep{2016A&A...595A.112G}, does provide more evidence to support this link. 

Finally, the presence of dust traps can reduce the efficiency of radial transport of dust and therefore ices through the disk \citep{2023arXiv231009077V}. This process has been shown to increase or decrease the local abundances of both CO and \ce{H_2O} within their respective sublimation fronts \citep{2019ApJ...883...98Z,2021ApJS..257....5Z,2020ApJ...903..124B,2023ApJ...957L..22B}.The same physics may also govern the observed gas-phase abundances of other ice-derived molecules in disks. 
Qualitatively, the disks with the brightest SO and \ce{CH_3OH} emissions are the most radially compact disks with a single dust trap: Oph-IRS~48 and HD~100453. When we go to disks such as HD~100546 and HD~169142, where there are multiple rings, these molecules are still present but are relatively dimmer. Then disks without central cavities and/or multiple dust traps, e.g, HD~163296 and HD~97048, do do not show this reservoir of low C/O gas. Therefore, the variations we see in disk chemistry may tie back to the disk physics, namely to efficiency of the radial transport and the presence disk sub-structures \citep{2021ApJ...921...84K,2023ApJ...954...66K}.

\subsection{Which processes set the COMs reservoir in disks?}

The thermal sublimation front(s) of COMs are observationally accessible in Herbig disks with ALMA, and there are now detections of \ce{CH_3OH} in five such disks which are all also SO-rich (six if the tentative detection in HD~135344B is included; \citealt{2021NatAs...5..684B, 2021A&A...651L...5V, 2023A&A...678A.146B, 2025arXiv250414023B} and this work). 
The presence of \ce{CH_3OH} (and isotopologues) in these relatively warm disks also provides compelling evidence for the inheritance of ices in disks from the earlier stages of star formation
\citep{2021NatAs...5..684B, 2025arXiv250414023B}.
These \ce{CH_3OH} detections are exclusively linked to Herbig transition disks, whereas detections of the other COMs, \ce{CH_3OCH_3} and \ce{CH_3CN} are not. 
As mentioned earlier, the in-situ formation of \ce{CH_3CN} in the disk can be enhanced in irradiated, C/O~$>$~1 environments \citep{2023NatAs...7...49C}, which may account for the variability in its detection statistics across disks, although, modeling is needed to confirm this. 

The variation in \ce{CH_3OCHO} and \ce{CH_3OCH_3} detections across \ce{CH_3OH}-rich—and, so far, \ce{CH_3OH}-poor—disks may also be due to in-situ disk chemical processes. This is evidenced by both the high abundances of these species relative to \ce{CH_3OH} in some disks and the variation in relative abundances of these two COMs when compared to other environments where this ratio is relatively constant \citep[e.g.;][]{2020A&A...641A..54C, 2023A&A...678A.137C}. 
\ce{CH_3OCHO} is detected in the transition disks Oph-IRS~48, HD~100546, and HD~100453 and, as discussed in \citet{2024AJ....167..165B}, the relative abundance of this species may be enhanced in the disk via the conversion of initially inherited \ce{CH_3OH} ice into more complex species via dissociation and subsequent radical–radical chemistry in the ice, which depend intricately on the input ionization  rate which may be system dependent \citep{2014A&A...563A..33W, 2014ApJ...790...97F}.  
 
Interestingly, aside from Oph-IRS~48, which has a unique molecular inventory, the detections of \ce{CH_3OCH_3} are in disks that show evidence for CO freeze-out and no abundant \ce{CH_3OH} gas \citep{2019A&A...629A..75B,2020ApJ...893..101L,2021ApJS..257....5Z}. 
\citet{2024ApJ...974...83Y} evoke the different binding energies and emitting layers in the disk of the different COMs relative to the optically thick dust surface, in addition to chemical processes, to explain the abundant \ce{CH_3OCH_3} in MWC~480 but lack of \ce{CH_3OH}. The detection of only \ce{CH_3OCH_3} in the HD~97048 transition disk adds further complexity to this problem as these sources have different inner disk structures. There is a pathway to form \ce{CH_3OCH_3} in the ice from \ce{CH_3} \citep[e.g.;][]{2014A&A...563A..33W, 2014ApJ...790...97F}, and \ce{CH_4} has a similar binding energy to CO \citep{2022ESC.....6..597M}. Therefore, \ce{CH_3OCH_3} could be preferentially enhanced in the colder and likely \ce{CH_4}-rich regions of both the MWC~480 and HD~97048 disks and then be transported to its sublimation front via radial drift. There is indeed evidence for radial drift locally enhancing the abundance of CO in the MWC~480 disk \citep{2021ApJS..257....5Z}. 
Even within this drift framework, the lack of \ce{CH_3OH} in both disks remains puzzling and may require evoking an in-situ chemical pathway to enhance the formation rate of \ce{CH_3OCH_3}, possibly in the gas-phase rather than the ices \citep{2022ApJS..259....1G,2023A&A...678A.137C}. 
This mix of COM detections and abundances in Herbig disks with different dust structures will be investigated further using gas-grain chemical models (Booth et al., in prep). Generally though, it is hard to draw robust conclusions and look for trends in COMs abundances with such a small sample size and given all of these disks are morphologically unique. 

% Perhaps a contribution to CH3OCH3 from gas-phase chemistry with higher UV? There is a nice discussion on CH3OCH3 chemistry in comparison with Garrod et al. (2022) models by Chen et al. (2023), her CoCCoA paper. Compared with other COMs, CH3OCH3 seems to have a larger gas-phase contribution

\subsection{Connecting disk chemistry to the giant exoplanet accretion reservoir}

\begin{figure*}[t!]
    \centering
    \includegraphics[trim={0cm 0cm 0cm 2.5cm},clip, width=0.85\hsize]{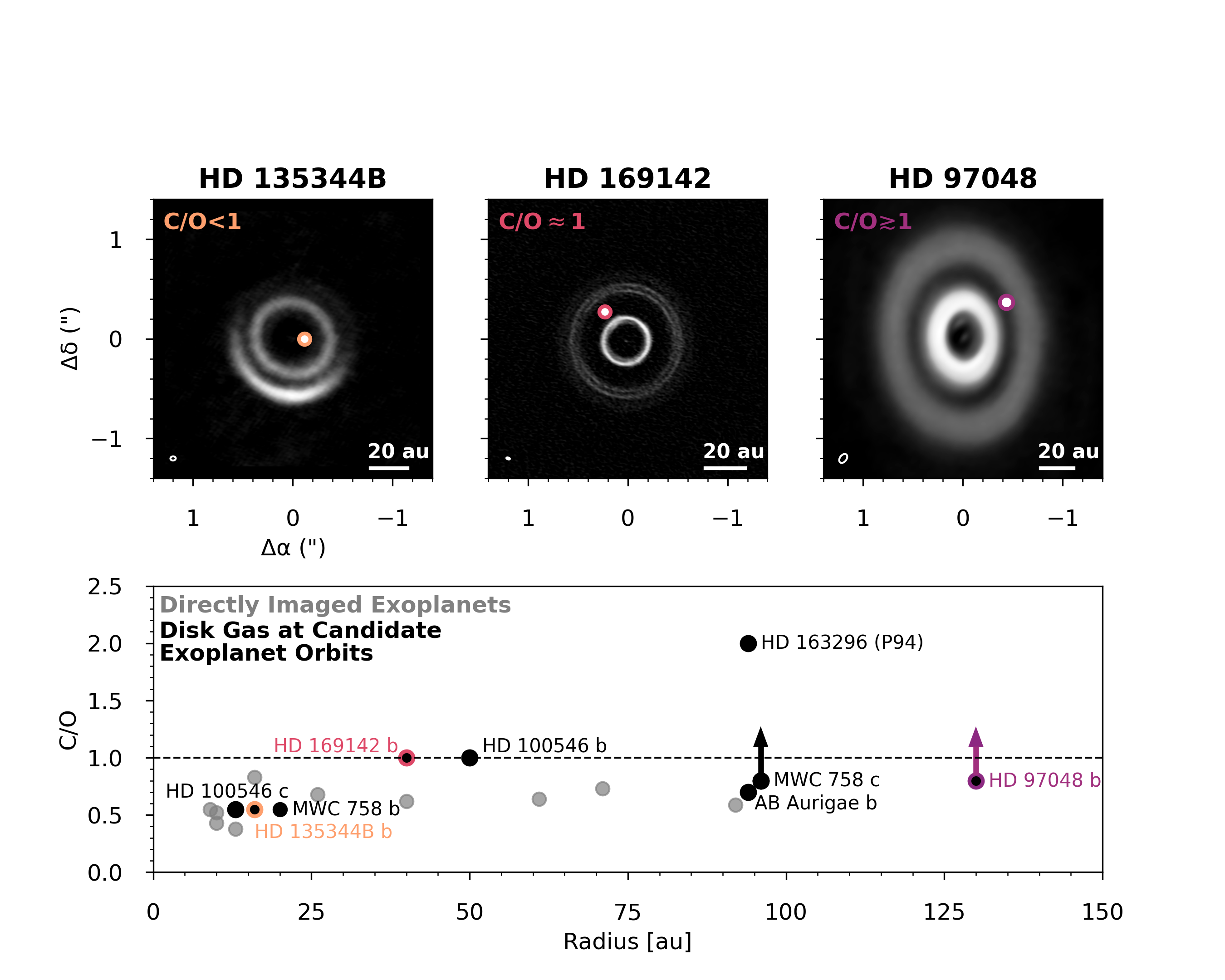}
    \caption{Top: Continuum images of the HD~135344B, HD~169142 and HD~97048 disks highlighting the locations of candidate giant exoplanets and the C/O of the disk gas at the approximate orbital location of these planets. 
    Bottom: A compilation of C/O ratios measured for directly imaged exoplanets around F, A and B-type stars in gray compared to the C/O constraints on the disk gas at the orbital radii of disk exoplanet candidates. The reported errors on exoplanet C/O are typically smaller than the plot markers and all exoplanet values, uncertainties and references are listed in Table~\ref{table:exoplanets}.
    % The exoplanets and references for this C/O ratios are: 
    % Beta Pic~b \citep[][]{2020A&A...633A.110G},
    % AF~Lep~b \citep{2025AJ....169...30B},
    % HD~206893~b \citep{2025AJ....169..175S},
    % 51~Eri~b \citep{2023A&A...673A..98B},
    % HR~8799~b, c, d, e \citep{2024A&A...687A.298N},
    % HD~95086~b \citep{2024A&A...690A.316M},
    % $\kappa$~And~b \citep{2024ApJ...970...71X}, and
    % HIP~65426~b \citep{2021A&A...648A..59P}.
    % For the disk exoplanet candidates we consider the systems and exoplanet candidates:  
    % HD~97048 (130~au; \citealt{2019NatAs...3.1109P}), HD~169142 (40~au; \citealt{2023MNRAS.522L..51H}), HD~135344B (16~au; \citealt{2025arXiv250706206S}) and MWC~758 (20~au and 96~au; \citealt{2023NatAs...7.1208W, 2019ApJ...883...37B}), HD~100546 (13~au and 50~au; \citealt{2019ApJ...883...37B,2015ApJ...814L..27C}), HD~163296 (86~au; \citealt{2018ApJ...860L..12T,2020ApJ...890L...9P, 2022ApJ...928....2I}) and AB~Aur (93~au; \citealt{2022NatAs...6..751C}).
    } 
    \label{exoplanets}
\end{figure*}

Chemical modeling has shown that the CS/SO ratio is sensitive to the underlying gas-phase C/O ratio in protoplanetary disks \citep{2018A&A...617A..28S,2020A&A...638A.110F,2021ApJS..257...12L}. In particular, radially anti-correlated CS and SO emissions can trace gas inside and outside of the \ce{H_2O} snowline \citep{2024MNRAS.534.3576K}. From the relatively more compact detections of SO compared to CS in 5 out of 6 disks, we can therefore infer that the former is tracing warm ($>$100 K) gas in a C/O~$<$~1 environment. This is a simple assumption and we note that disk specific modeling work is needed to recover the absolute C/H, O/H, and S/H ratios across disks \citep[e.g.;][]{2024MNRAS.534.3576K} to properly connect to the giant exoplanet population. 

A number of disks in our sample show evidence for embedded or accreting protoplanets, including HD~97048, HD~169142, HD~135344B, and MWC~758 \citep{2019NatAs...3.1109P, 2023MNRAS.522L..51H, 2025arXiv250706206S, 2023NatAs...7.1208W}. Using our CS, SO, and \ce{C_2H} measurements, we estimate the C/O regimes of the gas accessible to these planets.
For HD~169142~b, located beyond the \ce{H_2O} snowline, modeling indicates a gas-phase C/O$\approx$1 with only moderate C/H and O/H depletion—far less than the orders-of-magnitude depletion in Herbig disks such as HD~163296 and MWC~480 \citep{2021ApJS..257....7B}. A similar moderately high C/O reservoir may be available to the putative HD~97048~b at 130~au, which is co-spatial with CS (and weak SO) emission. However, the bright \ce{C_2H} and possible CO freeze-out in HD~97048 also allow for an elevated C/O$\approx$2, as seen in MWC~480 and HD~163296, so we conservatively infer C/O$\gtrsim$0.8 for any planet beyond the \ce{H_2O} snowline in this disk.
In contrast, HD~135344B~b orbits at 16~au inside the dust cavity, where SO tracing indicates a lower C/O$\approx$0.55 gas flowing through the cavity and available for accretion. In MWC~758, the candidate at $\sim$96~au would lie in a CS-rich region analogous to HD~97048, whereas the 20~au planet candidate \citep{2018A&A...611A..74R} occupies a lower C/O$\approx$0.55 environment similar to HD~135344B~b. In the subsequent comparison to exoplanets we additionally include candidates in HD~100546, HD~163296, and AB~Aur. In HD~100546, the proposed planets at 13~au and 50~au access gas with C/O$\approx$0.55 and 1.0, respectively \citep{2024MNRAS.528..388K, 2024A&A...687A.299L}. For HD~163296, the 94~au candidate—supported by multiple lines of evidence \citep{2018ApJ...860L..12T, 2020ApJ...890L...9P, 2022ApJ...928....2I}—resides in a C/O$\approx$2 regime \citep{2021ApJS..257....7B}. We exclude more distant or uncertain candidates (e.g., those at 48, 137, and 260~au) but note they would also fall within high-C/O gas. Finally, the $\sim$93~au candidate in AB~Aur accesses gas with C/O$\approx$0.7 \citep{2020A&A...642A..32R}.

In Figure~\ref{exoplanets} and compare the C/O constraints we have for disks to those for directly imaged exoplanets. For the exoplanets, we include only those detected via direct imaging around F-, A-, and B-type stars, with an upper limit on host star mass of 3~$\mathrm{M_{\odot}}$ and orbital radii less than 200~au. These 11 exoplanets were identified from the NASA Exoplanet Archive\footnote{\url{https://exoplanetarchive.ipac.caltech.edu/}} and the results from a C/O literature search are summarized in Table~\ref{table:exoplanets}.
Although relatively qualitative, Figure~\ref{exoplanets} shows clear overlap between the disk gas C/O constraints and the exoplanet atmospheric measurements particularly on scales $<$100~au. 
There are many caveats to this comparison (see \citet{2024ApJ...969L..21B} and \citet{2025arXiv250600669F} for lengthy discussions) but some deviations from the canonical picture from \citet{2011ApJ...743L..16O} include the migration history of the planet \citep{2022ApJ...937...36P}, the chemical evolution of the disk \citep{2020MNRAS.499.2229N}, the form of the accreted materials \citep{2023ApJ...952L..18C}, the potential for late-stage accretion \citep{2024ApJ...966...39S}, and the feedback of the forming planet on the disk \citep{2015ApJ...807....2C, 2025arXiv250703370C}. The main takeaway from our comparison is that, in the simplest scenario, this low C/O gas is available on 10's of au scales in the progenitor systems of giant exoplanet systems.

\section{Conclusion} \label{sec:conclusion} 

This paper presents the results of an ALMA molecular line survey of six transition disks around young F-, A-, and B-type stars significantly increasing the known molecular inventories of Herbig disks. Across this survey, along with detections reported by \citet{2025arXiv250414023B}, we identify 17 different molecular and isotopic species, ranging from the simple but rare CO isotope \ce{^{13}C^{18}O} to tracers of the gas-phase C/O ratio and a selection of large and complex organics. We also compare our sample to other Herbig disks from the literature to explore trends in line fluxes with system properties, examine the connections between disk structure and observable chemistry, and characterize the potential accretion reservoirs of giant planets forming in these disks. Our main conclusions are as follows:

\begin{itemize}
\item All disks in our survey show detections of \ce{^{13}C^{18}O}, CS, SO and \ce{H_2CO}, but beyond that our observations reveal a diversity in Herbig disk chemistry which is not necessarily disk mass dependent. When taking into account the gas masses of each disk, there is $>$2 order of magnitude spread in SO, \ce{SO_2}, \ce{H_2CO}, \ce{CH_3OH} and COMs line fluxes. In particular, the lowest mass disk in the sample – HD~100453 – shows the brightest reservoir of SO and \ce{CH_3OH}.

\item Molecule detections of note include the presence of \ce{SO_2} in the two SO-brightest disks, HD~100453 and CQ~Tau, compact \ce{H_2S} emission in the HD~169142 disk, \ce{^{33}SO} in the HD~100453 disk, and \ce{CH_3OCH_3} in the HD~97048 disk. We also report non-detections of HDO and SiO where observed.

\item When looking for trends in line fluxes for the Herbig disk population with bulk system properties, we find that tracers of C/O~$>$1 gas—CS, \ce{C_2H}, and \ce{CH_3CN}—are positively correlated with the disk \ce{C^{18}O} line flux. In contrast, SO and \ce{CH_3OH}, which are often associated with C/O$<$~1 gas, do not show any correlations with bulk system properties, but their brightnesses relative to CS and \ce{H_2CO} increase with decreasing disk mass/size and are only detected, so far, in Herbig transition disks. 

\item The simplest COM, \ce{CH_3OH}, is robustly detected in three disks and tentatively in a fourth, whereas the detections of \ce{CH_3CN}, \ce{CH_3OCHO} and \ce{CH_3OCH_3} are more scattered. These results further indicate that, although COM-rich ices can be inherited into disks, in-situ physical and/or chemical processes play a key role in setting the disk’s organic reservoir and in particular, the apparent lack of \ce{CH_3OH} in the disks with the detections of \ce{CH_3OCH_3} is still puzzling. 

\item Finally, we place our results in the context of the gas-giant exoplanet population, discussing the possible C/O regimes of protoplanets identified within Herbig disks in comparison to the directly imaged exoplanet population. We find that the orbital locations of these planets and the molecules observed in each disk correspond to a range of formation conditions. The recently identified HD~135344B~b is likely in a C/O$<$1 regime, whereas HD~169142~b has already been shown to occupy a region where C/O$\approx$1, and we further infer that HD~97048~b could have access to gas with C/O$\approx$0.8–2.
\end{itemize}

To conclude, disks around young intermediate-mass stars have revealed a rich observable chemistry that has not been readily accessible in lower-mass systems. Future high-angular-resolution and multi-line studies are necessary to maximize the potential of these disks both as tracers of the giant planet accretion reservoir and as laboratories to study the total, simple and complex, volatile abundances during this key stage of extrasolar system formation. Specifically, $\approx$0.1" line observations targeting both tracers of C/O and COMs are needed to understand the underlying emitting regions, excitation conditions, and optical depths of these species.

\begin{acknowledgements}
We thank the anonymous referee for the constructive feedback on our manuscript which improved the clarity of the final paper.
A.S.B. is supported by a Clay Postdoctoral Fellowship from the Smithsonian Astrophysical Observatory.
J.K.C. acknowledges support from the Kavli-Laukien
Origins of Life Fellowship at Harvard. K.I.Ö. acknowledges Simons Foundation grant No. 686302,
and an award from the Simons Foundation (grant No.
321183FY19).
E.v.D. and M.T. acknowledge support from the ERC grant 101019751
MOLDISK and the Danish National Research Foundation through the Center
of Excellence “InterCat” (DNRF150).
M.L. is funded by the European Union (ERC, UNVEIL, 101076613). Views and opinions expressed are however those of the author(s) only and do not necessarily reflect those of the European Union or the European Research Council. Neither the European Union nor the granting authority can be held responsible for them.
L.~E. acknowledges financial support from the Science and Technology Facilities Council (grant number ST/X001016/1).
Support for C.J.L. was provided by NASA through the NASA Hubble Fellowship grant No. HST-HF2-51535.001-A awarded by the Space Telescope Science Institute, which is operated by the Association of Universities for Research in Astronomy, Inc., for NASA, under contract NAS5-26555.
S.N. is grateful for support from Grants-in-Aid for JSPS (Japan Society for the Promotion of Science) Fellows Grant Number JP23KJ0329, MEXT/JSPS Grants-in-Aid for Scientific Research (KAKENHI) Grant Numbers JP23K13155 and JP24K00674, and Start-up Research Grant as one of The University of Tokyo Excellent Young Researcher 2024.
J.P. is supported by an STScI Postdoctoral Fellowship from the Space Telescope Science Institute.
C.W.~acknowledges financial support from the Science and Technology Facilities Council and UK Research and Innovation (grant numbers ST/X001016/1 and MR/T040726/1).
% to add: Lisa

This paper makes use of the following ALMA data: 2011.0.00318.S, 2012.1.00303.S, 2012.1.00613.S, 2013.1.00305.S, 2013.1.00498.S, 2015.1.00657.S, 2016.1.00826.S, 2016.1.00884.S, 2016.A.000026.S, 2016.1.00340.S, 2016.1.00344.S, 2017.1.01424.S, 2017.1.00492.S, 2017.1.00845.S, 2017.1.01404.S, 2017.1.01545.S, 2018.1.01429.S, 2018.1.01055.L, 2021.1.00690.S, 2021.1.00738.S, 2021.1.01123.L, 2023.1.00252.S, 2023.1.00628.S, 2024.1.00446.S. ALMA is a partnership of ESO (representing its member states), NSF (USA) and NINS (Japan), together with NRC (Canada), MOST and ASIAA (Taiwan), and KASI (Republic of Korea), in cooperation with the Republic of Chile. The Joint ALMA Observatory is operated by ESO, AUI/NRAO and NAOJ. 
We also include fluxes from to SMA data set 2020A-S018 and from IRAM NOEMA Interferometer programs S14AO and W18AB. 
The Submillimeter Array is a joint project between the Smithsonian
Astrophysical Observatory and the Academia Sinica Institute
of Astronomy and Astrophysics and is funded by the
Smithsonian Institution and the Academia Sinica. 
IRAM is supported by INSU/CNRS (France), MPG (Germany) and IGN (Spain)

This work has used the following additional software packages that have not been referred to in the main text: Astropy, IPython, Jupyter, Matplotlib, NumPy and SciPy \citep{Astropy,IPython,Jupyter,Matplotlib,NumPy,2020NatMe..17..261V}.
\end{acknowledgements}

\bibliography{sample631}{}

\begin{thebibliography}{}
\expandafter\ifx\csname natexlab\endcsname\relax\def\natexlab#1{#1}\fi
\providecommand{\url}[1]{\href{#1}{#1}}
\providecommand{\dodoi}[1]{doi:~\href{http://doi.org/#1}{\nolinkurl{#1}}}
\providecommand{\doeprint}[1]{\href{http://ascl.net/#1}{\nolinkurl{http://ascl.net/#1}}}
\providecommand{\doarXiv}[1]{\href{https://arxiv.org/abs/#1}{\nolinkurl{https://arxiv.org/abs/#1}}}

\bibitem[{{Adams} {et~al.}(2023){Adams}, {Meyer}, {Howe}, {Burningham},
  {Daemgen}, {Fortney}, {Line}, {Marley}, {Quanz}, \&
  {Todorov}}]{2023AJ....166..192A}
{Adams}, A.~D., {Meyer}, M.~R., {Howe}, A.~R., {et~al.} 2023, \aj, 166, 192,
  \dodoi{10.3847/1538-3881/acfb87}

\bibitem[{{Ag{\'u}ndez} {et~al.}(2018){Ag{\'u}ndez}, {Roueff}, {Le Petit}, \&
  {Le Bourlot}}]{2018A&A...616A..19A}
{Ag{\'u}ndez}, M., {Roueff}, E., {Le Petit}, F., \& {Le Bourlot}, J. 2018,
  \aap, 616, A19, \dodoi{10.1051/0004-6361/201732518}

\bibitem[{{Astropy Collaboration} {et~al.}(2022){Astropy Collaboration},
  {Price-Whelan}, {Lim}, {Earl}, {Starkman}, {Bradley}, {Shupe}, {Patil},
  {Corrales}, {Brasseur}, {N{\"o}the}, {Donath}, {Tollerud}, {Morris},
  {Ginsburg}, {Vaher}, {Weaver}, {Tocknell}, {Jamieson}, {van Kerkwijk},
  {Robitaille}, {Merry}, {Bachetti}, {G{\"u}nther}, {Aldcroft},
  {Alvarado-Montes}, {Archibald}, {B{\'o}di}, {Bapat}, {Barentsen},
  {Baz{\'a}n}, {Biswas}, {Boquien}, {Burke}, {Cara}, {Cara}, {Conroy},
  {Conseil}, {Craig}, {Cross}, {Cruz}, {D'Eugenio}, {Dencheva}, {Devillepoix},
  {Dietrich}, {Eigenbrot}, {Erben}, {Ferreira}, {Foreman-Mackey}, {Fox},
  {Freij}, {Garg}, {Geda}, {Glattly}, {Gondhalekar}, {Gordon}, {Grant},
  {Greenfield}, {Groener}, {Guest}, {Gurovich}, {Handberg}, {Hart},
  {Hatfield-Dodds}, {Homeier}, {Hosseinzadeh}, {Jenness}, {Jones}, {Joseph},
  {Kalmbach}, {Karamehmetoglu}, {Ka{\l}uszy{\'n}ski}, {Kelley}, {Kern},
  {Kerzendorf}, {Koch}, {Kulumani}, {Lee}, {Ly}, {Ma}, {MacBride}, {Maljaars},
  {Muna}, {Murphy}, {Norman}, {O'Steen}, {Oman}, {Pacifici}, {Pascual},
  {Pascual-Granado}, {Patil}, {Perren}, {Pickering}, {Rastogi}, {Roulston},
  {Ryan}, {Rykoff}, {Sabater}, {Sakurikar}, {Salgado}, {Sanghi}, {Saunders},
  {Savchenko}, {Schwardt}, {Seifert-Eckert}, {Shih}, {Jain}, {Shukla}, {Sick},
  {Simpson}, {Singanamalla}, {Singer}, {Singhal}, {Sinha}, {Sip{\H{o}}cz},
  {Spitler}, {Stansby}, {Streicher}, {{\v{S}}umak}, {Swinbank}, {Taranu},
  {Tewary}, {Tremblay}, {de Val-Borro}, {Van Kooten}, {Vasovi{\'c}}, {Verma},
  {de Miranda Cardoso}, {Williams}, {Wilson}, {Winkel}, {Wood-Vasey}, {Xue},
  {Yoachim}, {Zhang}, {Zonca}, \& {Astropy Project Contributors}}]{Astropy}
{Astropy Collaboration}, {Price-Whelan}, A.~M., {Lim}, P.~L., {et~al.} 2022,
  \apj, 935, 167, \dodoi{10.3847/1538-4357/ac7c74}

\bibitem[{{Bailey} {et~al.}(2014){Bailey}, {Meshkat}, {Reiter}, {Morzinski},
  {Males}, {Su}, {Hinz}, {Kenworthy}, {Stark}, {Mamajek}, {Briguglio}, {Close},
  {Follette}, {Puglisi}, {Rodigas}, {Weinberger}, \&
  {Xompero}}]{2014ApJ...780L...4B}
{Bailey}, V., {Meshkat}, T., {Reiter}, M., {et~al.} 2014, \apjl, 780, L4,
  \dodoi{10.1088/2041-8205/780/1/L4}

\bibitem[{{Baines} {et~al.}(2006){Baines}, {Oudmaijer}, {Porter}, \&
  {Pozzo}}]{2006MNRAS.367..737B}
{Baines}, D., {Oudmaijer}, R.~D., {Porter}, J.~M., \& {Pozzo}, M. 2006, \mnras,
  367, 737, \dodoi{10.1111/j.1365-2966.2006.10006.x}

\bibitem[{{Banzatti} {et~al.}(2020){Banzatti}, {Pascucci}, {Bosman}, {Pinilla},
  {Salyk}, {Herczeg}, {Pontoppidan}, {Vazquez}, {Watkins}, {Krijt}, {Hendler},
  \& {Long}}]{2020ApJ...903..124B}
{Banzatti}, A., {Pascucci}, I., {Bosman}, A.~D., {et~al.} 2020, \apj, 903, 124,
  \dodoi{10.3847/1538-4357/abbc1a}

\bibitem[{{Banzatti} {et~al.}(2023){Banzatti}, {Pontoppidan}, {Carr},
  {Jellison}, {Pascucci}, {Najita}, {Romero-Mirza}, {{\"O}berg}, {Kalyaan},
  {Pinilla}, {Krijt}, {Long}, {Lambrechts}, {Rosotti}, {Herczeg}, {Salyk},
  {Zhang}, {Bergin}, {Ballering}, {Meyer}, {Bruderer}, \& {Jdiscs
  Collaboration}}]{2023ApJ...957L..22B}
{Banzatti}, A., {Pontoppidan}, K.~M., {Carr}, J.~S., {et~al.} 2023, \apjl, 957,
  L22, \dodoi{10.3847/2041-8213/acf5ec}

\bibitem[{{Benisty} {et~al.}(2015){Benisty}, {Juhasz}, {Boccaletti},
  {Avenhaus}, {Milli}, {Thalmann}, {Dominik}, {Pinilla}, {Buenzli}, {Pohl},
  {Beuzit}, {Birnstiel}, {de Boer}, {Bonnefoy}, {Chauvin}, {Christiaens},
  {Garufi}, {Grady}, {Henning}, {Huelamo}, {Isella}, {Langlois}, {M{\'e}nard},
  {Mouillet}, {Olofsson}, {Pantin}, {Pinte}, \& {Pueyo}}]{2015A&A...578L...6B}
{Benisty}, M., {Juhasz}, A., {Boccaletti}, A., {et~al.} 2015, \aap, 578, L6,
  \dodoi{10.1051/0004-6361/201526011}

\bibitem[{{Bergin} {et~al.}(2024){Bergin}, {Booth}, {Colmenares}, \&
  {Ilee}}]{2024ApJ...969L..21B}
{Bergin}, E.~A., {Booth}, R.~A., {Colmenares}, M.~J., \& {Ilee}, J.~D. 2024,
  \apjl, 969, L21, \dodoi{10.3847/2041-8213/ad5839}

\bibitem[{{Blunt} {et~al.}(2023){Blunt}, {Balmer}, {Wang}, {Lacour}, {Petrus},
  {Bourdarot}, {Kammerer}, {Pourr{\'e}}, {Rickman}, {Shangguan},
  {Winterhalder}, {Abuter}, {Amorim}, {Asensio-Torres}, {Benisty}, {Berger},
  {Beust}, {Boccaletti}, {Bohn}, {Bonnefoy}, {Bonnet}, {Brandner},
  {Cantalloube}, {Caselli}, {Charnay}, {Chauvin}, {Chavez}, {Choquet},
  {Christiaens}, {Cl{\'e}net}, {Du Foresto}, {Cridland}, {Dembet}, {Drescher},
  {Duvert}, {Eckart}, {Eisenhauer}, {Feuchtgruber}, {Garcia}, {Garcia Lopez},
  {Gendron}, {Genzel}, {Gillessen}, {Girard}, {Haubois}, {Hei{\ss}el},
  {Henning}, {Hinkley}, {Hippler}, {Horrobin}, {Houll{\'e}}, {Hubert}, {Jocou},
  {Keppler}, {Kervella}, {Kreidberg}, {Lagrange}, {Lapeyr{\`e}re}, {Le
  Bouquin}, {L{\'e}na}, {Lutz}, {Maire}, {Mang}, {Marleau}, {M{\'e}rand},
  {Molli{\`e}re}, {Monnier}, {Mordasini}, {Mouillet}, {Nasedkin}, {Nowak},
  {Ott}, {Otten}, {Paladini}, {Paumard}, {Perraut}, {Perrin}, {Pfuhl}, {Pueyo},
  {Rameau}, {Rodet}, {Rustamkulov}, {Shimizu}, {Sing}, {Stolker},
  {Straubmeier}, {Sturm}, {Tacconi}, {van Dishoeck}, {Vigan}, {Vincent},
  {Ward-Duong}, {Widmann}, {Wieprecht}, {Wiezorrek}, {Woillez}, {Yazici},
  {Young}, \& {Exogravity Collaboration}}]{2023AJ....166..257B}
{Blunt}, S., {Balmer}, W.~O., {Wang}, J.~J., {et~al.} 2023, \aj, 166, 257,
  \dodoi{10.3847/1538-3881/ad06b7}

\bibitem[{{Boehler} {et~al.}(2018){Boehler}, {Ricci}, {Weaver}, {Isella},
  {Benisty}, {Carpenter}, {Grady}, {Shen}, {Tang}, \&
  {Perez}}]{2018ApJ...853..162B}
{Boehler}, Y., {Ricci}, L., {Weaver}, E., {et~al.} 2018, \apj, 853, 162,
  \dodoi{10.3847/1538-4357/aaa19c}

\bibitem[{{Booth} {et~al.}(2023{\natexlab{a}}){Booth}, {Ilee}, {Walsh}, {Kama},
  {Keyte}, {van Dishoeck}, \& {Nomura}}]{Booth23a}
{Booth}, A.~S., {Ilee}, J.~D., {Walsh}, C., {et~al.} 2023{\natexlab{a}}, \aap,
  669, A53, \dodoi{10.1051/0004-6361/202244472}

\bibitem[{{Booth} {et~al.}(2023{\natexlab{b}}){Booth}, {Law}, {Temmink},
  {Leemker}, \& {Mac{\'\i}as}}]{2023A&A...678A.146B}
{Booth}, A.~S., {Law}, C.~J., {Temmink}, M., {Leemker}, M., \& {Mac{\'\i}as},
  E. 2023{\natexlab{b}}, \aap, 678, A146, \dodoi{10.1051/0004-6361/202346974}

\bibitem[{{Booth} {et~al.}(2021{\natexlab{a}}){Booth}, {van der Marel},
  {Leemker}, {van Dishoeck}, \& {Ohashi}}]{2021A&A...651L...6B}
{Booth}, A.~S., {van der Marel}, N., {Leemker}, M., {van Dishoeck}, E.~F., \&
  {Ohashi}, S. 2021{\natexlab{a}}, \aap, 651, L6,
  \dodoi{10.1051/0004-6361/202141057}

\bibitem[{{Booth} {et~al.}(2019){Booth}, {Walsh}, \&
  {Ilee}}]{2019A&A...629A..75B}
{Booth}, A.~S., {Walsh}, C., \& {Ilee}, J.~D. 2019, \aap, 629, A75,
  \dodoi{10.1051/0004-6361/201834388}

\bibitem[{{Booth} {et~al.}(2021{\natexlab{b}}){Booth}, {Walsh}, {Terwisscha van
  Scheltinga}, {van Dishoeck}, {Ilee}, {Hogerheijde}, {Kama}, \&
  {Nomura}}]{2021NatAs...5..684B}
{Booth}, A.~S., {Walsh}, C., {Terwisscha van Scheltinga}, J., {et~al.}
  2021{\natexlab{b}}, Nature Astronomy, 5, 684,
  \dodoi{10.1038/s41550-021-01352-w}

\bibitem[{{Booth} {et~al.}(2024{\natexlab{a}}){Booth}, {Leemker}, {van
  Dishoeck}, {Evans}, {Ilee}, {Kama}, {Keyte}, {Law}, {van der Marel},
  {Nomura}, {Notsu}, {{\"O}berg}, {Temmink}, \& {Walsh}}]{2024AJ....167..164B}
{Booth}, A.~S., {Leemker}, M., {van Dishoeck}, E.~F., {et~al.}
  2024{\natexlab{a}}, \aj, 167, 164, \dodoi{10.3847/1538-3881/ad2700}

\bibitem[{{Booth} {et~al.}(2024{\natexlab{b}}){Booth}, {Temmink}, {van
  Dishoeck}, {Evans}, {Ilee}, {Kama}, {Keyte}, {Law}, {Leemker}, {van der
  Marel}, {Nomura}, {Notsu}, {{\"O}berg}, \& {Walsh}}]{2024AJ....167..165B}
{Booth}, A.~S., {Temmink}, M., {van Dishoeck}, E.~F., {et~al.}
  2024{\natexlab{b}}, \aj, 167, 165, \dodoi{10.3847/1538-3881/ad26ff}

\bibitem[{{Booth} {et~al.}(2025){Booth}, {W{\"o}lfer}, {Temmink}, {Calahan},
  {Evans}, {Law}, {Leemker}, {Notsu}, {{\"O}berg}, \&
  {Walsh}}]{2025arXiv250414023B}
{Booth}, A.~S., {W{\"o}lfer}, L., {Temmink}, M., {et~al.} 2025, \apjl, 986, L9,
  \dodoi{10.3847/2041-8213/adc7b2}

\bibitem[{{Bosman} {et~al.}(2018){Bosman}, {Walsh}, \& {van
  Dishoeck}}]{2018A&A...618A.182B}
{Bosman}, A.~D., {Walsh}, C., \& {van Dishoeck}, E.~F. 2018, \aap, 618, A182,
  \dodoi{10.1051/0004-6361/201833497}

\bibitem[{{Bosman} {et~al.}(2021){Bosman}, {Alarc{\'o}n}, {Bergin}, {Zhang},
  {van't Hoff}, {{\"O}berg}, {Guzm{\'a}n}, {Walsh}, {Aikawa}, {Andrews},
  {Bergner}, {Booth}, {Cataldi}, {Cleeves}, {Czekala}, {Furuya}, {Huang},
  {Ilee}, {Law}, {Le Gal}, {Liu}, {Long}, {Loomis}, {M{\'e}nard}, {Nomura},
  {Qi}, {Schwarz}, {Teague}, {Tsukagoshi}, {Yamato}, \&
  {Wilner}}]{2021ApJS..257....7B}
{Bosman}, A.~D., {Alarc{\'o}n}, F., {Bergin}, E.~A., {et~al.} 2021, \apjs, 257,
  7, \dodoi{10.3847/1538-4365/ac1435}

\bibitem[{{Brown-Sevilla} {et~al.}(2023){Brown-Sevilla}, {Maire},
  {Molli{\`e}re}, {Samland}, {Feldt}, {Brandner}, {Henning}, {Gratton},
  {Janson}, {Stolker}, {Hagelberg}, {Zurlo}, {Cantalloube}, {Boccaletti},
  {Bonnefoy}, {Chauvin}, {Desidera}, {D'Orazi}, {Lagrange}, {Langlois},
  {Menard}, {Mesa}, {Meyer}, {Pavlov}, {Petit}, {Rochat}, {Rouan}, {Schmidt},
  {Vigan}, \& {Weber}}]{2023A&A...673A..98B}
{Brown-Sevilla}, S.~B., {Maire}, A.~L., {Molli{\`e}re}, P., {et~al.} 2023,
  \aap, 673, A98, \dodoi{10.1051/0004-6361/202244826}

\bibitem[{{Brunken} {et~al.}(2022){Brunken}, {Booth}, {Leemker}, {Nazari}, {van
  der Marel}, \& {van Dishoeck}}]{2022A&A...659A..29B}
{Brunken}, N. G.~C., {Booth}, A.~S., {Leemker}, M., {et~al.} 2022, \aap, 659,
  A29, \dodoi{10.1051/0004-6361/202142981}

\bibitem[{{Calahan} {et~al.}(2021{\natexlab{a}}){Calahan}, {Bergin}, {Zhang},
  {Teague}, {Cleeves}, {Bergner}, {Blake}, {Cazzoletti}, {Guzm{\'a}n},
  {Hogerheijde}, {Huang}, {Kama}, {Loomis}, {{\"O}berg}, {Qi}, {van Dishoeck},
  {Terwisscha van Scheltinga}, {Walsh}, \& {Wilner}}]{2021ApJ...908....8C}
{Calahan}, J.~K., {Bergin}, E., {Zhang}, K., {et~al.} 2021{\natexlab{a}}, \apj,
  908, 8, \dodoi{10.3847/1538-4357/abd255}

\bibitem[{{Calahan} {et~al.}(2021{\natexlab{b}}){Calahan}, {Bergin}, {Zhang},
  {Schwarz}, {{\"O}berg}, {Guzm{\'a}n}, {Walsh}, {Aikawa}, {Alarc{\'o}n},
  {Andrews}, {Bae}, {Bergner}, {Booth}, {Bosman}, {Cataldi}, {Czekala},
  {Huang}, {Ilee}, {Law}, {Le Gal}, {Long}, {Loomis}, {M{\'e}nard}, {Nomura},
  {Qi}, {Teague}, {van't Hoff}, {Wilner}, \& {Yamato}}]{2021ApJS..257...17C}
{Calahan}, J.~K., {Bergin}, E.~A., {Zhang}, K., {et~al.} 2021{\natexlab{b}},
  \apjs, 257, 17, \dodoi{10.3847/1538-4365/ac143f}

\bibitem[{{Calahan} {et~al.}(2023){Calahan}, {Bergin}, {Bosman}, {Rich},
  {Andrews}, {Bergner}, {Cleeves}, {Guzm{\'a}n}, {Huang}, {Ilee}, {Law}, {Le
  Gal}, {{\"O}berg}, {Teague}, {Walsh}, {Wilner}, \&
  {Zhang}}]{2023NatAs...7...49C}
{Calahan}, J.~K., {Bergin}, E.~A., {Bosman}, A.~D., {et~al.} 2023, Nature
  Astronomy, 7, 49, \dodoi{10.1038/s41550-022-01831-8}

\bibitem[{{Carney} {et~al.}(2019){Carney}, {Hogerheijde}, {Guzm{\'a}n},
  {Walsh}, {{\"O}berg}, {Fayolle}, {Cleeves}, {Carpenter}, \&
  {Qi}}]{Carney2019}
{Carney}, M.~T., {Hogerheijde}, M.~R., {Guzm{\'a}n}, V.~V., {et~al.} 2019,
  \aap, 623, A124, \dodoi{10.1051/0004-6361/201834353}

\bibitem[{{Carson} {et~al.}(2013){Carson}, {Thalmann}, {Janson}, {Kozakis},
  {Bonnefoy}, {Biller}, {Schlieder}, {Currie}, {McElwain}, {Goto}, {Henning},
  {Brandner}, {Feldt}, {Kandori}, {Kuzuhara}, {Stevens}, {Wong}, {Gainey},
  {Fukagawa}, {Kuwada}, {Brandt}, {Kwon}, {Abe}, {Egner}, {Grady}, {Guyon},
  {Hashimoto}, {Hayano}, {Hayashi}, {Hayashi}, {Hodapp}, {Ishii}, {Iye},
  {Knapp}, {Kudo}, {Kusakabe}, {Matsuo}, {Miyama}, {Morino}, {Moro-Martin},
  {Nishimura}, {Pyo}, {Serabyn}, {Suto}, {Suzuki}, {Takami}, {Takato},
  {Terada}, {Tomono}, {Turner}, {Watanabe}, {Wisniewski}, {Yamada}, {Takami},
  {Usuda}, \& {Tamura}}]{2013ApJ...763L..32C}
{Carson}, J., {Thalmann}, C., {Janson}, M., {et~al.} 2013, \apjl, 763, L32,
  \dodoi{10.1088/2041-8205/763/2/L32}

\bibitem[{{CASA Team} {et~al.}(2022){CASA Team}, {Bean}, {Bhatnagar}, {Castro},
  {Donovan Meyer}, {Emonts}, {Garcia}, {Garwood}, {Golap}, {Gonzalez Villalba},
  {Harris}, {Hayashi}, {Hoskins}, {Hsieh}, {Jagannathan}, {Kawasaki},
  {Keimpema}, {Kettenis}, {Lopez}, {Marvil}, {Masters}, {McNichols},
  {Mehringer}, {Miel}, {Moellenbrock}, {Montesino}, {Nakazato}, {Ott}, {Petry},
  {Pokorny}, {Raba}, {Rau}, {Schiebel}, {Schweighart}, {Sekhar}, {Shimada},
  {Small}, {Steeb}, {Sugimoto}, {Suoranta}, {Tsutsumi}, {van Bemmel},
  {Verkouter}, {Wells}, {Xiong}, {Szomoru}, {Griffith}, {Glendenning}, \&
  {Kern}}]{2022PASP..134k4501C}
{CASA Team}, {Bean}, B., {Bhatnagar}, S., {et~al.} 2022, \pasp, 134, 114501,
  \dodoi{10.1088/1538-3873/ac9642}

\bibitem[{{Casassus} {et~al.}(2021){Casassus}, {Christiaens}, {C{\'a}rcamo},
  {P{\'e}rez}, {Weber}, {Ercolano}, {van der Marel}, {Pinte}, {Dong},
  {Baruteau}, {Cieza}, {van Dishoeck}, {Jordan}, {Price}, {Absil}, {Arce-Tord},
  {Faramaz}, {Flores}, \& {Reggiani}}]{2021MNRAS.507.3789C}
{Casassus}, S., {Christiaens}, V., {C{\'a}rcamo}, M., {et~al.} 2021, \mnras,
  507, 3789, \dodoi{10.1093/mnras/stab2359}

\bibitem[{{Cazzoletti} {et~al.}(2018{\natexlab{a}}){Cazzoletti}, {van
  Dishoeck}, {Visser}, {Facchini}, \& {Bruderer}}]{2018A&A...609A..93C}
{Cazzoletti}, P., {van Dishoeck}, E.~F., {Visser}, R., {Facchini}, S., \&
  {Bruderer}, S. 2018{\natexlab{a}}, \aap, 609, A93,
  \dodoi{10.1051/0004-6361/201731457}

\bibitem[{{Cazzoletti} {et~al.}(2018{\natexlab{b}}){Cazzoletti}, {van
  Dishoeck}, {Pinilla}, {Tazzari}, {Facchini}, {van der Marel}, {Benisty},
  {Garufi}, \& {P{\'e}rez}}]{2018A&A...619A.161C}
{Cazzoletti}, P., {van Dishoeck}, E.~F., {Pinilla}, P., {et~al.}
  2018{\natexlab{b}}, \aap, 619, A161, \dodoi{10.1051/0004-6361/201834006}

\bibitem[{{Chauvin} {et~al.}(2017){Chauvin}, {Desidera}, {Lagrange}, {Vigan},
  {Gratton}, {Langlois}, {Bonnefoy}, {Beuzit}, {Feldt}, {Mouillet}, {Meyer},
  {Cheetham}, {Biller}, {Boccaletti}, {D'Orazi}, {Galicher}, {Hagelberg},
  {Maire}, {Mesa}, {Olofsson}, {Samland}, {Schmidt}, {Sissa}, {Bonavita},
  {Charnay}, {Cudel}, {Daemgen}, {Delorme}, {Janin-Potiron}, {Janson},
  {Keppler}, {Le Coroller}, {Ligi}, {Marleau}, {Messina}, {Molli{\`e}re},
  {Mordasini}, {M{\"u}ller}, {Peretti}, {Perrot}, {Rodet}, {Rouan}, {Zurlo},
  {Dominik}, {Henning}, {Menard}, {Schmid}, {Turatto}, {Udry}, {Vakili}, {Abe},
  {Antichi}, {Baruffolo}, {Baudoz}, {Baudrand}, {Blanchard}, {Bazzon}, {Buey},
  {Carbillet}, {Carle}, {Charton}, {Cascone}, {Claudi}, {Costille}, {Deboulbe},
  {De Caprio}, {Dohlen}, {Fantinel}, {Feautrier}, {Fusco}, {Gigan}, {Giro},
  {Gisler}, {Gluck}, {Hubin}, {Hugot}, {Jaquet}, {Kasper}, {Madec}, {Magnard},
  {Martinez}, {Maurel}, {Le Mignant}, {M{\"o}ller-Nilsson}, {Llored}, {Moulin},
  {Orign{\'e}}, {Pavlov}, {Perret}, {Petit}, {Pragt}, {Puget}, {Rabou},
  {Ramos}, {Rigal}, {Rochat}, {Roelfsema}, {Rousset}, {Roux}, {Salasnich},
  {Sauvage}, {Sevin}, {Soenke}, {Stadler}, {Suarez}, {Weber}, {Wildi},
  {Antoniucci}, {Augereau}, {Baudino}, {Brandner}, {Engler}, {Girard}, {Gry},
  {Kral}, {Kopytova}, {Lagadec}, {Milli}, {Moutou}, {Schlieder},
  {Szul{\'a}gyi}, {Thalmann}, \& {Wahhaj}}]{2017A&A...605L...9C}
{Chauvin}, G., {Desidera}, S., {Lagrange}, A.~M., {et~al.} 2017, \aap, 605, L9,
  \dodoi{10.1051/0004-6361/201731152}

\bibitem[{{Chen} {et~al.}(2023){Chen}, {van Gelder}, {Nazari}, {Brogan}, {van
  Dishoeck}, {Linnartz}, {J{\o}rgensen}, {Hunter}, {Wilkins}, {Blake},
  {Caselli}, {Chuang}, {Codella}, {Cooke}, {Drozdovskaya}, {Garrod}, {Ioppolo},
  {Jin}, {Kulterer}, {Ligterink}, {Lipnicky}, {Loomis}, {Rachid}, {Spezzano},
  \& {McGuire}}]{2023A&A...678A.137C}
{Chen}, Y., {van Gelder}, M.~L., {Nazari}, P., {et~al.} 2023, \aap, 678, A137,
  \dodoi{10.1051/0004-6361/202346491}

\bibitem[{{Chrenko} {et~al.}(2025){Chrenko}, {Casassus}, \&
  {Chametla}}]{2025arXiv250703370C}
{Chrenko}, O., {Casassus}, S., \& {Chametla}, R.~O. 2025, \aap, 700, A82,
  \dodoi{10.1051/0004-6361/202554623}

\bibitem[{{Cleeves} {et~al.}(2015){Cleeves}, {Bergin}, \&
  {Harries}}]{2015ApJ...807....2C}
{Cleeves}, L.~I., {Bergin}, E.~A., \& {Harries}, T.~J. 2015, \apj, 807, 2,
  \dodoi{10.1088/0004-637X/807/1/2}

\bibitem[{{Coletta} {et~al.}(2020){Coletta}, {Fontani}, {Rivilla}, {Mininni},
  {Colzi}, {S{\'a}nchez-Monge}, \& {Beltr{\'a}n}}]{2020A&A...641A..54C}
{Coletta}, A., {Fontani}, F., {Rivilla}, V.~M., {et~al.} 2020, \aap, 641, A54,
  \dodoi{10.1051/0004-6361/202038212}

\bibitem[{{Collins} {et~al.}(2009){Collins}, {Grady}, {Hamaguchi},
  {Wisniewski}, {Brittain}, {Sitko}, {Carpenter}, {Williams}, {Mathews},
  {Williger}, {van Boekel}, {Carmona}, {Henning}, {van den Ancker}, {Meeus},
  {Chen}, {Petre}, \& {Woodgate}}]{2009ApJ...697..557C}
{Collins}, K.~A., {Grady}, C.~A., {Hamaguchi}, K., {et~al.} 2009, \apj, 697,
  557, \dodoi{10.1088/0004-637X/697/1/557}

\bibitem[{{Crossfield}(2023)}]{2023ApJ...952L..18C}
{Crossfield}, I. J.~M. 2023, \apjl, 952, L18, \dodoi{10.3847/2041-8213/ace35f}

\bibitem[{{Curone} {et~al.}(2025){Curone}, {Facchini}, {Andrews}, {Testi},
  {Benisty}, {Czekala}, {Huang}, {Ilee}, {Isella}, {Lodato}, {Loomis},
  {Stadler}, {Winter}, {Bae}, {Barraza-Alfaro}, {Cataldi}, {Cuello}, {Fasano},
  {Flock}, {Fukagawa}, {Galloway-Sprietsma}, {Garg}, {Hall}, {Izquierdo},
  {Kanagawa}, {Lesur}, {Longarini}, {Menard}, {Orihara}, {Pinte}, {Price},
  {Rosotti}, {Teague}, {Wafflard-Fernandez}, {Wilner}, {W{\"o}lfer}, {Yen},
  {Yoshida}, \& {Zawadzki}}]{2025ApJ...984L...9C}
{Curone}, P., {Facchini}, S., {Andrews}, S.~M., {et~al.} 2025, \apjl, 984, L9,
  \dodoi{10.3847/2041-8213/adc438}

\bibitem[{{Currie} {et~al.}(2023){Currie}, {Brandt}, {Brandt}, {Lacy},
  {Burrows}, {Guyon}, {Tamura}, {Liu}, {Sagynbayeva}, {Tobin}, {Chilcote},
  {Groff}, {Marois}, {Thompson}, {Murphy}, {Kuzuhara}, {Lawson}, {Lozi}, {Deo},
  {Vievard}, {Skaf}, {Uyama}, {Jovanovic}, {Martinache}, {Kasdin}, {Kudo},
  {McElwain}, {Janson}, {Wisniewski}, {Hodapp}, {Nishikawa}, {He{\l}miniak},
  {Kwon}, \& {Hayashi}}]{2023Sci...380..198C}
{Currie}, T., {Brandt}, G.~M., {Brandt}, T.~D., {et~al.} 2023, Science, 380,
  198, \dodoi{10.1126/science.abo6192}

\bibitem[{{Denis} {et~al.}(2025){Denis}, {Vigan}, {Costes}, {Chauvin},
  {Radcliffe}, {Ravet}, {Balmer}, {Palma-Bifani}, {Petrus}, {Parmentier},
  {Martos}, {Simonnin}, {Bonnefoy}, {Cadet}, {Forveille}, {Charnay}, {Kiefer},
  {Lagrange}, {Chiavassa}, {Stolker}, {Lavail}, {Godoy}, {Janson}, {Pourcelot},
  {Delorme}, {Rickman}, {Cont}, {Reiners}, {De Rosa}, {Anwand-Heerwart},
  {Charles}, {Costille}, {El Morsy}, {Garcia}, {Houll{\'e}}, {Lopez}, {Murray},
  {Muslimov}, {Otten}, {Paufique}, {Phillips}, {Seemann}, {Viret}, \&
  {Zins}}]{2025A&A...696A...6D}
{Denis}, A., {Vigan}, A., {Costes}, J., {et~al.} 2025, \aap, 696, A6,
  \dodoi{10.1051/0004-6361/202453108}

\bibitem[{{Dong} {et~al.}(2018){Dong}, {Liu}, {Eisner}, {Andrews}, {Fung},
  {Zhu}, {Chiang}, {Hashimoto}, {Liu}, {Casassus}, {Esposito}, {Hasegawa},
  {Muto}, {Pavlyuchenkov}, {Wilner}, {Akiyama}, {Tamura}, \&
  {Wisniewski}}]{2018ApJ...860..124D}
{Dong}, R., {Liu}, S.-y., {Eisner}, J., {et~al.} 2018, \apj, 860, 124,
  \dodoi{10.3847/1538-4357/aac6cb}

\bibitem[{{Drozdovskaya} {et~al.}(2018){Drozdovskaya}, {van Dishoeck},
  {J{\o}rgensen}, {Calmonte}, {van der Wiel}, {Coutens}, {Calcutt},
  {M{\"u}ller}, {Bjerkeli}, {Persson}, {Wampfler}, \&
  {Altwegg}}]{2018MNRAS.476.4949D}
{Drozdovskaya}, M.~N., {van Dishoeck}, E.~F., {J{\o}rgensen}, J.~K., {et~al.}
  2018, \mnras, 476, 4949, \dodoi{10.1093/mnras/sty462}

\bibitem[{{Dutrey} {et~al.}(2024){Dutrey}, {Chapillon}, {Guilloteau}, {Tang},
  {Boccaletti}, {Bouscasse}, {Collin-Dufresne}, {Di Folco}, {Fuente},
  {Pi{\'e}tu}, {Rivi{\`e}re-Marichalar}, \& {Semenov}}]{Dutrey24}
{Dutrey}, A., {Chapillon}, E., {Guilloteau}, S., {et~al.} 2024, \aap, 689, L7,
  \dodoi{10.1051/0004-6361/202451299}

\bibitem[{{Evans} {et~al.}(2025{\natexlab{a}}){Evans}, {Booth}, {Walsh},
  {Ilee}, {Keyte}, {Law}, {Leemker}, {Notsu}, {{\"O}berg}, {Temmink}, \& {van
  der Marel}}]{2025ApJ...982...62E}
{Evans}, L., {Booth}, A.~S., {Walsh}, C., {et~al.} 2025{\natexlab{a}}, \apj,
  982, 62, \dodoi{10.3847/1538-4357/adb287}

\bibitem[{{Evans} {et~al.}(2025{\natexlab{b}}){Evans}, {Booth}, {Walsh},
  {Ilee}, {Keyte}, {Law}, {Leemker}, {Notsu}, {{\"O}berg}, {Temmink}, \& {van
  der Marel}}]{2025arXiv250204957E}
---. 2025{\natexlab{b}}, arXiv e-prints, arXiv:2502.04957,
  \dodoi{10.48550/arXiv.2502.04957}

\bibitem[{{Fedele} \& {Favre}(2020)}]{2020A&A...638A.110F}
{Fedele}, D., \& {Favre}, C. 2020, \aap, 638, A110,
  \dodoi{10.1051/0004-6361/202037927}

\bibitem[{{Fedele} {et~al.}(2017){Fedele}, {Carney}, {Hogerheijde}, {Walsh},
  {Miotello}, {Klaassen}, {Bruderer}, {Henning}, \& {van
  Dishoeck}}]{2017A&A...600A..72F}
{Fedele}, D., {Carney}, M., {Hogerheijde}, M.~R., {et~al.} 2017, \aap, 600,
  A72, \dodoi{10.1051/0004-6361/201629860}

\bibitem[{{Feinstein} {et~al.}(2025){Feinstein}, {Booth}, {Bergner},
  {Lothringer}, {Matthews}, {Welbanks}, {Miguel}, {Bitsch}, {Eriksson}, {Kirk},
  {Pelletier}, {Penzlin}, {Piette}, {Piaulet-Ghorayeb}, {Schwarz}, {Turrini},
  {Acu{\~n}a-Aguirre}, {Ahrer}, {Barber}, {Brande}, {Chakrabarty},
  {Crossfield}, {Marleau}, {Huang}, {Johansen}, {Kreidberg}, {Livingston},
  {Luque}, {Oreshenko}, {Pacetti}, {Perotti}, {Polman}, {Prinoth}, {Semenov},
  {Simon}, {Teske}, \& {Whiteford}}]{2025arXiv250600669F}
{Feinstein}, A.~D., {Booth}, R.~A., {Bergner}, J.~B., {et~al.} 2025, arXiv
  e-prints, arXiv:2506.00669, \dodoi{10.48550/arXiv.2506.00669}

\bibitem[{{Follette} {et~al.}(2023){Follette}, {Close}, {Males}, {Ward-Duong},
  {Balmer}, {Redai}, {Morales}, {Sarosi}, {Dacus}, {De Rosa}, {Garcia Toro},
  {Leonard}, {Macintosh}, {Morzinski}, {Mullen}, {Palmo}, {Saitoti}, {Spiro},
  {Treiber}, {Wagner}, {Wang}, {Wang}, {Watson}, \&
  {Weinberger}}]{2023AJ....165..225F}
{Follette}, K.~B., {Close}, L.~M., {Males}, J.~R., {et~al.} 2023, \aj, 165,
  225, \dodoi{10.3847/1538-3881/acc183}

\bibitem[{{Francis} \& {van der Marel}(2020)}]{2020ApJ...892..111F}
{Francis}, L., \& {van der Marel}, N. 2020, \apj, 892, 111,
  \dodoi{10.3847/1538-4357/ab7b63}

\bibitem[{{Fuente} {et~al.}(2010){Fuente}, {Cernicharo}, {Ag{\'u}ndez},
  {Bern{\'e}}, {Goicoechea}, {Alonso-Albi}, \&
  {Marcelino}}]{2010A&A...524A..19F}
{Fuente}, A., {Cernicharo}, J., {Ag{\'u}ndez}, M., {et~al.} 2010, \aap, 524,
  A19, \dodoi{10.1051/0004-6361/201014905}

\bibitem[{{Furuya} \& {Aikawa}(2014)}]{2014ApJ...790...97F}
{Furuya}, K., \& {Aikawa}, Y. 2014, \apj, 790, 97,
  \dodoi{10.1088/0004-637X/790/2/97}

\bibitem[{{Garg} {et~al.}(2022){Garg}, {Pinte}, {Hammond}, {Teague}, {Hilder},
  {Price}, {Calcino}, {Christiaens}, \& {Poblete}}]{2022MNRAS.517.5942G}
{Garg}, H., {Pinte}, C., {Hammond}, I., {et~al.} 2022, \mnras, 517, 5942,
  \dodoi{10.1093/mnras/stac3039}

\bibitem[{{Garrod} {et~al.}(2022){Garrod}, {Jin}, {Matis}, {Jones}, {Willis},
  \& {Herbst}}]{2022ApJS..259....1G}
{Garrod}, R.~T., {Jin}, M., {Matis}, K.~A., {et~al.} 2022, \apjs, 259, 1,
  \dodoi{10.3847/1538-4365/ac3131}

\bibitem[{{Ginski} {et~al.}(2016){Ginski}, {Stolker}, {Pinilla}, {Dominik},
  {Boccaletti}, {de Boer}, {Benisty}, {Biller}, {Feldt}, {Garufi}, {Keller},
  {Kenworthy}, {Maire}, {M{\'e}nard}, {Mesa}, {Milli}, {Min}, {Pinte}, {Quanz},
  {van Boekel}, {Bonnefoy}, {Chauvin}, {Desidera}, {Gratton}, {Girard},
  {Keppler}, {Kopytova}, {Lagrange}, {Langlois}, {Rouan}, \&
  {Vigan}}]{2016A&A...595A.112G}
{Ginski}, C., {Stolker}, T., {Pinilla}, P., {et~al.} 2016, \aap, 595, A112,
  \dodoi{10.1051/0004-6361/201629265}

\bibitem[{{Ginski} {et~al.}(2024){Ginski}, {Garufi}, {Benisty}, {Tazaki},
  {Dominik}, {Ribas}, {Engler}, {Birnstiel}, {Chauvin}, {Columba}, {Facchini},
  {Goncharov}, {Hagelberg}, {Henning}, {Hogerheijde}, {van Holstein}, {Huang},
  {Muto}, {Pinilla}, {Kanagawa}, {Kim}, {Kurtovic}, {Langlois}, {Manara},
  {Milli}, {Momose}, {Orihara}, {Pawellek}, {Pinte}, {Rab}, {Schmidt}, {Snik},
  {Wahhaj}, {Williams}, \& {Zurlo}}]{2024A&A...685A..52G}
{Ginski}, C., {Garufi}, A., {Benisty}, M., {et~al.} 2024, \aap, 685, A52,
  \dodoi{10.1051/0004-6361/202244005}

\bibitem[{{Godoy} {et~al.}(2024){Godoy}, {Choquet}, {Serabyn}, {Danielski},
  {Stolker}, {Charnay}, {Hinkley}, {Lagage}, {Ressler}, {Tremblin}, \&
  {Vigan}}]{2024A&A...689A.185G}
{Godoy}, N., {Choquet}, E., {Serabyn}, E., {et~al.} 2024, \aap, 689, A185,
  \dodoi{10.1051/0004-6361/202449951}

\bibitem[{{GRAVITY Collaboration} {et~al.}(2020){GRAVITY Collaboration},
  {Nowak}, {Lacour}, {Molli{\`e}re}, {Wang}, {Charnay}, {van Dishoeck},
  {Abuter}, {Amorim}, {Berger}, {Beust}, {Bonnefoy}, {Bonnet}, {Brandner},
  {Buron}, {Cantalloube}, {Collin}, {Chapron}, {Cl{\'e}net}, {Coud{\'e} Du
  Foresto}, {de Zeeuw}, {Dembet}, {Dexter}, {Duvert}, {Eckart}, {Eisenhauer},
  {F{\"o}rster Schreiber}, {F{\'e}dou}, {Garcia Lopez}, {Gao}, {Gendron},
  {Genzel}, {Gillessen}, {Hau{\ss}mann}, {Henning}, {Hippler}, {Hubert},
  {Jocou}, {Kervella}, {Lagrange}, {Lapeyr{\`e}re}, {Le Bouquin}, {L{\'e}na},
  {Maire}, {Ott}, {Paumard}, {Paladini}, {Perraut}, {Perrin}, {Pueyo}, {Pfuhl},
  {Rabien}, {Rau}, {Rodr{\'\i}guez-Coira}, {Rousset}, {Scheithauer},
  {Shangguan}, {Straub}, {Straubmeier}, {Sturm}, {Tacconi}, {Vincent},
  {Widmann}, {Wieprecht}, {Wiezorrek}, {Woillez}, {Yazici}, \&
  {Ziegler}}]{2020A&A...633A.110G}
{GRAVITY Collaboration}, {Nowak}, M., {Lacour}, S., {et~al.} 2020, \aap, 633,
  A110, \dodoi{10.1051/0004-6361/201936898}

\bibitem[{{Guzm{\'a}n} {et~al.}(2021){Guzm{\'a}n}, {Bergner}, {Law},
  {{\"O}berg}, {Walsh}, {Cataldi}, {Aikawa}, {Bergin}, {Czekala}, {Huang},
  {Andrews}, {Loomis}, {Zhang}, {Le Gal}, {Alarc{\'o}n}, {Ilee}, {Teague},
  {Cleeves}, {Wilner}, {Long}, {Schwarz}, {Bosman}, {P{\'e}rez}, {M{\'e}nard},
  \& {Liu}}]{2021ApJS..257....6G}
{Guzm{\'a}n}, V.~V., {Bergner}, J.~B., {Law}, C.~J., {et~al.} 2021, \apjs, 257,
  6, \dodoi{10.3847/1538-4365/ac1440}

\bibitem[{{Guzm{\'a}n-D{\'\i}az} {et~al.}(2021){Guzm{\'a}n-D{\'\i}az},
  {Mendigut{\'\i}a}, {Montesinos}, {Oudmaijer}, {Vioque}, {Rodrigo}, {Solano},
  {Meeus}, \& {Marcos-Arenal}}]{2021A&A...650A.182G}
{Guzm{\'a}n-D{\'\i}az}, J., {Mendigut{\'\i}a}, I., {Montesinos}, B., {et~al.}
  2021, \aap, 650, A182, \dodoi{10.1051/0004-6361/202039519}

\bibitem[{{Hammond} {et~al.}(2023){Hammond}, {Christiaens}, {Price}, {Toci},
  {Pinte}, {Juillard}, \& {Garg}}]{2023MNRAS.522L..51H}
{Hammond}, I., {Christiaens}, V., {Price}, D.~J., {et~al.} 2023, \mnras, 522,
  L51, \dodoi{10.1093/mnrasl/slad027}

\bibitem[{{Hammond} {et~al.}(2022){Hammond}, {Christiaens}, {Price},
  {Ubeira-Gabellini}, {Baird}, {Calcino}, {Benisty}, {Lodato}, {Testi},
  {Pinte}, {Toci}, \& {Fedele}}]{2022MNRAS.515.6109H}
---. 2022, \mnras, 515, 6109, \dodoi{10.1093/mnras/stac2119}

\bibitem[{Harris {et~al.}(2020)Harris, Millman, van~der Walt, Gommers,
  Virtanen, Cournapeau, Wieser, Taylor, Berg, Smith, Kern, Picus, Hoyer, van
  Kerkwijk, Brett, Haldane, del R{\'{i}}o, Wiebe, Peterson,
  G{\'{e}}rard-Marchant, Sheppard, Reddy, Weckesser, Abbasi, Gohlke, \&
  Oliphant}]{NumPy}
Harris, C.~R., Millman, K.~J., van~der Walt, S.~J., {et~al.} 2020, Nature, 585,
  357, \dodoi{10.1038/s41586-020-2649-2}

\bibitem[{{Hinkley} {et~al.}(2023){Hinkley}, {Lacour}, {Marleau}, {Lagrange},
  {Wang}, {Kammerer}, {Cumming}, {Nowak}, {Rodet}, {Stolker}, {Balmer}, {Ray},
  {Bonnefoy}, {Molli{\`e}re}, {Lazzoni}, {Kennedy}, {Mordasini}, {Abuter},
  {Aigrain}, {Amorim}, {Asensio-Torres}, {Babusiaux}, {Benisty}, {Berger},
  {Beust}, {Blunt}, {Boccaletti}, {Bohn}, {Bonnet}, {Bourdarot}, {Brandner},
  {Cantalloube}, {Caselli}, {Charnay}, {Chauvin}, {Chomez}, {Choquet},
  {Christiaens}, {Cl{\'e}net}, {Coud{\'e} du Foresto}, {Cridland}, {Delorme},
  {Dembet}, {Drescher}, {Duvert}, {Eckart}, {Eisenhauer}, {Feuchtgruber},
  {Galland}, {Garcia}, {Garcia Lopez}, {Gardner}, {Gendron}, {Genzel},
  {Gillessen}, {Girard}, {Grandjean}, {Haubois}, {Hei{\ss}el}, {Henning},
  {Hippler}, {Horrobin}, {Houll{\'e}}, {Hubert}, {Jocou}, {Keppler},
  {Kervella}, {Kreidberg}, {Lapeyr{\`e}re}, {Le Bouquin}, {L{\'e}na}, {Lutz},
  {Maire}, {Mang}, {M{\'e}rand}, {Meunier}, {Monnier}, {Mouillet}, {Nasedkin},
  {Ott}, {Otten}, {Paladini}, {Paumard}, {Perraut}, {Perrin}, {Philipot},
  {Pfuhl}, {Pourr{\'e}}, {Pueyo}, {Rameau}, {Rickman}, {Rubini}, {Rustamkulov},
  {Samland}, {Shangguan}, {Shimizu}, {Sing}, {Straubmeier}, {Sturm}, {Tacconi},
  {van Dishoeck}, {Vigan}, {Vincent}, {Ward-Duong}, {Widmann}, {Wieprecht},
  {Wiezorrek}, {Woillez}, {Yazici}, {Young}, \& {Zicher}}]{2023A&A...671L...5H}
{Hinkley}, S., {Lacour}, S., {Marleau}, G.~D., {et~al.} 2023, \aap, 671, L5,
  \dodoi{10.1051/0004-6361/202244727}

\bibitem[{Hunter(2007)}]{Matplotlib}
Hunter, J.~D. 2007, Computing in Science \& Engineering, 9, 90,
  \dodoi{10.1109/MCSE.2007.55}

\bibitem[{{Hunter} {et~al.}(2023){Hunter}, {Indebetouw}, {Brogan}, {Berry},
  {Chang}, {Francke}, {Geers}, {G{\'o}mez}, {Hibbard}, {Humphreys}, {Kent},
  {Kepley}, {Kunneriath}, {Lipnicky}, {Loomis}, {Mason}, {Masters}, {Maud},
  {Muders}, {Sabater}, {Sugimoto}, {Sz{\H{u}}cs}, {Vasiliev}, {Videla},
  {Villard}, {Williams}, {Xue}, \& {Yoon}}]{2023PASP..135g4501H}
{Hunter}, T.~R., {Indebetouw}, R., {Brogan}, C.~L., {et~al.} 2023, \pasp, 135,
  074501, \dodoi{10.1088/1538-3873/ace216}

\bibitem[{{Ilee} {et~al.}(2021){Ilee}, {Walsh}, {Booth}, {Aikawa}, {Andrews},
  {Bae}, {Bergin}, {Bergner}, {Bosman}, {Cataldi}, {Cleeves}, {Czekala},
  {Guzm{\'a}n}, {Huang}, {Law}, {Le Gal}, {Loomis}, {M{\'e}nard}, {Nomura},
  {{\"O}berg}, {Qi}, {Schwarz}, {Teague}, {Tsukagoshi}, {Wilner}, {Yamato}, \&
  {Zhang}}]{2021ApJS..257....9I}
{Ilee}, J.~D., {Walsh}, C., {Booth}, A.~S., {et~al.} 2021, \apjs, 257, 9,
  \dodoi{10.3847/1538-4365/ac1441}

\bibitem[{{Izquierdo} {et~al.}(2022){Izquierdo}, {Facchini}, {Rosotti}, {van
  Dishoeck}, \& {Testi}}]{2022ApJ...928....2I}
{Izquierdo}, A.~F., {Facchini}, S., {Rosotti}, G.~P., {van Dishoeck}, E.~F., \&
  {Testi}, L. 2022, \apj, 928, 2, \dodoi{10.3847/1538-4357/ac474d}

\bibitem[{{Janson} {et~al.}(2019){Janson}, {Asensio-Torres}, {Andr{\'e}},
  {Bonnefoy}, {Delorme}, {Reffert}, {Desidera}, {Langlois}, {Chauvin},
  {Gratton}, {Bohn}, {Eriksson}, {Marleau}, {Mamajek}, {Vigan}, \&
  {Carson}}]{2019A&A...626A..99J}
{Janson}, M., {Asensio-Torres}, R., {Andr{\'e}}, D., {et~al.} 2019, \aap, 626,
  A99, \dodoi{10.1051/0004-6361/201935687}

\bibitem[{{Johnson} {et~al.}(2010){Johnson}, {Aller}, {Howard}, \&
  {Crepp}}]{2010PASP..122..905J}
{Johnson}, J.~A., {Aller}, K.~M., {Howard}, A.~W., \& {Crepp}, J.~R. 2010,
  \pasp, 122, 905, \dodoi{10.1086/655775}

\bibitem[{{Kalyaan} {et~al.}(2021){Kalyaan}, {Pinilla}, {Krijt}, {Mulders}, \&
  {Banzatti}}]{2021ApJ...921...84K}
{Kalyaan}, A., {Pinilla}, P., {Krijt}, S., {Mulders}, G.~D., \& {Banzatti}, A.
  2021, \apj, 921, 84, \dodoi{10.3847/1538-4357/ac1e96}

\bibitem[{{Kalyaan} {et~al.}(2023){Kalyaan}, {Pinilla}, {Krijt}, {Banzatti},
  {Rosotti}, {Mulders}, {Lambrechts}, {Long}, \&
  {Herczeg}}]{2023ApJ...954...66K}
{Kalyaan}, A., {Pinilla}, P., {Krijt}, S., {et~al.} 2023, \apj, 954, 66,
  \dodoi{10.3847/1538-4357/ace535}

\bibitem[{{Kama} {et~al.}(2016){Kama}, {Bruderer}, {van Dishoeck},
  {Hogerheijde}, {Folsom}, {Miotello}, {Fedele}, {Belloche}, {G{\"u}sten}, \&
  {Wyrowski}}]{2016A&A...592A..83K}
{Kama}, M., {Bruderer}, S., {van Dishoeck}, E.~F., {et~al.} 2016, \aap, 592,
  A83, \dodoi{10.1051/0004-6361/201526991}

\bibitem[{{Keyte} {et~al.}(2024{\natexlab{a}}){Keyte}, {Kama}, {Booth}, {Law},
  \& {Leemker}}]{2024MNRAS.534.3576K}
{Keyte}, L., {Kama}, M., {Booth}, A.~S., {Law}, C.~J., \& {Leemker}, M.
  2024{\natexlab{a}}, \mnras, 534, 3576, \dodoi{10.1093/mnras/stae2314}

\bibitem[{{Keyte} {et~al.}(2024{\natexlab{b}}){Keyte}, {Kama}, {Chuang},
  {Cleeves}, {Drozdovskaya}, {Furuya}, {Rawlings}, \&
  {Shorttle}}]{2024MNRAS.528..388K}
{Keyte}, L., {Kama}, M., {Chuang}, K.-J., {et~al.} 2024{\natexlab{b}}, \mnras,
  528, 388, \dodoi{10.1093/mnras/stae019}

\bibitem[{Kluyver {et~al.}(2016)Kluyver, Ragan-Kelley, P{\'e}rez, Granger,
  Bussonnier, Frederic, Kelley, Hamrick, Grout, Corlay, Ivanov, Avila, Abdalla,
  \& Willing}]{Jupyter}
Kluyver, T., Ragan-Kelley, B., P{\'e}rez, F., {et~al.} 2016, in Positioning and
  Power in Academic Publishing: Players, Agents and Agendas, ed. F.~Loizides \&
  B.~Schmidt, IOS Press, 87 -- 90

\bibitem[{{Konopacky} {et~al.}(2016){Konopacky}, {Rameau}, {Duch{\^e}ne},
  {Filippazzo}, {Giorla Godfrey}, {Marois}, {Nielsen}, {Pueyo}, {Rafikov},
  {Rice}, {Wang}, {Ammons}, {Bailey}, {Barman}, {Bulger}, {Bruzzone},
  {Chilcote}, {Cotten}, {Dawson}, {De Rosa}, {Doyon}, {Esposito}, {Fitzgerald},
  {Follette}, {Goodsell}, {Graham}, {Greenbaum}, {Hibon}, {Hung}, {Ingraham},
  {Kalas}, {Lafreni{\`e}re}, {Larkin}, {Macintosh}, {Maire}, {Marchis},
  {Marley}, {Matthews}, {Metchev}, {Millar-Blanchaer}, {Oppenheimer}, {Palmer},
  {Patience}, {Perrin}, {Poyneer}, {Rajan}, {Rantakyr{\"o}}, {Savransky},
  {Schneider}, {Sivaramakrishnan}, {Song}, {Soummer}, {Thomas}, {Wallace},
  {Ward-Duong}, {Wiktorowicz}, \& {Wolff}}]{2016ApJ...829L...4K}
{Konopacky}, Q.~M., {Rameau}, J., {Duch{\^e}ne}, G., {et~al.} 2016, \apjl, 829,
  L4, \dodoi{10.3847/2041-8205/829/1/L4}

\bibitem[{{Krijt} {et~al.}(2023){Krijt}, {Kama}, {McClure}, {Teske}, {Bergin},
  {Shorttle}, {Walsh}, \& {Raymond}}]{2023ASPC..534.1031K}
{Krijt}, S., {Kama}, M., {McClure}, M., {et~al.} 2023, in Astronomical Society
  of the Pacific Conference Series, Vol. 534, Protostars and Planets VII, ed.
  S.~{Inutsuka}, Y.~{Aikawa}, T.~{Muto}, K.~{Tomida}, \& M.~{Tamura}, 1031,
  \dodoi{10.48550/arXiv.2203.10056}

\bibitem[{{Kuzuhara} {et~al.}(2022){Kuzuhara}, {Currie}, {Takarada}, {Brandt},
  {Sato}, {Uyama}, {Janson}, {Chilcote}, {Tobin}, {Lawson}, {Hori}, {Guyon},
  {Groff}, {Lozi}, {Vievard}, {Sahoo}, {Deo}, {Jovanovic}, {Ahn}, {Martinache},
  {Skaf}, {Akiyama}, {Norris}, {Bonnefoy}, {He{\l}miniak}, {Kudo}, {McElwain},
  {Samland}, {Wagner}, {Wisniewski}, {Knapp}, {Kwon}, {Nishikawa}, {Serabyn},
  {Hayashi}, \& {Tamura}}]{2022ApJ...934L..18K}
{Kuzuhara}, M., {Currie}, T., {Takarada}, T., {et~al.} 2022, \apjl, 934, L18,
  \dodoi{10.3847/2041-8213/ac772f}

\bibitem[{{Lafreni{\`e}re} {et~al.}(2011){Lafreni{\`e}re}, {Jayawardhana},
  {Janson}, {Helling}, {Witte}, \& {Hauschildt}}]{2011ApJ...730...42L}
{Lafreni{\`e}re}, D., {Jayawardhana}, R., {Janson}, M., {et~al.} 2011, \apj,
  730, 42, \dodoi{10.1088/0004-637X/730/1/42}

\bibitem[{{Lagage} {et~al.}(2006){Lagage}, {Doucet}, {Pantin}, {Habart},
  {Duch{\^e}ne}, {M{\'e}nard}, {Pinte}, {Charnoz}, \&
  {Pel}}]{2006Sci...314..621L}
{Lagage}, P.-O., {Doucet}, C., {Pantin}, E., {et~al.} 2006, Science, 314, 621,
  \dodoi{10.1126/science.1131436}

\bibitem[{{Lagrange} {et~al.}(2010){Lagrange}, {Bonnefoy}, {Chauvin}, {Apai},
  {Ehrenreich}, {Boccaletti}, {Gratadour}, {Rouan}, {Mouillet}, {Lacour}, \&
  {Kasper}}]{2010Sci...329...57L}
{Lagrange}, A.~M., {Bonnefoy}, M., {Chauvin}, G., {et~al.} 2010, Science, 329,
  57, \dodoi{10.1126/science.1187187}

\bibitem[{{Landman} {et~al.}(2024){Landman}, {Stolker}, {Snellen}, {Costes},
  {de Regt}, {Zhang}, {Gandhi}, {Molliere}, {Kesseli}, {Vigan}, \&
  {Sanchez-L{\'o}pez}}]{2024A&A...682A..48L}
{Landman}, R., {Stolker}, T., {Snellen}, I.~A.~G., {et~al.} 2024, \aap, 682,
  A48, \dodoi{10.1051/0004-6361/202347846}

\bibitem[{{Law} {et~al.}(2023){Law}, {Booth}, \&
  {{\"O}berg}}]{2023ApJ...952L..19L}
{Law}, C.~J., {Booth}, A.~S., \& {{\"O}berg}, K.~I. 2023, \apjl, 952, L19,
  \dodoi{10.3847/2041-8213/acdfd0}

\bibitem[{{Law} {et~al.}(2021){Law}, {Teague}, {Loomis}, {Bae}, {{\"O}berg},
  {Czekala}, {Andrews}, {Aikawa}, {Alarc{\'o}n}, {Bergin}, {Bergner}, {Booth},
  {Bosman}, {Calahan}, {Cataldi}, {Cleeves}, {Furuya}, {Guzm{\'a}n}, {Huang},
  {Ilee}, {Le Gal}, {Liu}, {Long}, {M{\'e}nard}, {Nomura}, {P{\'e}rez}, {Qi},
  {Schwarz}, {Soto}, {Tsukagoshi}, {Yamato}, {van't Hoff}, {Walsh}, {Wilner},
  \& {Zhang}}]{2021ApJS..257....4L}
{Law}, C.~J., {Teague}, R., {Loomis}, R.~A., {et~al.} 2021, \apjs, 257, 4,
  \dodoi{10.3847/1538-4365/ac1439}

\bibitem[{{Law} {et~al.}(2022){Law}, {Crystian}, {Teague}, {{\"O}berg}, {Rich},
  {Andrews}, {Bae}, {Flaherty}, {Guzm{\'a}n}, {Huang}, {Ilee}, {Kastner},
  {Loomis}, {Long}, {P{\'e}rez}, {P{\'e}rez}, {Qi}, {Rosotti},
  {Ru{\'\i}z-Rodr{\'\i}guez}, {Tsukagoshi}, \& {Wilner}}]{2022ApJ...932..114L}
{Law}, C.~J., {Crystian}, S., {Teague}, R., {et~al.} 2022, \apj, 932, 114,
  \dodoi{10.3847/1538-4357/ac6c02}

\bibitem[{{Law} {et~al.}(2025){Law}, {Le Gal}, {Yamato}, {Zhang}, {Guzm{\'a}n},
  {Hern{\'a}ndez-Vera}, {Cleeves}, {Guidi}, \& {Booth}}]{2025ApJ...985...84L}
{Law}, C.~J., {Le Gal}, R., {Yamato}, Y., {et~al.} 2025, \apj, 985, 84,
  \dodoi{10.3847/1538-4357/adc304}

\bibitem[{{Le Gal} {et~al.}(2021){Le Gal}, {{\"O}berg}, {Teague}, {Loomis},
  {Law}, {Walsh}, {Bergin}, {M{\'e}nard}, {Wilner}, {Andrews}, {Aikawa},
  {Booth}, {Cataldi}, {Bergner}, {Bosman}, {Cleeves}, {Czekala}, {Furuya},
  {Guzm{\'a}n}, {Huang}, {Ilee}, {Nomura}, {Qi}, {Schwarz}, {Tsukagoshi},
  {Yamato}, \& {Zhang}}]{2021ApJS..257...12L}
{Le Gal}, R., {{\"O}berg}, K.~I., {Teague}, R., {et~al.} 2021, \apjs, 257, 12,
  \dodoi{10.3847/1538-4365/ac2583}

\bibitem[{{Leemker} {et~al.}(2023){Leemker}, {Booth}, {van Dishoeck}, {van der
  Marel}, {Tabone}, {Ligterink}, {Brunken}, \&
  {Hogerheijde}}]{2023arXiv230300768L}
{Leemker}, M., {Booth}, A.~S., {van Dishoeck}, E.~F., {et~al.} 2023, \aap, 673,
  A7, \dodoi{10.1051/0004-6361/202245662}

\bibitem[{{Leemker} {et~al.}(2024){Leemker}, {Booth}, {van Dishoeck},
  {W{\"o}lfer}, \& {Dent}}]{2024A&A...687A.299L}
{Leemker}, M., {Booth}, A.~S., {van Dishoeck}, E.~F., {W{\"o}lfer}, L., \&
  {Dent}, B. 2024, \aap, 687, A299, \dodoi{10.1051/0004-6361/202349072}

\bibitem[{{Leemker} {et~al.}(2021){Leemker}, {van't Hoff}, {Trapman}, {van
  Gelder}, {Hogerheijde}, {Ru{\'\i}z-Rodr{\'\i}guez}, \& {van
  Dishoeck}}]{2021A&A...646A...3L}
{Leemker}, M., {van't Hoff}, M.~L.~R., {Trapman}, L., {et~al.} 2021, \aap, 646,
  A3, \dodoi{10.1051/0004-6361/202039387}

\bibitem[{{Leemker} {et~al.}(2022){Leemker}, {Booth}, {van Dishoeck},
  {P{\'e}rez-S{\'a}nchez}, {Szul{\'a}gyi}, {Bosman}, {Bruderer}, {Facchini},
  {Hogerheijde}, {Paneque-Carre{\~n}o}, \& {Sturm}}]{2022A&A...663A..23L}
{Leemker}, M., {Booth}, A.~S., {van Dishoeck}, E.~F., {et~al.} 2022, \aap, 663,
  A23, \dodoi{10.1051/0004-6361/202243229}

\bibitem[{{Loomis} {et~al.}(2015){Loomis}, {Cleeves}, {{\"O}berg}, {Guzman}, \&
  {Andrews}}]{2015ApJ...809L..25L}
{Loomis}, R.~A., {Cleeves}, L.~I., {{\"O}berg}, K.~I., {Guzman}, V.~V., \&
  {Andrews}, S.~M. 2015, \apjl, 809, L25, \dodoi{10.1088/2041-8205/809/2/L25}

\bibitem[{{Loomis} {et~al.}(2018){Loomis}, {{\"O}berg}, {Andrews}, {Walsh},
  {Czekala}, {Huang}, \& {Rosenfeld}}]{Loomis2018}
{Loomis}, R.~A., {{\"O}berg}, K.~I., {Andrews}, S.~M., {et~al.} 2018, \aj, 155,
  182, \dodoi{10.3847/1538-3881/aab604}

\bibitem[{{Loomis} {et~al.}(2020){Loomis}, {{\"O}berg}, {Andrews}, {Bergin},
  {Bergner}, {Blake}, {Cleeves}, {Czekala}, {Huang}, {Le Gal}, {M{\'e}nard},
  {Pegues}, {Qi}, {Walsh}, {Williams}, \& {Wilner}}]{2020ApJ...893..101L}
---. 2020, \apj, 893, 101, \dodoi{10.3847/1538-4357/ab7cc8}

\bibitem[{{Loomis} {et~al.}(2025){Loomis}, {Facchini}, {Benisty}, {Curone},
  {Ilee}, {Cataldi}, {Yen}, {Teague}, {Pinte}, {Huang}, {Garg}, {Orihara},
  {Czekala}, {Zawadzki}, {Andrews}, {Wilner}, {Bae}, {Barraza-Alfaro},
  {Fasano}, {Flock}, {Fukagawa}, {Galloway-Sprietsma}, {Izquierdo}, {Kanagawa},
  {Lesur}, {Longarini}, {Menard}, {Price}, {Rosotti}, {Stadler},
  {Wafflard-Fernandez}, {W{\"o}lfer}, \& {Yoshida}}]{2025ApJ...984L...7L}
{Loomis}, R.~A., {Facchini}, S., {Benisty}, M., {et~al.} 2025, \apjl, 984, L7,
  \dodoi{10.3847/2041-8213/adc43a}

\bibitem[{{Mac{\'\i}as} {et~al.}(2019){Mac{\'\i}as}, {Espaillat}, {Osorio},
  {Anglada}, {Torrelles}, {Carrasco-Gonz{\'a}lez}, {Flock}, {Linz}, {Bertrang},
  {Henning}, {G{\'o}mez}, {Calvet}, \& {Dent}}]{2019ApJ...881..159M}
{Mac{\'\i}as}, E., {Espaillat}, C.~C., {Osorio}, M., {et~al.} 2019, \apj, 881,
  159, \dodoi{10.3847/1538-4357/ab31a2}

\bibitem[{{Macintosh} {et~al.}(2015){Macintosh}, {Graham}, {Barman}, {De Rosa},
  {Konopacky}, {Marley}, {Marois}, {Nielsen}, {Pueyo}, {Rajan}, {Rameau},
  {Saumon}, {Wang}, {Patience}, {Ammons}, {Arriaga}, {Artigau}, {Beckwith},
  {Brewster}, {Bruzzone}, {Bulger}, {Burningham}, {Burrows}, {Chen}, {Chiang},
  {Chilcote}, {Dawson}, {Dong}, {Doyon}, {Draper}, {Duch{\^e}ne}, {Esposito},
  {Fabrycky}, {Fitzgerald}, {Follette}, {Fortney}, {Gerard}, {Goodsell},
  {Greenbaum}, {Hibon}, {Hinkley}, {Cotten}, {Hung}, {Ingraham},
  {Johnson-Groh}, {Kalas}, {Lafreniere}, {Larkin}, {Lee}, {Line}, {Long},
  {Maire}, {Marchis}, {Matthews}, {Max}, {Metchev}, {Millar-Blanchaer},
  {Mittal}, {Morley}, {Morzinski}, {Murray-Clay}, {Oppenheimer}, {Palmer},
  {Patel}, {Perrin}, {Poyneer}, {Rafikov}, {Rantakyr{\"o}}, {Rice}, {Rojo},
  {Rudy}, {Ruffio}, {Ruiz}, {Sadakuni}, {Saddlemyer}, {Salama}, {Savransky},
  {Schneider}, {Sivaramakrishnan}, {Song}, {Soummer}, {Thomas}, {Vasisht},
  {Wallace}, {Ward-Duong}, {Wiktorowicz}, {Wolff}, \&
  {Zuckerman}}]{2015Sci...350...64M}
{Macintosh}, B., {Graham}, J.~R., {Barman}, T., {et~al.} 2015, Science, 350,
  64, \dodoi{10.1126/science.aac5891}

\bibitem[{{M{\^a}lin} {et~al.}(2024){M{\^a}lin}, {Boccaletti}, {Perrot},
  {Baudoz}, {Rouan}, {Lagage}, {Waters}, {G{\"u}del}, {Henning},
  {Vandenbussche}, {Absil}, {Barrado}, {Cossou}, {Decin}, {Glauser}, {Pye},
  {Olofsson}, {Glasse}, {Lahuis}, {Patapis}, {Royer}, {Scheithauer},
  {Whiteford}, {Serabyn}, {Choquet}, {Colina}, {Ostlin}, {Ray}, \&
  {Wright}}]{2024A&A...690A.316M}
{M{\^a}lin}, M., {Boccaletti}, A., {Perrot}, C., {et~al.} 2024, \aap, 690,
  A316, \dodoi{10.1051/0004-6361/202450470}

\bibitem[{{Marois} {et~al.}(2008){Marois}, {Macintosh}, {Barman}, {Zuckerman},
  {Song}, {Patience}, {Lafreni{\`e}re}, \& {Doyon}}]{2008Sci...322.1348M}
{Marois}, C., {Macintosh}, B., {Barman}, T., {et~al.} 2008, Science, 322, 1348,
  \dodoi{10.1126/science.1166585}

\bibitem[{{Marois} {et~al.}(2010){Marois}, {Zuckerman}, {Konopacky},
  {Macintosh}, \& {Barman}}]{2010Natur.468.1080M}
{Marois}, C., {Zuckerman}, B., {Konopacky}, Q.~M., {Macintosh}, B., \&
  {Barman}, T. 2010, \nat, 468, 1080, \dodoi{10.1038/nature09684}

\bibitem[{{McMullin} {et~al.}(2007){McMullin}, {Waters}, {Schiebel}, {Young},
  \& {Golap}}]{2007ASPC..376..127M}
{McMullin}, J.~P., {Waters}, B., {Schiebel}, D., {Young}, W., \& {Golap}, K.
  2007, in Astronomical Society of the Pacific Conference Series, Vol. 376,
  Astronomical Data Analysis Software and Systems XVI, ed. R.~A. {Shaw},
  F.~{Hill}, \& D.~J. {Bell}, 127

\bibitem[{{Mesa} {et~al.}(2023){Mesa}, {Gratton}, {Kervella}, {Bonavita},
  {Desidera}, {D'Orazi}, {Marino}, {Zurlo}, \&
  {Rigliaco}}]{2023A&A...672A..93M}
{Mesa}, D., {Gratton}, R., {Kervella}, P., {et~al.} 2023, \aap, 672, A93,
  \dodoi{10.1051/0004-6361/202345865}

\bibitem[{{Minissale} {et~al.}(2022){Minissale}, {Aikawa}, {Bergin}, {Bertin},
  {Brown}, {Cazaux}, {Charnley}, {Coutens}, {Cuppen}, {Guzman}, {Linnartz},
  {McCoustra}, {Rimola}, {Schrauwen}, {Toubin}, {Ugliengo}, {Watanabe},
  {Wakelam}, \& {Dulieu}}]{2022ESC.....6..597M}
{Minissale}, M., {Aikawa}, Y., {Bergin}, E., {et~al.} 2022, ACS Earth and Space
  Chemistry, 6, 597, \dodoi{10.1021/acsearthspacechem.1c00357}

\bibitem[{{M{\"u}ller} {et~al.}(2005){M{\"u}ller}, {Schl{\"o}der}, {Stutzki},
  \& {Winnewisser}}]{2005JMoSt.742..215M}
{M{\"u}ller}, H. S.~P., {Schl{\"o}der}, F., {Stutzki}, J., \& {Winnewisser}, G.
  2005, Journal of Molecular Structure, 742, 215,
  \dodoi{10.1016/j.molstruc.2005.01.027}

\bibitem[{{M{\"u}ller} {et~al.}(2001){M{\"u}ller}, {Thorwirth}, {Roth}, \&
  {Winnewisser}}]{2001A&A...370L..49M}
{M{\"u}ller}, H.~S.~P., {Thorwirth}, S., {Roth}, D.~A., \& {Winnewisser}, G.
  2001, \aap, 370, L49, \dodoi{10.1051/0004-6361:20010367}

\bibitem[{{Nasedkin} {et~al.}(2024){Nasedkin}, {Molli{\`e}re}, {Lacour},
  {Nowak}, {Kreidberg}, {Stolker}, {Wang}, {Balmer}, {Kammerer}, {Shangguan},
  {Abuter}, {Amorim}, {Asensio-Torres}, {Benisty}, {Berger}, {Beust}, {Blunt},
  {Boccaletti}, {Bonnefoy}, {Bonnet}, {Bordoni}, {Bourdarot}, {Brandner},
  {Cantalloube}, {Caselli}, {Charnay}, {Chauvin}, {Chavez}, {Choquet},
  {Christiaens}, {Cl{\'e}net}, {Coud{\'e} Du Foresto}, {Cridland}, {Davies},
  {Dembet}, {Dexter}, {Drescher}, {Duvert}, {Eckart}, {Eisenhauer},
  {F{\"o}rster Schreiber}, {Garcia}, {Garcia Lopez}, {Gendron}, {Genzel},
  {Gillessen}, {Girard}, {Grant}, {Haubois}, {Hei{\ss}el}, {Henning},
  {Hinkley}, {Hippler}, {Houll{\'e}}, {Hubert}, {Jocou}, {Keppler}, {Kervella},
  {Kurtovic}, {Lagrange}, {Lapeyr{\`e}re}, {Le Bouquin}, {Lutz}, {Maire},
  {Mang}, {Marleau}, {M{\'e}rand}, {Monnier}, {Mordasini}, {Ott}, {Otten},
  {Paladini}, {Paumard}, {Perraut}, {Perrin}, {Pfuhl}, {Pourr{\'e}}, {Pueyo},
  {Ribeiro}, {Rickman}, {Ruffio}, {Rustamkulov}, {Shimizu}, {Sing}, {Stadler},
  {Straub}, {Straubmeier}, {Sturm}, {Tacconi}, {van Dishoeck}, {Vigan},
  {Vincent}, {von Fellenberg}, {Widmann}, {Winterhalder}, {Woillez}, {Yazici},
  \& {Gravity Collaboration}}]{2024A&A...687A.298N}
{Nasedkin}, E., {Molli{\`e}re}, P., {Lacour}, S., {et~al.} 2024, \aap, 687,
  A298, \dodoi{10.1051/0004-6361/202449328}

\bibitem[{{Nielsen} {et~al.}(2019){Nielsen}, {De Rosa}, {Macintosh}, {Wang},
  {Ruffio}, {Chiang}, {Marley}, {Saumon}, {Savransky}, {Ammons}, {Bailey},
  {Barman}, {Blain}, {Bulger}, {Burrows}, {Chilcote}, {Cotten}, {Czekala},
  {Doyon}, {Duch{\^e}ne}, {Esposito}, {Fabrycky}, {Fitzgerald}, {Follette},
  {Fortney}, {Gerard}, {Goodsell}, {Graham}, {Greenbaum}, {Hibon}, {Hinkley},
  {Hirsch}, {Hom}, {Hung}, {Dawson}, {Ingraham}, {Kalas}, {Konopacky},
  {Larkin}, {Lee}, {Lin}, {Maire}, {Marchis}, {Marois}, {Metchev},
  {Millar-Blanchaer}, {Morzinski}, {Oppenheimer}, {Palmer}, {Patience},
  {Perrin}, {Poyneer}, {Pueyo}, {Rafikov}, {Rajan}, {Rameau}, {Rantakyr{\"o}},
  {Ren}, {Schneider}, {Sivaramakrishnan}, {Song}, {Soummer}, {Tallis},
  {Thomas}, {Ward-Duong}, \& {Wolff}}]{2019AJ....158...13N}
{Nielsen}, E.~L., {De Rosa}, R.~J., {Macintosh}, B., {et~al.} 2019, \aj, 158,
  13, \dodoi{10.3847/1538-3881/ab16e9}

\bibitem[{{Notsu} {et~al.}(2020){Notsu}, {Eistrup}, {Walsh}, \&
  {Nomura}}]{2020MNRAS.499.2229N}
{Notsu}, S., {Eistrup}, C., {Walsh}, C., \& {Nomura}, H. 2020, \mnras, 499,
  2229, \dodoi{10.1093/mnras/staa2944}

\bibitem[{{{\"O}berg} {et~al.}(2015){{\"O}berg}, {Guzm{\'a}n}, {Furuya}, {Qi},
  {Aikawa}, {Andrews}, {Loomis}, \& {Wilner}}]{2015Natur.520..198O}
{{\"O}berg}, K.~I., {Guzm{\'a}n}, V.~V., {Furuya}, K., {et~al.} 2015, \nat,
  520, 198, \dodoi{10.1038/nature14276}

\bibitem[{{{\"O}berg} {et~al.}(2011){{\"O}berg}, {Murray-Clay}, \&
  {Bergin}}]{2011ApJ...743L..16O}
{{\"O}berg}, K.~I., {Murray-Clay}, R., \& {Bergin}, E.~A. 2011, \apjl, 743,
  L16, \dodoi{10.1088/2041-8205/743/1/L16}

\bibitem[{{{\"O}berg} {et~al.}(2021){{\"O}berg}, {Guzm{\'a}n}, {Walsh},
  {Aikawa}, {Bergin}, {Law}, {Loomis}, {Alarc{\'o}n}, {Andrews}, {Bae},
  {Bergner}, {Boehler}, {Booth}, {Bosman}, {Calahan}, {Cataldi}, {Cleeves},
  {Czekala}, {Furuya}, {Huang}, {Ilee}, {Kurtovic}, {Le Gal}, {Liu}, {Long},
  {M{\'e}nard}, {Nomura}, {P{\'e}rez}, {Qi}, {Schwarz}, {Sierra}, {Teague},
  {Tsukagoshi}, {Yamato}, {van't Hoff}, {Waggoner}, {Wilner}, \&
  {Zhang}}]{2021ApJS..257....1O}
{{\"O}berg}, K.~I., {Guzm{\'a}n}, V.~V., {Walsh}, C., {et~al.} 2021, \apjs,
  257, 1, \dodoi{10.3847/1538-4365/ac1432}

\bibitem[{{Pacetti} {et~al.}(2022){Pacetti}, {Turrini}, {Schisano}, {Molinari},
  {Fonte}, {Politi}, {Hennebelle}, {Klessen}, {Testi}, \&
  {Lebreuilly}}]{2022ApJ...937...36P}
{Pacetti}, E., {Turrini}, D., {Schisano}, E., {et~al.} 2022, \apj, 937, 36,
  \dodoi{10.3847/1538-4357/ac8b11}

\bibitem[{{Pacheco-V{\'a}zquez} {et~al.}(2016){Pacheco-V{\'a}zquez}, {Fuente},
  {Baruteau}, {Bern{\'e}}, {Ag{\'u}ndez}, {Neri}, {Goicoechea}, {Cernicharo},
  \& {Bachiller}}]{abaur_chem}
{Pacheco-V{\'a}zquez}, S., {Fuente}, A., {Baruteau}, C., {et~al.} 2016, \aap,
  589, A60, \dodoi{10.1051/0004-6361/201527089}

\bibitem[{{Paneque-Carre{\~n}o} {et~al.}(2025){Paneque-Carre{\~n}o},
  {Miotello}, {van Dishoeck}, {Rosotti}, \& {Tabone}}]{2025A&A...698A.231P}
{Paneque-Carre{\~n}o}, T., {Miotello}, A., {van Dishoeck}, E.~F., {Rosotti},
  G., \& {Tabone}, B. 2025, \aap, 698, A231,
  \dodoi{10.1051/0004-6361/202451862}

\bibitem[{{Pegues} {et~al.}(2020){Pegues}, {{\"O}berg}, {Bergner}, {Loomis},
  {Qi}, {Le Gal}, {Cleeves}, {Guzm{\'a}n}, {Huang}, {J{\o}rgensen}, {Andrews},
  {Blake}, {Carpenter}, {Schwarz}, {Williams}, \&
  {Wilner}}]{2020ApJ...890..142P}
{Pegues}, J., {{\"O}berg}, K.~I., {Bergner}, J.~B., {et~al.} 2020, \apj, 890,
  142, \dodoi{10.3847/1538-4357/ab64d9}

\bibitem[{{Pegues} {et~al.}(2023){Pegues}, {{\"O}berg}, {Qi}, {Andrews},
  {Huang}, {Law}, {Le Gal}, {Matr{\`a}}, \& {Wilner}}]{2023ApJ...948...57P}
{Pegues}, J., {{\"O}berg}, K.~I., {Qi}, C., {et~al.} 2023, \apj, 948, 57,
  \dodoi{10.3847/1538-4357/acbf31}

\bibitem[{P\'erez \& Granger(2007)}]{IPython}
P\'erez, F., \& Granger, B.~E. 2007, Computing in Science and Engineering, 9,
  21, \dodoi{10.1109/MCSE.2007.53}

\bibitem[{{P{\'e}rez} {et~al.}(2019){P{\'e}rez}, {Casassus}, {Baruteau},
  {Dong}, {Hales}, \& {Cieza}}]{2019AJ....158...15P}
{P{\'e}rez}, S., {Casassus}, S., {Baruteau}, C., {et~al.} 2019, \aj, 158, 15,
  \dodoi{10.3847/1538-3881/ab1f88}

\bibitem[{{Phuong} {et~al.}(2018){Phuong}, {Chapillon}, {Majumdar}, {Dutrey},
  {Guilloteau}, {Pi{\'e}tu}, {Wakelam}, {Diep}, {Tang}, {Beck}, \&
  {Bary}}]{2018A&A...616L...5P}
{Phuong}, N.~T., {Chapillon}, E., {Majumdar}, L., {et~al.} 2018, \aap, 616, L5,
  \dodoi{10.1051/0004-6361/201833766}

\bibitem[{{Pinte} {et~al.}(2019){Pinte}, {van der Plas}, {M{\'e}nard}, {Price},
  {Christiaens}, {Hill}, {Mentiplay}, {Ginski}, {Choquet}, {Boehler},
  {Duch{\^e}ne}, {Perez}, \& {Casassus}}]{2019NatAs...3.1109P}
{Pinte}, C., {van der Plas}, G., {M{\'e}nard}, F., {et~al.} 2019, Nature
  Astronomy, 3, 1109, \dodoi{10.1038/s41550-019-0852-6}

\bibitem[{{Pinte} {et~al.}(2020){Pinte}, {Price}, {M{\'e}nard}, {Duch{\^e}ne},
  {Christiaens}, {Andrews}, {Huang}, {Hill}, {van der Plas}, {Perez}, {Isella},
  {Boehler}, {Dent}, {Mentiplay}, \& {Loomis}}]{2020ApJ...890L...9P}
{Pinte}, C., {Price}, D.~J., {M{\'e}nard}, F., {et~al.} 2020, \apjl, 890, L9,
  \dodoi{10.3847/2041-8213/ab6dda}

\bibitem[{{Rameau} {et~al.}(2013){Rameau}, {Chauvin}, {Lagrange}, {Boccaletti},
  {Quanz}, {Bonnefoy}, {Girard}, {Delorme}, {Desidera}, {Klahr}, {Mordasini},
  {Dumas}, \& {Bonavita}}]{2013ApJ...772L..15R}
{Rameau}, J., {Chauvin}, G., {Lagrange}, A.~M., {et~al.} 2013, \apjl, 772, L15,
  \dodoi{10.1088/2041-8205/772/2/L15}

\bibitem[{{Reffert} {et~al.}(2015){Reffert}, {Bergmann}, {Quirrenbach},
  {Trifonov}, \& {K{\"u}nstler}}]{2015A&A...574A.116R}
{Reffert}, S., {Bergmann}, C., {Quirrenbach}, A., {Trifonov}, T., \&
  {K{\"u}nstler}, A. 2015, \aap, 574, A116, \dodoi{10.1051/0004-6361/201322360}

\bibitem[{{Reggiani} {et~al.}(2018){Reggiani}, {Christiaens}, {Absil}, {Mawet},
  {Huby}, {Choquet}, {Gomez Gonzalez}, {Ruane}, {Femenia}, {Serabyn},
  {Matthews}, {Barraza}, {Carlomagno}, {Defr{\`e}re}, {Delacroix}, {Habraken},
  {Jolivet}, {Karlsson}, {Orban de Xivry}, {Piron}, {Surdej}, {Vargas Catalan},
  \& {Wertz}}]{2018A&A...611A..74R}
{Reggiani}, M., {Christiaens}, V., {Absil}, O., {et~al.} 2018, \aap, 611, A74,
  \dodoi{10.1051/0004-6361/201732016}

\bibitem[{{Rivi{\`e}re-Marichalar} {et~al.}(2020){Rivi{\`e}re-Marichalar},
  {Fuente}, {Le Gal}, {Baruteau}, {Neri}, {Navarro-Almaida},
  {Trevi{\~n}o-Morales}, {Mac{\'\i}as}, {Bachiller}, \&
  {Osorio}}]{2020A&A...642A..32R}
{Rivi{\`e}re-Marichalar}, P., {Fuente}, A., {Le Gal}, R., {et~al.} 2020, \aap,
  642, A32, \dodoi{10.1051/0004-6361/202038549}

\bibitem[{{Rivi{\`e}re-Marichalar} {et~al.}(2022){Rivi{\`e}re-Marichalar},
  {Fuente}, {Esplugues}, {Wakelam}, {le Gal}, {Baruteau}, {Ribas},
  {Mac{\'\i}as}, {Neri}, \& {Navarro-Almaida}}]{2022A&A...665A..61R}
{Rivi{\`e}re-Marichalar}, P., {Fuente}, A., {Esplugues}, G., {et~al.} 2022,
  \aap, 665, A61, \dodoi{10.1051/0004-6361/202142906}

\bibitem[{{Rosotti} {et~al.}(2020){Rosotti}, {Benisty}, {Juh{\'a}sz}, {Teague},
  {Clarke}, {Dominik}, {Dullemond}, {Klaassen}, {Matr{\`a}}, \&
  {Stolker}}]{2020MNRAS.491.1335R}
{Rosotti}, G.~P., {Benisty}, M., {Juh{\'a}sz}, A., {et~al.} 2020, \mnras, 491,
  1335, \dodoi{10.1093/mnras/stz3090}

\bibitem[{{Sainsbury-Martinez} \& {Walsh}(2024)}]{2024ApJ...966...39S}
{Sainsbury-Martinez}, F., \& {Walsh}, C. 2024, \apj, 966, 39,
  \dodoi{10.3847/1538-4357/ad28b3}

\bibitem[{{Sappey} {et~al.}(2025){Sappey}, {Konopacky}, {{\'O}}, {Barman},
  {Ruffio}, {Wang}, {Theissen}, {Finnerty}, {Xuan}, {Hortsman}, {Mawet},
  {Zhang}, {Inglis}, {Wallack}, {Sanghi}, {Baker}, {Bartos}, {Blake}, {Bond},
  {Calvin}, {Cetre}, {Delorme}, {Doppmann}, {Echeverri}, {Fitzgerald}, {Hsu},
  {Jovanovic}, {Liberman}, {L{\'o}pez}, {Martin}, {Morris}, {Pezzato-Rovner},
  {Phillips}, {Ruane}, {Schofield}, {Skemer}, {Venenciano}, {Wallace}, {Wang},
  {Wizinowich}, \& {Xin}}]{2025AJ....169..175S}
{Sappey}, B., {Konopacky}, Q., {{\'O}}, C. R.~D., {et~al.} 2025, \aj, 169, 175,
  \dodoi{10.3847/1538-3881/adad70}

\bibitem[{{Semenov} {et~al.}(2018){Semenov}, {Favre}, {Fedele}, {Guilloteau},
  {Teague}, {Henning}, {Dutrey}, {Chapillon}, {Hersant}, \&
  {Pi{\'e}tu}}]{2018A&A...617A..28S}
{Semenov}, D., {Favre}, C., {Fedele}, D., {et~al.} 2018, \aap, 617, A28,
  \dodoi{10.1051/0004-6361/201832980}

\bibitem[{{Smirnov-Pinchukov} {et~al.}(2022){Smirnov-Pinchukov}, {Mo{\'o}r},
  {Semenov}, {{\'A}brah{\'a}m}, {Henning}, {K{\'o}sp{\'a}l}, {Hughes}, \& {di
  Folco}}]{2022MNRAS.510.1148S}
{Smirnov-Pinchukov}, G.~V., {Mo{\'o}r}, A., {Semenov}, D.~A., {et~al.} 2022,
  \mnras, 510, 1148, \dodoi{10.1093/mnras/stab3146}

\bibitem[{{Speedie} {et~al.}(2025){Speedie}, {Dong}, {Teague}, {Segura-Cox},
  {Pineda}, {Calcino}, {Longarini}, {Hall}, {Tang}, {Hashimoto},
  {Paneque-Carre{\~n}o}, {Lodato}, \& {Veronesi}}]{2025ApJ...981L..30S}
{Speedie}, J., {Dong}, R., {Teague}, R., {et~al.} 2025, \apjl, 981, L30,
  \dodoi{10.3847/2041-8213/adb7d5}

\bibitem[{{Squicciarini} {et~al.}(2025){Squicciarini}, {Mazoyer}, {Lagrange},
  {Chomez}, {Delorme}, {Flasseur}, {Kiefer}, {Bergeon}, {Albert}, \&
  {Meunier}}]{2025A&A...693A..54S}
{Squicciarini}, V., {Mazoyer}, J., {Lagrange}, A.~M., {et~al.} 2025, \aap, 693,
  A54, \dodoi{10.1051/0004-6361/202452310}

\bibitem[{{Stapper} {et~al.}(2025{\natexlab{a}}){Stapper}, {Hogerheijde}, {van
  Dishoeck}, {Booth}, {Grant}, \& {van Terwisga}}]{2025A&A...693A..49S}
{Stapper}, L.~M., {Hogerheijde}, M.~R., {van Dishoeck}, E.~F., {et~al.}
  2025{\natexlab{a}}, \aap, 693, A49, \dodoi{10.1051/0004-6361/202450678}

\bibitem[{{Stapper} {et~al.}(2022){Stapper}, {Hogerheijde}, {van Dishoeck}, \&
  {Mentel}}]{2022A&A...658A.112S}
{Stapper}, L.~M., {Hogerheijde}, M.~R., {van Dishoeck}, E.~F., \& {Mentel}, R.
  2022, \aap, 658, A112, \dodoi{10.1051/0004-6361/202142164}

\bibitem[{{Stapper} {et~al.}(2025{\natexlab{b}}){Stapper}, {Hogerheijde}, {van
  Dishoeck}, {Vioque}, {Williams}, \& {Ginski}}]{2025A&A...693A.286S}
{Stapper}, L.~M., {Hogerheijde}, M.~R., {van Dishoeck}, E.~F., {et~al.}
  2025{\natexlab{b}}, \aap, 693, A286, \dodoi{10.1051/0004-6361/202450260}

\bibitem[{{Stapper} {et~al.}(2024){Stapper}, {Hogerheijde}, {van Dishoeck},
  {Lin}, {Ahmadi}, {Booth}, {Grant}, {Immer}, {Leemker}, \&
  {P{\'e}rez-S{\'a}nchez}}]{2024A&A...682A.149S}
---. 2024, \aap, 682, A149, \dodoi{10.1051/0004-6361/202347271}

\bibitem[{{Stolker} {et~al.}(2016){Stolker}, {Dominik}, {Avenhaus}, {Min}, {de
  Boer}, {Ginski}, {Schmid}, {Juhasz}, {Bazzon}, {Waters}, {Garufi},
  {Augereau}, {Benisty}, {Boccaletti}, {Henning}, {Langlois}, {Maire},
  {M{\'e}nard}, {Meyer}, {Pinte}, {Quanz}, {Thalmann}, {Beuzit}, {Carbillet},
  {Costille}, {Dohlen}, {Feldt}, {Gisler}, {Mouillet}, {Pavlov}, {Perret},
  {Petit}, {Pragt}, {Rochat}, {Roelfsema}, {Salasnich}, {Soenke}, \&
  {Wildi}}]{2016A&A...595A.113S}
{Stolker}, T., {Dominik}, C., {Avenhaus}, H., {et~al.} 2016, \aap, 595, A113,
  \dodoi{10.1051/0004-6361/201528039}

\bibitem[{{Stolker} {et~al.}(2025){Stolker}, {Samland}, {Waters}, {van den
  Ancker}, {Balmer}, {Lacour}, {Sitko}, {Wang}, {Nowak}, {Maire}, {Kammerer},
  {Otten}, {Abuter}, {Amorim}, {Benisty}, {Berger}, {Beust}, {Blunt},
  {Boccaletti}, {Bonnefoy}, {Bonnet}, {Bordoni}, {Bourdarot}, {Brandner},
  {Cantalloube}, {Caselli}, {Charnay}, {Chauvin}, {Chavez}, {Chomez},
  {Choquet}, {Christiaens}, {Cl{\'e}net}, {Coud{\'e} du Foresto}, {Cridland},
  {Davies}, {Dembet}, {Dexter}, {Dominik}, {Drescher}, {Duvert}, {Eckart},
  {Eisenhauer}, {F{\"o}rster Schreiber}, {Garcia}, {Garcia Lopez}, {Gardner},
  {Gendron}, {Genzel}, {Gillessen}, {Girard}, {Grant}, {Haubois}, {Hei{\ss}el},
  {Henning}, {Hinkley}, {Hippler}, {Houll{\'e}}, {Hubert}, {Jocou}, {Keppler},
  {Kervella}, {Kreidberg}, {Kurtovic}, {Lagrange}, {Lapeyr{\`e}re}, {Le
  Bouquin}, {Lutz}, {Mang}, {Marleau}, {M{\'e}rand}, {Min}, {Molli{\`e}re},
  {Monnier}, {Mordasini}, {Mouillet}, {Nasedkin}, {Ott}, {Paladini}, {Paumard},
  {Perraut}, {Perrin}, {Pfuhl}, {Pourr{\'e}}, {Pueyo}, {Quanz}, {Ribeiro},
  {Rickman}, {Rustamkulov}, {Shangguan}, {Shimizu}, {Sing}, {Stadler},
  {Straub}, {Straubmeier}, {Sturm}, {Tacconi}, {van Dishoeck}, {Vigan},
  {Vincent}, {von Fellenberg}, {Widmann}, {Winterhalder}, {Woillez}, \&
  {Yazici}}]{2025arXiv250706206S}
{Stolker}, T., {Samland}, M., {Waters}, L.~B.~F.~M., {et~al.} 2025, \aap, 700,
  A21, \dodoi{10.1051/0004-6361/202555064}

\bibitem[{Teague(2019)}]{GoFish}
Teague, R. 2019, The Journal of Open Source Software, 4, 1632,
  \dodoi{10.21105/joss.01632}

\bibitem[{{Teague} {et~al.}(2018){Teague}, {Bae}, {Bergin}, {Birnstiel}, \&
  {Foreman-Mackey}}]{2018ApJ...860L..12T}
{Teague}, R., {Bae}, J., {Bergin}, E.~A., {Birnstiel}, T., \& {Foreman-Mackey},
  D. 2018, \apjl, 860, L12, \dodoi{10.3847/2041-8213/aac6d7}

\bibitem[{{Teague} {et~al.}(2016){Teague}, {Guilloteau}, {Semenov}, {Henning},
  {Dutrey}, {Pi{\'e}tu}, {Birnstiel}, {Chapillon}, {Hollenbach}, \&
  {Gorti}}]{2016A&A...592A..49T}
{Teague}, R., {Guilloteau}, S., {Semenov}, D., {et~al.} 2016, \aap, 592, A49,
  \dodoi{10.1051/0004-6361/201628550}

\bibitem[{{Teague} {et~al.}(2025){Teague}, {Benisty}, {Facchini}, {Fukagawa},
  {Pinte}, {Andrews}, {Bae}, {Barraza-Alfaro}, {Cataldi}, {Cuello}, {Curone},
  {Czekala}, {Fasano}, {Flock}, {Galloway-Sprietsma}, {Garg}, {Hall},
  {Hammond}, {Hilder}, {Huang}, {Ilee}, {Izquierdo}, {Kanagawa}, {Lesur},
  {Lodato}, {Longarini}, {Loomis}, {Masset}, {Menard}, {Orihara}, {Price},
  {Rosotti}, {Stadler}, {Testi}, {Yen}, {Wafflard-Fernandez}, {Wilner},
  {Winter}, {W{\"o}lfer}, {Yoshida}, \& {Zawadzki}}]{2025ApJ...984L...6T}
{Teague}, R., {Benisty}, M., {Facchini}, S., {et~al.} 2025, \apjl, 984, L6,
  \dodoi{10.3847/2041-8213/adc43b}

\bibitem[{{Temmink} {et~al.}(2023){Temmink}, {Booth}, {van der Marel}, \& {van
  Dishoeck}}]{Temmink2023}
{Temmink}, M., {Booth}, A.~S., {van der Marel}, N., \& {van Dishoeck}, E.~F.
  2023, \aap, 675, A131, \dodoi{10.1051/0004-6361/202346272}

\bibitem[{{Temmink} {et~al.}(2025){Temmink}, {Booth}, {Leemker}, {van der
  Marel}, {van Dishoeck}, {Evans}, {Keyte}, {Law}, {Notsu}, {{\"O}berg}, \&
  {Walsh}}]{2025A&A...693A101T}
{Temmink}, M., {Booth}, A.~S., {Leemker}, M., {et~al.} 2025, \aap, 693, A101,
  \dodoi{10.1051/0004-6361/202452175}

\bibitem[{{Terwisscha van Scheltinga} {et~al.}(2021){Terwisscha van
  Scheltinga}, {Hogerheijde}, {Cleeves}, {Loomis}, {Walsh}, {{\"O}berg},
  {Bergin}, {Bergner}, {Blake}, {Calahan}, {Cazzoletti}, {van Dishoeck},
  {Guzm{\'a}n}, {Huang}, {Kama}, {Qi}, {Teague}, \&
  {Wilner}}]{2021ApJ...906..111T}
{Terwisscha van Scheltinga}, J., {Hogerheijde}, M.~R., {Cleeves}, L.~I.,
  {et~al.} 2021, \apj, 906, 111, \dodoi{10.3847/1538-4357/abc9ba}

\bibitem[{{Tobin} {et~al.}(2024){Tobin}, {Currie}, {Li}, {Chilcote}, {Brandt},
  {Lacy}, {Kuzuhara}, {Vincent}, {El Morsy}, {Deo}, {Williams}, {Guyon},
  {Lozi}, {Vievard}, {Skaf}, {Ahn}, {Groff}, {Kasdin}, {Uyama}, {Tamura},
  {Gibbs}, {Lewis}, {Bowens-Rubin}, {Salama}, {An}, \&
  {Chen}}]{2024AJ....167..205T}
{Tobin}, T.~L., {Currie}, T., {Li}, Y., {et~al.} 2024, \aj, 167, 205,
  \dodoi{10.3847/1538-3881/ad3077}

\bibitem[{{Trapman} {et~al.}(2025){Trapman}, {Longarini}, {Rosotti}, {Andrews},
  {Bae}, {Barraza-Alfaro}, {Benisty}, {Cataldi}, {Curone}, {Czekala},
  {Facchini}, {Fasano}, {Flock}, {Fukagawa}, {Galloway-Sprietsma}, {Garg},
  {Hall}, {Huang}, {Ilee}, {Izquierdo}, {Kanagawa}, {Lesur}, {Lodato},
  {Loomis}, {Orihara}, {Paneque-Carreno}, {Pinte}, {Price}, {Stadler},
  {Teague}, {van Terwisga}, {Testi}, {Yen}, {Wafflard-Fernandez}, {Wilner},
  {Winter}, {W{\"o}lfer}, {Yoshida}, {Zawadzki}, \&
  {Zhang}}]{2025ApJ...984L..18T}
{Trapman}, L., {Longarini}, C., {Rosotti}, G.~P., {et~al.} 2025, \apjl, 984,
  L18, \dodoi{10.3847/2041-8213/adc430}

\bibitem[{{Ubeira Gabellini} {et~al.}(2019){Ubeira Gabellini}, {Miotello},
  {Facchini}, {Ragusa}, {Lodato}, {Testi}, {Benisty}, {Bruderer}, {T.
  Kurtovic}, {Andrews}, {Carpenter}, {Corder}, {Dipierro}, {Ercolano},
  {Fedele}, {Guidi}, {Henning}, {Isella}, {Kwon}, {Linz}, {McClure}, {Perez},
  {Ricci}, {Rosotti}, {Tazzari}, \& {Wilner}}]{2019MNRAS.486.4638U}
{Ubeira Gabellini}, M.~G., {Miotello}, A., {Facchini}, S., {et~al.} 2019,
  \mnras, 486, 4638, \dodoi{10.1093/mnras/stz1138}

\bibitem[{{Uyama} {et~al.}(2020){Uyama}, {Muto}, {Mawet}, {Christiaens},
  {Hashimoto}, {Kudo}, {Kuzuhara}, {Ruane}, {Beichman}, {Absil}, {Akiyama},
  {Bae}, {Bottom}, {Choquet}, {Currie}, {Dong}, {Follette}, {Fukagawa},
  {Guidi}, {Huby}, {Kwon}, {Mayama}, {Meshkat}, {Reggiani}, {Ricci}, {Serabyn},
  {Tamura}, {Testi}, {Wallack}, {Williams}, \& {Zhu}}]{2020AJ....159..118U}
{Uyama}, T., {Muto}, T., {Mawet}, D., {et~al.} 2020, \aj, 159, 118,
  \dodoi{10.3847/1538-3881/ab7006}

\bibitem[{{van der Marel} {et~al.}(2021{\natexlab{a}}){van der Marel}, {Booth},
  {Leemker}, {van Dishoeck}, \& {Ohashi}}]{2021A&A...651L...5V}
{van der Marel}, N., {Booth}, A.~S., {Leemker}, M., {van Dishoeck}, E.~F., \&
  {Ohashi}, S. 2021{\natexlab{a}}, \aap, 651, L5,
  \dodoi{10.1051/0004-6361/202141051}

\bibitem[{{van der Marel} {et~al.}(2021{\natexlab{b}}){van der Marel},
  {Bosman}, {Krijt}, {Mulders}, \& {Bergner}}]{2021A&A...653L...9V}
{van der Marel}, N., {Bosman}, A.~D., {Krijt}, S., {Mulders}, G.~D., \&
  {Bergner}, J.~B. 2021{\natexlab{b}}, \aap, 653, L9,
  \dodoi{10.1051/0004-6361/202141786}

\bibitem[{{van der Marel} {et~al.}(2016{\natexlab{a}}){van der Marel},
  {Cazzoletti}, {Pinilla}, \& {Garufi}}]{2016ApJ...832..178V}
{van der Marel}, N., {Cazzoletti}, P., {Pinilla}, P., \& {Garufi}, A.
  2016{\natexlab{a}}, \apj, 832, 178, \dodoi{10.3847/0004-637X/832/2/178}

\bibitem[{{van der Marel} \& {Pinilla}(2023)}]{2023arXiv231009077V}
{van der Marel}, N., \& {Pinilla}, P. 2023, arXiv e-prints, arXiv:2310.09077,
  \dodoi{10.48550/arXiv.2310.09077}

\bibitem[{{van der Marel} {et~al.}(2016{\natexlab{b}}){van der Marel}, {van
  Dishoeck}, {Bruderer}, {Andrews}, {Pontoppidan}, {Herczeg}, {van Kempen}, \&
  {Miotello}}]{2016A&A...585A..58V}
{van der Marel}, N., {van Dishoeck}, E.~F., {Bruderer}, S., {et~al.}
  2016{\natexlab{b}}, \aap, 585, A58, \dodoi{10.1051/0004-6361/201526988}

\bibitem[{{van der Plas} {et~al.}(2017){van der Plas}, {Wright}, {M{\'e}nard},
  {Casassus}, {Canovas}, {Pinte}, {Maddison}, {Maaskant}, {Avenhaus}, {Cieza},
  {Perez}, \& {Ubach}}]{2017A&A...597A..32V}
{van der Plas}, G., {Wright}, C.~M., {M{\'e}nard}, F., {et~al.} 2017, \aap,
  597, A32, \dodoi{10.1051/0004-6361/201629523}

\bibitem[{{Vigan} {et~al.}(2021){Vigan}, {Fontanive}, {Meyer}, {Biller},
  {Bonavita}, {Feldt}, {Desidera}, {Marleau}, {Emsenhuber}, {Galicher}, {Rice},
  {Forgan}, {Mordasini}, {Gratton}, {Le Coroller}, {Maire}, {Cantalloube},
  {Chauvin}, {Cheetham}, {Hagelberg}, {Lagrange}, {Langlois}, {Bonnefoy},
  {Beuzit}, {Boccaletti}, {D'Orazi}, {Delorme}, {Dominik}, {Henning}, {Janson},
  {Lagadec}, {Lazzoni}, {Ligi}, {Menard}, {Mesa}, {Messina}, {Moutou},
  {M{\"u}ller}, {Perrot}, {Samland}, {Schmid}, {Schmidt}, {Sissa}, {Turatto},
  {Udry}, {Zurlo}, {Abe}, {Antichi}, {Asensio-Torres}, {Baruffolo}, {Baudoz},
  {Baudrand}, {Bazzon}, {Blanchard}, {Bohn}, {Brown Sevilla}, {Carbillet},
  {Carle}, {Cascone}, {Charton}, {Claudi}, {Costille}, {De Caprio},
  {Delboulb{\'e}}, {Dohlen}, {Engler}, {Fantinel}, {Feautrier}, {Fusco},
  {Gigan}, {Girard}, {Giro}, {Gisler}, {Gluck}, {Gry}, {Hubin}, {Hugot},
  {Jaquet}, {Kasper}, {Le Mignant}, {Llored}, {Madec}, {Magnard}, {Martinez},
  {Maurel}, {M{\"o}ller-Nilsson}, {Mouillet}, {Moulin}, {Orign{\'e}}, {Pavlov},
  {Perret}, {Petit}, {Pragt}, {Puget}, {Rabou}, {Ramos}, {Rickman}, {Rigal},
  {Rochat}, {Roelfsema}, {Rousset}, {Roux}, {Salasnich}, {Sauvage}, {Sevin},
  {Soenke}, {Stadler}, {Suarez}, {Wahhaj}, {Weber}, \&
  {Wildi}}]{2021A&A...651A..72V}
{Vigan}, A., {Fontanive}, C., {Meyer}, M., {et~al.} 2021, \aap, 651, A72,
  \dodoi{10.1051/0004-6361/202038107}

\bibitem[{{Vioque} {et~al.}(2018){Vioque}, {Oudmaijer}, {Baines},
  {Mendigut{\'\i}a}, \& {P{\'e}rez-Mart{\'\i}nez}}]{2018A&A...620A.128V}
{Vioque}, M., {Oudmaijer}, R.~D., {Baines}, D., {Mendigut{\'\i}a}, I., \&
  {P{\'e}rez-Mart{\'\i}nez}, R. 2018, \aap, 620, A128,
  \dodoi{10.1051/0004-6361/201832870}

\bibitem[{Virtanen {et~al.}(2020)Virtanen, Gommers, Oliphant, Haberland, Reddy,
  Cournapeau, Burovski, Peterson, Weckesser, Bright, van~der Walt, Brett,
  Wilson, Millman, Mayorov, Nelson, Jones, Kern, Larson, Carey, Polat, Feng,
  Moore, VanderPlas, Laxalde, Perktold, Cimrman, Henriksen, Quintero, Harris,
  Archibald, Ribeiro, Pedregosa, van Mulbregt, \&
  Contributors}]{2020SciPy-NMeth}
Virtanen, P., Gommers, R., Oliphant, T.~E., {et~al.} 2020, Nature Methods, 17,
  261, \dodoi{10.1038/s41592-019-0686-2}

\bibitem[{{Virtanen} {et~al.}(2020){Virtanen}, {Gommers}, {Oliphant},
  {Haberland}, {Reddy}, {Cournapeau}, {Burovski}, {Peterson}, {Weckesser},
  {Bright}, {van der Walt}, {Brett}, {Wilson}, {Millman}, {Mayorov}, {Nelson},
  {Jones}, {Kern}, {Larson}, {Carey}, {Polat}, {Feng}, {Moore}, {VanderPlas},
  {Laxalde}, {Perktold}, {Cimrman}, {Henriksen}, {Quintero}, {Harris},
  {Archibald}, {Ribeiro}, {Pedregosa}, {van Mulbregt}, \& {SciPy 1. 0
  Contributors}}]{2020NatMe..17..261V}
{Virtanen}, P., {Gommers}, R., {Oliphant}, T.~E., {et~al.} 2020, Nature
  Methods, 17, 261, \dodoi{10.1038/s41592-019-0686-2}

\bibitem[{{Wagner} {et~al.}(2015){Wagner}, {Apai}, {Kasper}, \&
  {Robberto}}]{2015ApJ...813L...2W}
{Wagner}, K., {Apai}, D., {Kasper}, M., \& {Robberto}, M. 2015, \apjl, 813, L2,
  \dodoi{10.1088/2041-8205/813/1/L2}

\bibitem[{{Wagner} {et~al.}(2018){Wagner}, {Dong}, {Sheehan}, {Apai}, {Kasper},
  {McClure}, {Morzinski}, {Close}, {Males}, {Hinz}, {Quanz}, \&
  {Fung}}]{2018ApJ...854..130W}
{Wagner}, K., {Dong}, R., {Sheehan}, P., {et~al.} 2018, \apj, 854, 130,
  \dodoi{10.3847/1538-4357/aaa767}

\bibitem[{{Wagner} {et~al.}(2023){Wagner}, {Stone}, {Skemer}, {Ertel}, {Dong},
  {Apai}, {Spalding}, {Leisenring}, {Sitko}, {Kratter}, {Barman}, {Marley},
  {Miles}, {Boccaletti}, {Assani}, {Bayyari}, {Uyama}, {Woodward}, {Hinz},
  {Briesemeister}, {Lawson}, {M{\'e}nard}, {Pantin}, {Russell}, {Skrutskie}, \&
  {Wisniewski}}]{2023NatAs...7.1208W}
{Wagner}, K., {Stone}, J., {Skemer}, A., {et~al.} 2023, Nature Astronomy, 7,
  1208, \dodoi{10.1038/s41550-023-02028-3}

\bibitem[{{Walsh} {et~al.}(2016){Walsh}, {Juh{\'a}sz}, {Meeus}, {Dent}, {Maud},
  {Aikawa}, {Millar}, \& {Nomura}}]{2016ApJ...831..200W}
{Walsh}, C., {Juh{\'a}sz}, A., {Meeus}, G., {et~al.} 2016, \apj, 831, 200,
  \dodoi{10.3847/0004-637X/831/2/200}

\bibitem[{{Walsh} {et~al.}(2010){Walsh}, {Millar}, \&
  {Nomura}}]{2010ApJ...722.1607W}
{Walsh}, C., {Millar}, T.~J., \& {Nomura}, H. 2010, \apj, 722, 1607,
  \dodoi{10.1088/0004-637X/722/2/1607}

\bibitem[{{Walsh} {et~al.}(2014){Walsh}, {Millar}, {Nomura}, {Herbst}, {Widicus
  Weaver}, {Aikawa}, {Laas}, \& {Vasyunin}}]{2014A&A...563A..33W}
{Walsh}, C., {Millar}, T.~J., {Nomura}, H., {et~al.} 2014, \aap, 563, A33,
  \dodoi{10.1051/0004-6361/201322446}

\bibitem[{{Wang} {et~al.}(2023){Wang}, {Wang}, {Ruffio}, {Blake}, {Mawet},
  {Baker}, {Bartos}, {Bond}, {Calvin}, {Cetre}, {Delorme}, {Doppmann},
  {Echeverri}, {Finnerty}, {Fitzgerald}, {Jovanovic}, {Lopez}, {Martin},
  {Morris}, {Pezzato}, {Ragland}, {Ruane}, {Sappey}, {Schofield}, {Skemer},
  {Venenciano}, {Wallace}, {Wizinowich}, {Xuan}, {Bryan}, {Roy}, \&
  {Wallack}}]{2023AJ....165....4W}
{Wang}, J., {Wang}, J.~J., {Ruffio}, J.-B., {et~al.} 2023, \aj, 165, 4,
  \dodoi{10.3847/1538-3881/ac9f19}

\bibitem[{{Ward-Duong} {et~al.}(2021){Ward-Duong}, {Patience}, {Follette}, {De
  Rosa}, {Rameau}, {Marley}, {Saumon}, {Nielsen}, {Rajan}, {Greenbaum}, {Lee},
  {Wang}, {Czekala}, {Duch{\^e}ne}, {Macintosh}, {Ammons}, {Bailey}, {Barman},
  {Bulger}, {Chen}, {Chilcote}, {Cotten}, {Doyon}, {Esposito}, {Fitzgerald},
  {Gerard}, {Goodsell}, {Graham}, {Hibon}, {Hom}, {Hung}, {Ingraham}, {Kalas},
  {Konopacky}, {Larkin}, {Maire}, {Marchis}, {Marois}, {Metchev},
  {Millar-Blanchaer}, {Oppenheimer}, {Palmer}, {Perrin}, {Poyneer}, {Pueyo},
  {Rantakyr{\"o}}, {Ren}, {Ruffio}, {Savransky}, {Schneider},
  {Sivaramakrishnan}, {Song}, {Soummer}, {Tallis}, {Thomas}, {Wallace},
  {Wiktorowicz}, \& {Wolff}}]{2021AJ....161....5W}
{Ward-Duong}, K., {Patience}, J., {Follette}, K., {et~al.} 2021, \aj, 161, 5,
  \dodoi{10.3847/1538-3881/abc263}

\bibitem[{{Wilson}(1999)}]{1999RPPh...62..143W}
{Wilson}, T.~L. 1999, Reports on Progress in Physics, 62, 143,
  \dodoi{10.1088/0034-4885/62/2/002}

\bibitem[{{Winter} {et~al.}(2025){Winter}, {Benisty}, {Izquierdo}, {Lodato},
  {Teague}, {Kimmig}, {Andrews}, {Bae}, {Barraza-Alfaro}, {Cuello}, {Curone},
  {Czekala}, {Facchini}, {Fasano}, {Hall}, {Hardiman}, {Hilder}, {Ilee},
  {Fukagawa}, {Longarini}, {M{\'e}nard}, {Orihara}, {Pinte}, {Price},
  {Rosotti}, {Stadler}, {Wilner}, {W{\"o}lfer}, {Yen}, {Yoshida}, \&
  {Zawadzki}}]{2025arXiv250711669W}
{Winter}, A.~J., {Benisty}, M., {Izquierdo}, A.~F., {et~al.} 2025, arXiv
  e-prints, arXiv:2507.11669, \dodoi{10.48550/arXiv.2507.11669}

\bibitem[{{W{\"o}lfer} {et~al.}(2021){W{\"o}lfer}, {Facchini}, {Kurtovic},
  {Teague}, {van Dishoeck}, {Benisty}, {Ercolano}, {Lodato}, {Miotello},
  {Rosotti}, {Testi}, \& {Ubeira Gabellini}}]{2021A&A...648A..19W}
{W{\"o}lfer}, L., {Facchini}, S., {Kurtovic}, N.~T., {et~al.} 2021, \aap, 648,
  A19, \dodoi{10.1051/0004-6361/202039469}

\bibitem[{{W{\"o}lfer} {et~al.}(2025){W{\"o}lfer}, {Barraza-Alfaro}, {Teague},
  {Curone}, {Benisty}, {Fukagawa}, {Bae}, {Cataldi}, {Czekala}, {Facchini},
  {Fasano}, {Flock}, {Galloway-Sprietsma}, {Garg}, {Hall}, {Huang}, {Ilee},
  {Izquierdo}, {Kanagawa}, {Lesur}, {Longarini}, {Loomis}, {Menard}, {Nath},
  {Orihara}, {Pinte}, {Price}, {Rosotti}, {Stadler}, {Wafflard-Fernandez},
  {Winter}, {Yen}, {Yoshida}, \& {Zawadzki}}]{2025ApJ...984L..22W}
{W{\"o}lfer}, L., {Barraza-Alfaro}, M., {Teague}, R., {et~al.} 2025, \apjl,
  984, L22, \dodoi{10.3847/2041-8213/adc42c}

\bibitem[{{Wolthoff} {et~al.}(2022){Wolthoff}, {Reffert}, {Quirrenbach},
  {Jones}, {Wittenmyer}, \& {Jenkins}}]{2022A&A...661A..63W}
{Wolthoff}, V., {Reffert}, S., {Quirrenbach}, A., {et~al.} 2022, \aap, 661,
  A63, \dodoi{10.1051/0004-6361/202142501}

\bibitem[{{Xuan} {et~al.}(2024){Xuan}, {Hsu}, {Finnerty}, {Wang}, {Ruffio},
  {Zhang}, {Knutson}, {Mawet}, {Mamajek}, {Inglis}, {Wallack}, {Bryan},
  {Blake}, {Molli{\`e}re}, {Hejazi}, {Baker}, {Bartos}, {Calvin}, {Cetre},
  {Delorme}, {Doppmann}, {Echeverri}, {Fitzgerald}, {Jovanovic}, {Liberman},
  {L{\'o}pez}, {Morris}, {Pezzato}, {Sappey}, {Schofield}, {Skemer}, {Wallace},
  {Wang}, {Agrawal}, \& {Horstman}}]{2024ApJ...970...71X}
{Xuan}, J.~W., {Hsu}, C.-C., {Finnerty}, L., {et~al.} 2024, \apj, 970, 71,
  \dodoi{10.3847/1538-4357/ad4796}

\bibitem[{{Yamato} {et~al.}(2024){Yamato}, {Aikawa}, {Guzm{\'a}n}, {Furuya},
  {Notsu}, {Cataldi}, {{\"O}berg}, {Qi}, {Law}, {Huang}, {Teague}, \& {Le
  Gal}}]{2024ApJ...974...83Y}
{Yamato}, Y., {Aikawa}, Y., {Guzm{\'a}n}, V.~V., {et~al.} 2024, \apj, 974, 83,
  \dodoi{10.3847/1538-4357/ad6981}

\bibitem[{{Zagaria} {et~al.}(2025){Zagaria}, {Jiang}, {Cataldi}, {Facchini},
  {Benisty}, {Aikawa}, {Andrews}, {Bae}, {Barraza-Alfaro}, {Curone}, {Czekala},
  {Fasano}, {Hall}, {Hammond}, {Huang}, {Ilee}, {Izquierdo}, {Lawrence},
  {Lodato}, {M{\'e}nard}, {Pinte}, {Rosotti}, {Stadler}, {Teague}, {Testi},
  {Wilner}, {Winter}, \& {Yoshida}}]{2025arXiv250616481Z}
{Zagaria}, F., {Jiang}, H., {Cataldi}, G., {et~al.} 2025, \apj, 989, 30,
  \dodoi{10.3847/1538-4357/ade683}

\bibitem[{{Zhang} {et~al.}(2017){Zhang}, {Bergin}, {Blake}, {Cleeves}, \&
  {Schwarz}}]{2017NatAs...1E.130Z}
{Zhang}, K., {Bergin}, E.~A., {Blake}, G.~A., {Cleeves}, L.~I., \& {Schwarz},
  K.~R. 2017, Nature Astronomy, 1, 0130, \dodoi{10.1038/s41550-017-0130}

\bibitem[{{Zhang} {et~al.}(2019){Zhang}, {Bergin}, {Schwarz}, {Krijt}, \&
  {Ciesla}}]{2019ApJ...883...98Z}
{Zhang}, K., {Bergin}, E.~A., {Schwarz}, K., {Krijt}, S., \& {Ciesla}, F. 2019,
  \apj, 883, 98, \dodoi{10.3847/1538-4357/ab38b9}

\bibitem[{{Zhang} {et~al.}(2020){Zhang}, {Bosman}, \&
  {Bergin}}]{2020ApJ...891L..16Z}
{Zhang}, K., {Bosman}, A.~D., \& {Bergin}, E.~A. 2020, \apjl, 891, L16,
  \dodoi{10.3847/2041-8213/ab77ca}

\bibitem[{{Zhang} {et~al.}(2021){Zhang}, {Booth}, {Law}, {Bosman}, {Schwarz},
  {Bergin}, {{\"O}berg}, {Andrews}, {Guzm{\'a}n}, {Walsh}, {Qi}, {van't Hoff},
  {Long}, {Wilner}, {Huang}, {Czekala}, {Ilee}, {Cataldi}, {Bergner}, {Aikawa},
  {Teague}, {Bae}, {Loomis}, {Calahan}, {Alarc{\'o}n}, {M{\'e}nard}, {Le Gal},
  {Sierra}, {Yamato}, {Nomura}, {Tsukagoshi}, {P{\'e}rez}, {Trapman}, {Liu}, \&
  {Furuya}}]{2021ApJS..257....5Z}
{Zhang}, K., {Booth}, A.~S., {Law}, C.~J., {et~al.} 2021, \apjs, 257, 5,
  \dodoi{10.3847/1538-4365/ac1580}

\bibitem[{{Zhang} {et~al.}(2024){Zhang}, {Xuan}, {Mawet}, {Wang}, {Hsu},
  {Ruffio}, {Knutson}, {Inglis}, {Blake}, {Chachan}, {Horstman}, {Baker},
  {Bartos}, {Calvin}, {Cetre}, {Delorme}, {Doppmann}, {Echeverri}, {Finnerty},
  {Fitzgerald}, {Jovanovic}, {Liberman}, {L{\'o}pez}, {Morris}, {Pezzato},
  {Sappey}, {Schofield}, {Skemer}, {Wallace}, {Wang}, \& {Do
  {\'O}}}]{2024AJ....168..131Z}
{Zhang}, Y., {Xuan}, J.~W., {Mawet}, D., {et~al.} 2024, \aj, 168, 131,
  \dodoi{10.3847/1538-3881/ad6609}

\end{thebibliography}
\bibliographystyle{aasjournal}

\begin{appendix}
\section{Observation details}
\begin{table*}[h!]
\centering
\caption{Observation details of 2023.1.00252.S}
% \Small
\begin{tabular}{cccccccccc}  
\hline 
Target   & Setting & Date & Antennas & Baselines & Integration & AR & MRS & PWV & Flux Cal. \\ 
   &  &  &  &  (m) &  (minutes) & (") &  (") & (mm) & \\ \hline  \hline 
HD 97048      & A       & 2024-05-12 & 45 & 15.1 - 783.1  & 38.81 & 0.3    & 4.2     & 1.3 & 	J1427-4206 	\\
              &         & 2024-05-13 & 46 & 15.1 - 783.1  & 38.81 & 0.3    & 4.2     & 1.2 & 	J1427-4206\\
              & B       & 2024-05-23 & 43 & 15.1 - 740.7  & 34.78 & 0.3    & 4.0     & 1.0 & J1427-4206 \\ \hline
CQ Tau        & B       & 2024-09-04 & 45 & 15.0 - 649.6  & 22.68 & 0.4    & 4.5     & 0.7 & 	J0423-0120\\ \hline
HD 135344B    & A       & 2024-08-31 & 46 & 15.1 - 783.1  & 35.78 & 0.3    & 4.5     & 1.4 & 	J1256-0547\\
                     &  & 2024-08-31 & 47 & 15.1 - 783.1  & 35.78 & 0.3    & 4.5     & 1.1 &	J1517-2422 \\
            &  B        & 2024-09-15 & 44 & 15.0 - 499.8  & 42.34 & 0.5    & 5.7     & 0.7 & 	J1924-2914 \\
                    &   & 2024-09-16 & 36 & 15.0 - 483.7  & 38.81 & 0.5    & 6.3     & 0.7 & 	J1256-0547\\ \hline
HD 169142 & A           & 2024-09-02 & 47 & 15.1 - 783.1  & 35.28 & 0.3    & 4.5     & 1.2 & 	J1924-2914\\
 &                      & 2024-09-02 & 28 & 15.1 - 783.1  & 35.28 & 0.3    & 4.5     & 1.9 & 	J2253+1608 \\
 &                      & 2024-09-04 & 35 & 15.1 - 696.8  & 35.28 & 0.4    & 4.8     &  0.8 & 	J1924-2914\\
 &                    B & 2024-09-01 & 47 &  15.1 - 783.1 & 41.33 & 0.3 &   4.4      & 1.1 & J1924-2914 \\
                    &   & 2024-09-01 & 46 &  15.1 - 783.1 & 41.33 &  0.3 &   4.4      &   1.2 & 	J2253+1608\\ \hline
MWC 758 & B             & 2024-09-04 &  45 & 15.0 - 649.6  & 23.18  & 0.4 &   4.5 & 0.7 & 	J0423-0120 \\  \hline
HD 100435 & A           & 2024-05-14 &  46  & 15.1 - 783.1  & 39.82 & 0.3 & 4.1   & 1.5 & 	J1256-0547 \\  
         &              & 2024-05-14 & 46   & 15.1 - 783.1  & 39.82 &  0.3 & 4.1   & 1.4 & 	J1427-4206 \\ 
                    & B & 2024-05-13  & 47   & 15.1- 783.1  &  17.64 & 0.3 & 4.2   &1.5 & 	J1256-0547 \\ 
 & &   2024-09-04 & 36   & 15.0 - 649.6  & 47.38 &  0.4 &  5.3   & 0.9 & 	J1037-2934\\ 
 && 2024-09-28  & 43  & 15.0 - 499.8   &   47.38  &  0.5 &  6.1  & 0.5 & 	J1058+0133 \\ 
          \hline      
\end{tabular}
\label{obstable}
\end{table*}

\begin{table*}[h!]
\centering
\caption{Observation details of 2023.1.00252.S}
\begin{tabular}{cccc|cccc}
\hline 
Setting & SPW & Central Freq. (sky) & Bandwidth & Setting & SPW & Central Freq. (sky) & Bandwidth \\
        &     & [GHz]           &  [GHz] &         &     &  [GHz]          &  [GHz] \\ \hline \hline
A       & 0   & 291.206             & 0.469     & B       & 0   & 299.386             & 0.224     \\
        & 1   & 292.386             & 0.234     &         & 1   & 300.825             & 0.234     \\
        & 2   & 291.886             & 0.234     &         & 2   & 300.235             & 0.234     \\
        & 3   & 292.756             & 0.234     &         & 3   & 300.485             & 0.234     \\
        & 4   & 293.086             & 0.234     &         & 4   & 302.735             & 0.469     \\
        & 5   & 294.102             & 0.469     &         & 5   & 301.285             & 0.234     \\
        & 6   & 303.985             & 0.234     &         & 6   & 312.135             & 0.234     \\
        & 7   & 303.535             & 0.469     &         & 7   & 311.785             & 0.234     \\
        & 8   & 304.252             & 0.234     &         & 8   & 312.485             & 0.469     \\
        & 9   & 305.785             & 0.234     &         & 9   & 313.665             & 0.234     \\
        & 10  & 306.385             & 0.234     &         & 10  & 314.785             & 0.469     \\
        & 11  & 304.885             & 0.234     &         & 11  & 314.185             & 0.234     \\
        & 12  & 305.485             & 0.234     &         &     &                     &           \\ \hline
\end{tabular}
\label{spwtable}
\end{table*}

\begin{table}[]
\centering
\caption{Molecular line properties}
\begin{tabular}{cccccccc} 
\hline
                     & Transition          & Frequency   & log$\mathrm{_{10}}$($A_{ul}$)  & $E_{up}$ & $g_{ul}$ & Setting \\ 
                     &                     & [GHz]   & {[}s$^{-1}${]} & {[}K{]}   \\ \hline \hline       
\ce{^{13}C^{18}O}    & $J=3-2$               & 314.119653  & -5.723         & 30.2     & 7        & B       \\
CS      & $J=6-5$               & 293.912087   & -3.283       & 49.4     & 13       & A       \\
% SO      & 8(7)- 7(6)         & 304.077844  & -3.44263       & 62.1     & 17       & A       \\
SO      & $J=7_7-6_6$         & 301.286124  & -3.465       & 71.0     & 15       & B       \\
\ce{^{33}SO} & $J=8_7-7_6$, $F=\frac{15}{2}-\frac{13}{2}$ & 301.0758637 & -3.470 & 61.5 & 16         & B \\
\ce{^{33}SO} & $J=8_7-7_6$, $F=\frac{13}{2}-\frac{11}{2}$ & 301.0743451 & -3.468 & 61.5 & 14         & B \\
\ce{^{33}SO} & $J=8_7-7_6$, $F=\frac{17}{2}-\frac{15}{2}$ & 301.0768858 & -3.465 & 61.5 &  18       & B \\
\ce{^{33}SO} & $J=8_7-7_6$, $F=\frac{19}{2}-\frac{17}{2}$ & 301.0772241 & -3.455 &  61.5 & 20         & B \\
\ce{SO_2}     & $J=17_{1,17}-16_{0,16}$  & 313.660852 & -3.426       & 136.1    & 35       & B       \\
% SO2     & 19(3,17)-19(2,18)  & 299.3168185 & -3.691         & 197.0    & 39       & B       \\
% SO2     & 22(4,18)-22(3,19)  & 312.5425195 & -3.54987       & 272.8    & 42       & B       \\
% SO2     & 12(6,6)-13(5,9)    & 312.2584237 & -4.38673       & 160      & 25       & B       \\   
\ce{H_2CS}   & $J=9_{1,9}-8_{1,8}$     & 304.307709 & -3.380       & 86.0     & 57       & A       \\
\ce{H_2CS}   & $J=9_{1,8}-8_{1,7}$     & 313.716648 & -3.340        & 88.5     & 57       & B       \\
\ce{H_2S}     & $J=3_{3,0}-3_{2,1}$      & 300.505560   & -3.989       & 168.9    & 21       & B       \\
\ce{H_2CO}    & $J=4_{1,3}-3_{1,2}$      & 300.836636  & -3.144         & 47.8     & 27       & B       \\
% \ce{H_2CO}    & 4(2,2)-3(2,1)      & 291.948067  & -3.280         & 82.1     & 9        & A       \\
% \ce{H_2CO}    & 4(3,2)-3(3,1)      & 291.380442  & -3.517         & 140.9    & 27       & A       \\
% \ce{H_2CO}    & 4(3,1)-3(3,0)      & 291.384362  & -3.517         & 140.9    & 27       & A        \\
% \ce{H_2^{13}CO}  & 4(1,3)-3(1,2)      & 293.126515  & -3.178         & 47.0     & 27       & A     \\
\ce{CH_3OH}    & $J=6_{1,5}-6_{0,6}$      & 311.852612  & -3.466         & 63.7     & 52       & B       \\
\ce{HC_3N}    & $J=32-31$              & 291.068427  & -2.706      & 230.5    & 65       & A       \\
\ce{HC_3N}    & $J=33-22$              &300.159647   & -2.666       &244.9    & 67       & B       \\
\ce{c-C_3H_2} & $J=5_{4,2}-4_{3,1}$   & 300.191718 & -3.160 & 45.3 & 11 & A   \\ 
\ce{c-C_3H_2} & $J=6_{3,3}-5_{4,2}$   & 312.709574 & -3.180 & 60.3 & 13           &      B              \\
\ce{CH_3CN}   & $J=17_0-16_0$        & 312.687743  & -2.737         & 135      & 70       & B        \\
\ce{CH_3OCHO} & $J=27_{1,27}-26_{1,26}$ A & 291.111871 & -3.428&  198.6  &110 & A \\ 
\ce{CH_3OCHO} & $J=27_{0,27}-26_{0,26}$ A & 291.111891  & -3.428&  198.6& 110  & A \\ 
\ce{CH_3OCHO} & $J=27_{1,27}-26_{1,26}$ E & 291.111192  & -3.428 &  198.6  & 110 & A \\ 
\ce{CH_3OCHO} & $J=27_{0,27}-26_{0,26}$ E & 291.111212 & -3.428&   198.6 &  110 & A \\ 
\ce{CH_3OCH_3} & $J=16_{1,16}-15{0,15}$ EA & 292.412244 & -3.705 &120.3 &  198  & A \\
\ce{CH_3OCH_3} & $J=16_{1,16}-15{0,15}$ EA & 292.412244 &-3.706 &120.3 &  132   & A \\
\ce{CH_3OCH_3}& $J=16_{1,16}-15{0,15}$ EE & 292.412416 & -3.705 &120.3 &  528   & A \\
\ce{CH_3OCH_3} & $J=16_{1,16}-15{0,15}$ AA & 292.412588 & -3.705 &120.3 &  330   & A \\
HDO & $J=6_{2,5}-5_{3,2}$       & 313.750620           & -4.426               & 553.7                & 13 & B    \\
SiO     & $J=7-6$ & 303.926812  & -2.835       & 58.3     & 15       & A       \\ \hline  
\end{tabular}
\tablecomments{All molecular data are taken from The Cologne Database for Molecular Spectroscopy \citep[CDMS][]{2001A&A...370L..49M,2005JMoSt.742..215M}.}
\label{molecules}
\end{table}

\clearpage
\newpage

\onecolumngrid

\section{Quantifying weak-line detections}

A number of molecules are detected in these data only in the disk-integrated fluxes and not the intensity maps, and vice versa. We use matched filtering \citep{Loomis2018} and the spectral stacking tool \textit{GoFish} \citep{GoFish} to quantify the significance of these identifications and to check for any additional weak detections. We generated the impulse response spectra using boolean Keplerian masks and found that using strong line detections as filters did not improve these results significantly. In the disks where we resolve multiple beams across the CO gas disk we generate radial emission profiles with \texttt{GoFish} where the annuli have widths approximately equal to 1/4 of the beam major axis for each image cube and for the other disks we generate a disk integrated stacked spectra. Due to the moderate spectral and spatial resolution of the data and the near face-on nature of a number of the disks in our survey the effective signal-to-noise boost gained from these techniques will be limited.  

\begin{figure*}[h!]
    \centering
    \includegraphics[width=\linewidth]{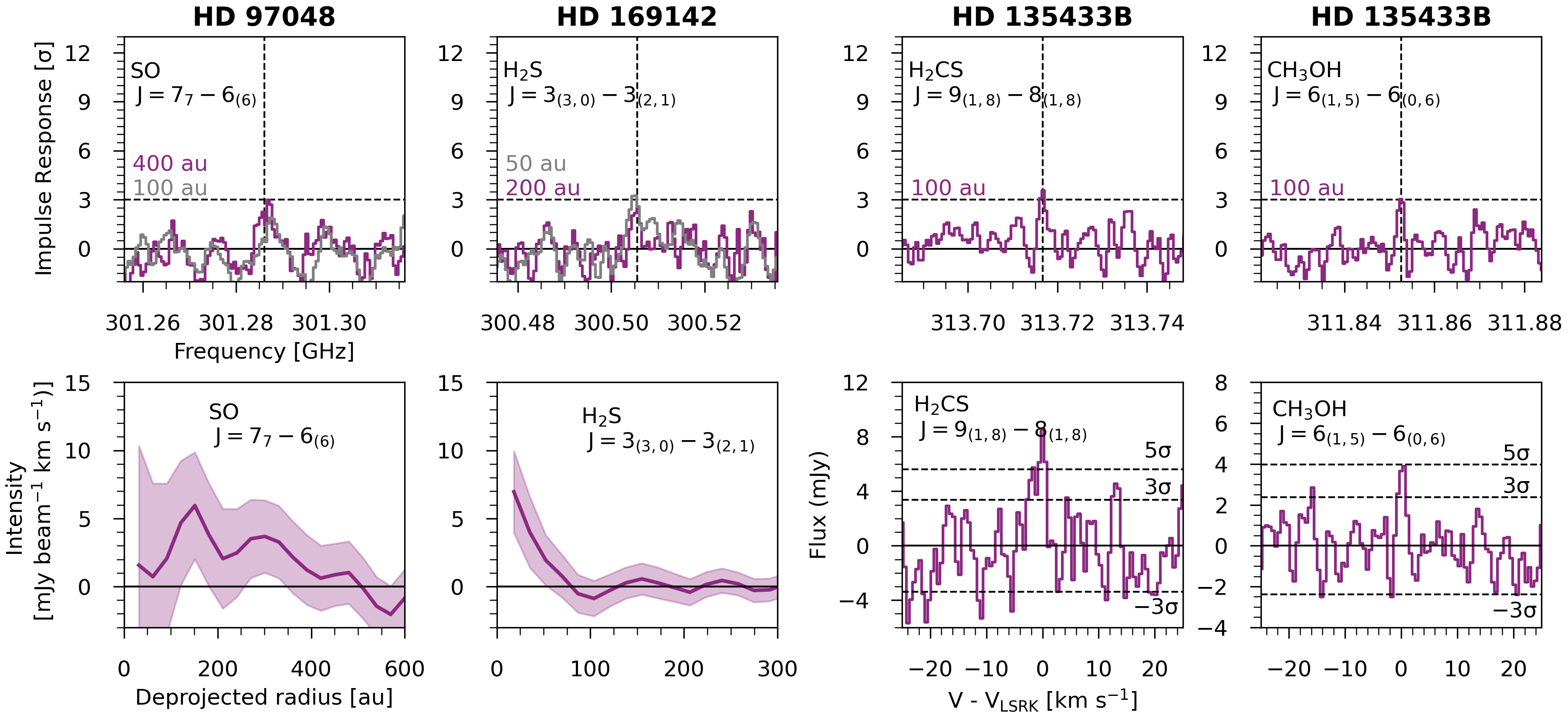}
    \caption{Top: Matched filter responses confirming the tentatively detected molecules in the survey. Each column signifies a different disk as noted in the plot titles and the radial size of the Keplerian mask(s) is stated in each panel where purple is a detection and grey a non-detection.
    Bottom: \textit{GoFish} radial profiles or disk integrated fluxes  generated over a 0 to 100~au area. } 
    \label{fig:filter}
\end{figure*}

For the sulphur-bearing molecules, we confirm the previously mentioned weak detections of SO, \ce{H_2S}, and \ce{H_2CS}, as shown in Figure~\ref{fig:filter}, and see no evidence for additional \ce{SO_2} detections. We find a 3$\sigma$ response for SO in HD~97048 when using an extended 400~au filter, and, this signal becomes insignificant when considering a more compact emission morphology. This is consistent with the \texttt{GoFish} radial profile which shows an extended and potentially ringed emission structure. 
For the \ce{H_2S} in the HD~169142 disk the signal is only significant in the impulse response when considering a compact 50~au filter and compact \ce{H_2S} emission is also recovered with \texttt{GoFish}.  Two lines of \ce{H_2CS} are detected with matched filtering in the HD~135344B disk (only one is shown here) and although the disk-integrated line fluxes are $<$3~$\sigma$, the detection of two lines with matched filtering and \texttt{GoFish} as well as significant signal in the integrated intensity maps leads us to consider this species a detection too. 

For the large organic molecules we searched for detections of \ce{c-C_3H_2}, \ce{HC_3N}, \ce{CH_3CN}, \ce{CH_3OH}, \ce{CH_3OCH_3}, and \ce{CH_3OCHO}. 
%\ce{c-C_3H_2} is detected in HD~169142 where although there is significant signal in the disk integrated line flux the images do not show significant emission. This indicates weak and extended emission which, tracks with the lack of a matched filter detection with a 50~au filter compared to a $>$5~$\sigma$ detection with a 200~au filter and this is consistent with the detection reported in \citealt{2023A&A...678A.146B}. 
There are no additional detections of \ce{c-C_3H_2} and \ce{HC_3N} and for \ce{CH_3CN}, the tentative detection in HD~97048 is not confirmed with matched filtering - likely due to the compact nature of the emission. For \ce{CH_3OH}, with matched filtering and \texttt{GoFish} we find evidence for tentative detection of \ce{CH_3OH} in HD~135344B as shown in Figure~\ref{fig:filter} increasing the detection statistic of \ce{CH_3OH} to five out of the total six disks surveyed. For the COMs larger than \ce{CH_3OH}, beyond the detection of \ce{CH_3OCHO} in HD~100453 \citep{2025arXiv250414023B}, there is only evidence for \ce{CH_3OCH_3} in HD~97048 in the image plane and similar to \ce{CH_3CN} this is not confirmed with matched filtering.

\clearpage
\newpage

\onecolumngrid

\section{Complementary literature data of Herbig disk integrated line fluxes.}

\begin{figure*}[h!]
    \centering
    \includegraphics[width=0.8\linewidth]{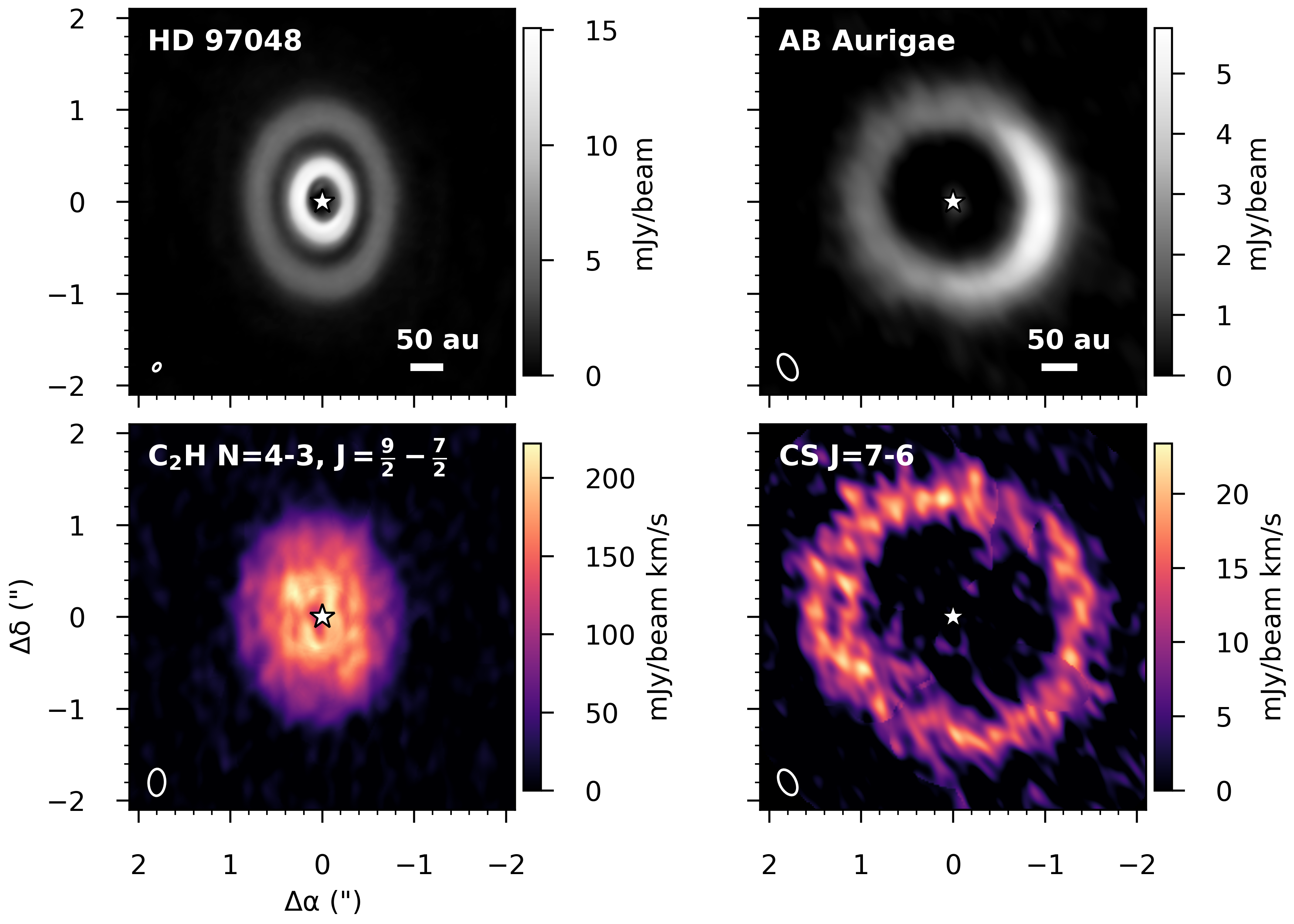}
\caption{Continuum and integrated intensity maps of archival line detections in the HD~97048 and AB~Aurigae disks. Where the HD~97048 \ce{C_2H} data are from 2017.1.00845.S and the AB~Aurigae continuum and CS data are from  2012.1.00303.S.}
    \label{fig:extra_maps}
\end{figure*}

To supplement our analysis we include reported literature and unpublished ALMA archive detections of \ce{^{13}C^{18}O}, CS, SO, \ce{H_2CO}, \ce{CH_3OH}, \ce{HC_3N}, \ce{CH_3CN} and \ce{C_2H} in Herbig disks where these additional data are listed in Table~\ref{table:literature_data}. This is not necessarily an exhaustive list of these molecule detections/upper-limits in Herbig disks and we only consider disks with reported \ce{C^{18}O} detections and those within 200~pc. In Table~\ref{table:literature_data} we also note the excitation temperatures assumed when converting line fluxes to our benchmark transitions where applicable. Where multiple transitions are detected in a disk we aim to pick the line with the closest properties to the ones detected in this work to mitigate the uncertainties in this flux conversion. The transitions for each molecule are as follows: $\mathrm{^{13}C^{18}O}$ J=3–2, CS J=6–5, SO $\mathrm{J=7_7-6_6}$, $\mathrm{H_2CO}$ $\mathrm{J=4_{(1,3)}-3_{(1,2)}}$, $\mathrm{CH_3OH}$ $\mathrm{J=6_{(1,5)}-6_{(0,6)}}$, $\mathrm{HC_3N}$ J=32–31, $\mathrm{CH_3CN}$ $\mathrm{J=17_{0}-16_{0}}$, and $\mathrm{C_2H}$ $N=2-1$, $=\frac{5}{2}-\frac{3}{2}$. For the Oph-IRS~48 disk the $J=2-1$ flux is determined from the $J=3-2$ line flux  \citep{2021A&A...651L...6B} under the assumption of optically thick line emission and a temperature of 30~K. For the \ce{^{13}C^{18}O} line flux for the HD~163296 disk we assume optically thin emission and a temperature of 30~K however, under an optically thick assumption, the inferred strength of J=3–2 line would be $\approx$30\% dimmer. For CS we use 25~K as reported for the HD~163296 disk by \citet{2025ApJ...985...84L} and for SO we assume 20~K, as measured for AB~Aurigae \citep{Dutrey24}, for all disks but Oph-IRS~48 disk where 60~K has been measured for \ce{SO_2} \citep{2025A&A...693A101T}. We use 30~K for \ce{H_2CO} for all disks \citep{2025ApJ...982...62E} aside from Oph~IRS~48 where we use 150~K \citet{2025A&A...693A101T} and we also use 150~K for \ce{CH_3OH} in all disks \citep{2025ApJ...982...62E}. For the other organics, \ce{HC_3N} and \ce{CH_3CN}, and \ce{C_2H}, we fix the temperature to 40~K following \citet{2021ApJS..257....9I} and \citet{2021ApJS..257....6G}. We also show the integrated intensity maps of the \ce{C_2H} emission in the HD~97048 disk and the CS emission in the AB~Aurigae disk in Figure~\ref{fig:extra_maps} which from the unpublished archival ALMA programs 2017.1.00845.S (P.I. Bergin) and 2012.1.00303.S (P.I. Tang) respectively. 
The unpublished HD~142527 data are part of a paper in preparation -  Temmink et al. (in prep.). The full comparison of all of these data, both presented in this work and from the literature, to bulk system properties are shown in Figure~\ref{fig:scatter_all} where we also include the comparison to the \ce{C^{18}O} fluxes.  

\begin{ThreePartTable}
\centering
\begin{TableNotes}
\item{$^{1}$ estimated from the reported error on the disk integrated flux at a similar frequency to our target transitions. \\
$^{2}$ obtained from M. Temmink via private communication with paper in preparation.}
\end{TableNotes}
\begin{longtable*}{ccccccc}
\caption{List of collated literature data on Herbig disks and line flux conversions} \\  \hline 
Molecule & Source    & Transition & Flux$\pm$Error  & Assumed $\mathrm{T_{ex}}$ & Scaled Flux  &  Reference \\ 
 &     &  &  [mJy km/s] &  [K] &  [mJy km/s] &  \\ \hline \hline 
\ce{^{13}C^{18}O}    
    & HD 142527 & $J=3-2$      &    297$\pm$30    & -     & -         &  \citet{Temmink2023} \\
    & HD 163296 & $J=2-1$      &    $370\pm30$      & 30    &  1130     &  \citet{2020ApJ...891L..16Z}      \\
    & MWC 480   & $J=3-2$      &    422$\pm$34      & -     & -         &  \citet{2020ApJ...893..101L}      \\ \hline 

CS  
    &    AB~Aurigae &  $J=7-6$     & 440$\pm$44    &   25   & 460      &  2012.1.00303.S (P.I. Tang)         \\ 
    &    HD~100546  &  $J=7-6$     & 932$\pm$28    &   25   & 970      &  \citet{2024AJ....167..164B}         \\ 
    &    HD~139614  &  $J=5-4$     & 1370$\pm$140  &   25   & 1610     &  \citet{2022MNRAS.510.1148S}          \\ 
    &    HD~142527  &  $J=7-6$     & 1100$\pm$100          &   25   & 1150     &  \citet{Temmink2023}                    \\ 
    &    HD~142666  &  $J=5-4$     & 237$\pm$25    &   25   & 280      &  \citet{2022MNRAS.510.1148S}         \\ 
    &    HD~145718  &  $J=5-4$     & 141$\pm$18    &   25   & 170      &  \citet{2022MNRAS.510.1148S}         \\ 
    &    HD~163296  &  $J=5-4$     & 1760$\pm$190  &   25   & 2070     &  \citet{2025ApJ...985...84L}         \\ 
    &    MWC~480    &  $J=6-5$     & 810$\pm$20    &   -    &  -       &  \citet{2020ApJ...893..101L}          \\ 
    &    Oph-IRS~48 &  $J=7-6$     & 40$\pm$17     &   25   & 42       &  \citet{2024AJ....167..165B}       \\  \hline 

SO  
    &    AB~Aurigae & $J=5_6-4_5$   &  330              &  20       &  290      & \citet{Dutrey24}  \\ 
    &    HD~100546  & $J=7_6-6_6$   &  124$\pm$12       &  -        &    -      &  \citet{Booth23a} \\ 
    &    MWC~480    & $J=7_6-6_6$   &  $<66^{1}$        &  -        &    -      &  \citet{2020ApJ...893..101L}  \\ 
    &    HD~142527  & $J=3_2–2_1$   &  $174\pm17$  &  20       & 520   & 2023.1.00628.S (P.I. Temmink)$^{2}$    \\ 
    &    HD~163296  & $J=5-4$       &  $<45$            &  20       & $<$130   &  \citet{2021ApJS..257...12L}  \\    
    &    Oph-IRS~48 & $J=7_8-6_7$   & 1070$\pm$13       &  60       &  870      &  \citet{2024AJ....167..165B}  \\  \hline 

\ce{H_2CO} 
    &    AB~Aurigae &  $J=3_{0,3}-2_{0,2}$  & 220$\pm$10    &   30      & 880           &  \citet{abaur_chem}               \\ 
    &    HD~100546  & $J=5_{1,5}-4_{1,4}$   & 3065$\pm$30   &   30      & 2430          &  \citet{2024AJ....167..164B}      \\ 
    &    HD~142527  & $J=4_{0,3}–3_{0,3}$   & 1200$\pm$120   &   30      & 2430          &  \citet{Temmink2023}              \\ 
    &    HD~163296  & $J=3_{0,3}-2_{0,2}$   & 813$\pm$78    &   30      & 3260          &  \citet{2021ApJS..257....6G}     \\ 
    &    MWC~480    & $J=4_{1,3}–3_{1,2}$   & 342$\pm$22    &    -      &    -          &  \citet{2020ApJ...893..101L}      \\ 
    &    Oph-IRS~48 & $J=5_{1,5}-4_{1,4}$   & 824$\pm$21    &   150     & 440           &   \citet{2024AJ....167..165B}     \\  \hline 
 
\ce{CH_3OH}    
    &    HD~142527  &  $J=7_{1,6}–7_{0,7}$          &  $<39$        &   150     &   $<37$        &   \citet{Temmink2023}  \\ 
    &    HD~100546  &  $J=7_0-6_0$                  &  39$\pm$12    &   150     &   69        &  \citet{2024AJ....167..164B} \\ 
    &    HD~163296  &  $J=2_{1,1}-2_{0,2}$          &  $<26$         &   150     &  $<54$        &  \citet{Carney2019} \\ 
    &    MWC~480    &  $J=6_{(1,5)}-6_{(0,6)}$      &   $<$66$^{1}$ &   -       &    -          &  \citet{2020ApJ...893..101L}          \\ 
    &    Oph-IRS~48 &  $J=7_0-6_0$                  & 182$\pm$18    &  150      &  320        &   \citet{2024AJ....167..165B}           \\  \hline 

\ce{HC_3N}      
    &    HD~100546  &  $J=38-37$    &  $<23$             &  40       &   $<$120        &  \citet{2024AJ....167..164B}          \\ 
    &    HD~163296  &  $J=29–28$    &  197$\pm$3.0       &  40       &   106           &  \citet{2021ApJS..257....9I}         \\ 
    &    MWC~480    &  $J=32-31$    &  85$\pm$19         &  -        &    -         &  \citet{2020ApJ...893..101L}          \\
    &    Oph-IRS~48 &  $J=38-37$    &  $<66$             &  40       &   $<340$         &   \citet{2024AJ....167..165B}           \\  \hline 

\ce{CH_3CN}  
    &    HD~100546  & $J=19_0-18_0$     & $<24$         &  40   &  $<35$    &  \citet{2024AJ....167..164B}  \\ 
    &    HD~163296  & $J=12_0-11_0$     &  23$\pm$1     &  40   &  12       &  \citet{2021ApJS..257....9I}   \\ 
    &    MWC~480    & $J=16_0–15_0$     &  50$\pm$10    &  40   &  30       &  \citet{2020ApJ...893..101L}    \\ 
    &    Oph-IRS~48 & $J=19_0-18_0$     &  $<75$        &  40   &  $<109$   &  \citet{2024AJ....167..165B}    \\  \hline 
    
\ce{C_2H}      
    &    HD~97048   & $N=4-3$, $J=\frac{9}{2}-\frac{7}{2}$  &  8180$\pm$820       & 40   & 1122   &  2017.1.00845.S (P.I. Bergin) \\ 
    &    HD~100453  & $N=2-1$, $J=\frac{5}{2}-\frac{3}{2}$  &  $<$29.1          & -     & -     &  \citet{2022MNRAS.510.1148S} \\ 
    &    HD~100546  & $N=4-3$, $J=\frac{9}{2}-\frac{7}{2}$  &  1601$\pm$25      & 40    & 219   &  \citet{2024AJ....167..164B} \\ 
    &    HD~139614  & $N=2-1$, $J=\frac{5}{2}-\frac{3}{2}$  &  484$\pm$51       & -     & -     &  \citet{2022MNRAS.510.1148S} \\  
    &    HD~142666  & $N=2-1$, $J=\frac{5}{2}-\frac{3}{2}$  & 115$\pm$15        & -     & -     &  \citet{2022MNRAS.510.1148S} \\  
    &    HD~142527  & $N=4-3$, $J=\frac{9}{2}-\frac{7}{2}$  & 933$\pm$93        & 40    & 171   &   2024.1.00446.S (P.I. Temmink)$^{2}$  \\ 
    &    HD~145718  & $N=2-1$, $J=\frac{5}{2}-\frac{3}{2}$  &  $<$63.4          & -     & -     &  \citet{2022MNRAS.510.1148S} \\ 
    &    HD~163296  & $N=3-2$, $J=\frac{5}{2}-\frac{3}{2}$  & $3230\pm400$      & 40    & 587   &  \citet{2021ApJS..257....6G}  \\ 
    &    MWC~480    & $N=3-2$, $J=\frac{5}{2}-\frac{3}{2}$  & $1418\pm73$       & 40    & 250   &   \citet{2021ApJS..257....6G} \\ 
    &    Oph-IRS~48 & $N=4-3$, $J=\frac{9}{2}-\frac{7}{2}$  &   $<53$           & 40    & $<7$  &   \citet{2024AJ....167..165B} \\  \hline 
    
\insertTableNotes
\end{longtable*}
\label{table:literature_data}
\end{ThreePartTable}

\begin{figure*}[h!]
\centering
    \includegraphics[width=0.95\linewidth]{Herbigs_scatter_bulk_more_c18o.png}
    \caption{Disk integrated line fluxes for Herbig disks normalized to 150 pc and compared to different bulk parameters of the systems. The purple points are the disks within this survey and the orange points are from the literature (see Table~\ref{table:literature_data}. The triangles denote 3$\sigma$ upper-limits. Numbers in the bottom right of each panel show the number of detections and the resulting Pearson correlation coefficient (r) and p-value for each set of variables where panels with p$>$0.05 are in gray as well as molecules with $<$5 detections.}
    \label{fig:scatter_all}
\end{figure*}

\clearpage
\newpage
\onecolumngrid

\section{Comparison to T-Tauri disks}

\begin{figure*}[h!]
    \centering
    \includegraphics[width=0.7\hsize]{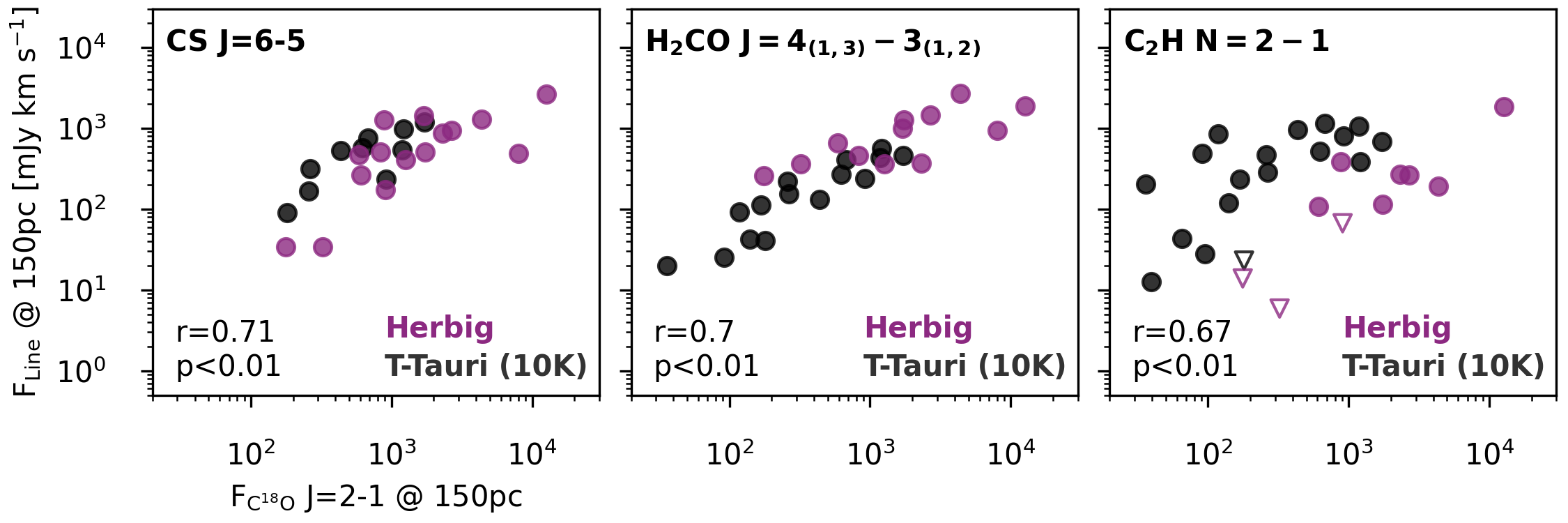}
    
    \vspace{0.25cm}
    \includegraphics[width=0.7\hsize]{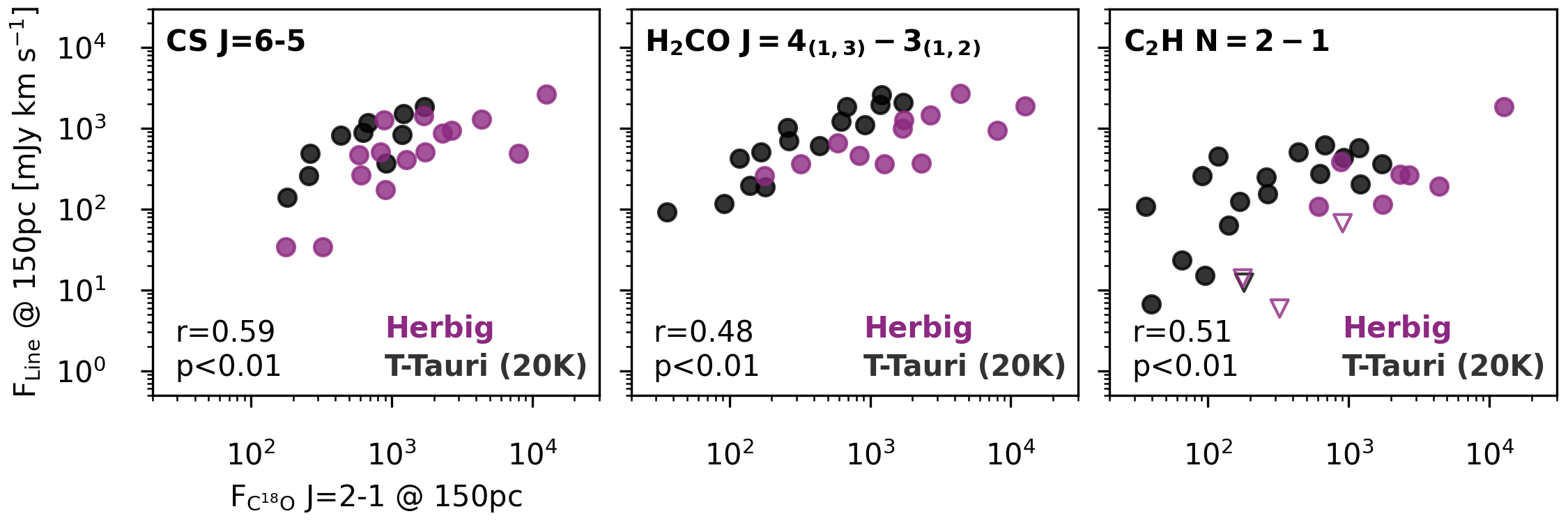}
    
    \vspace{0.25cm}
    
    \includegraphics[width=0.7\hsize]{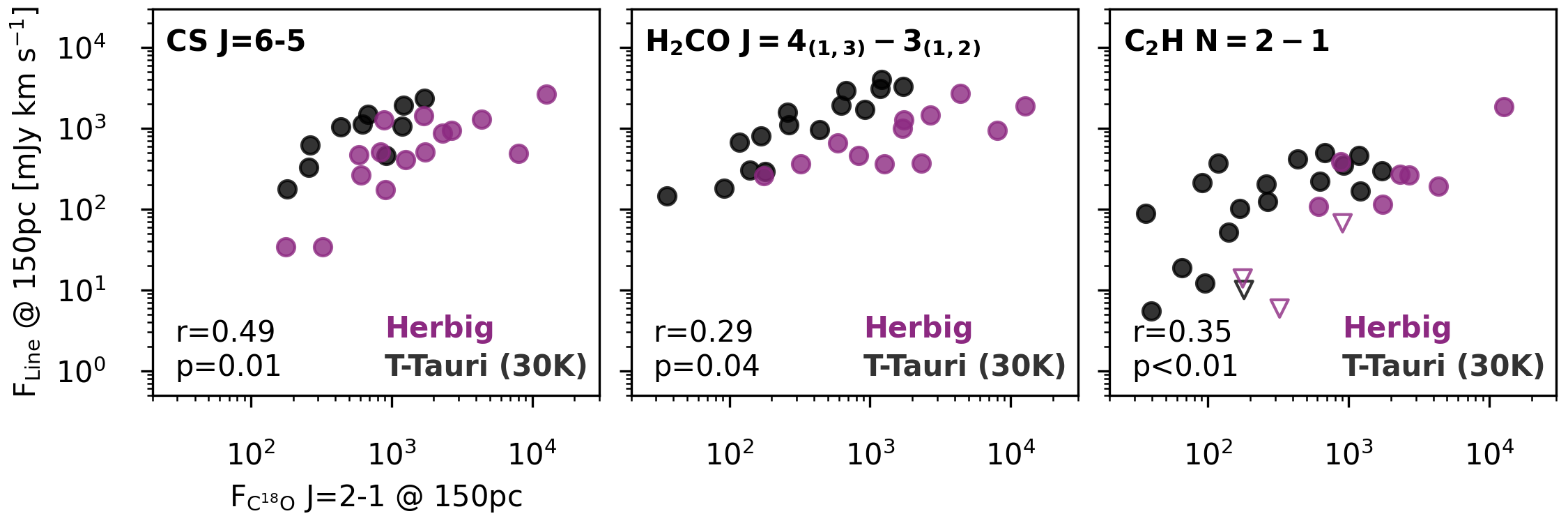}
    
    \label{fig:ttauris}
    \caption{Disk integrated line fluxes normalized to 150 pc for the Herbig sample (purple) presented here and the T-Tauri sample (black) compared to the \ce{C^{18}O} $J-2-1$ line flux. The T-Tauri fluxes are scaled assuming a range of gas temperatures where 20~K is likely the most realistic. Numbers in the bottom left of each panel show the resulting Pearson correlation coefficient (r) and p-value for each set of variables for all data points.}
\end{figure*}

\onecolumngrid
In Figure~\ref{fig:ttauris} we qualitatively compare the line fluxes of \ce{CS}, \ce{H_2CO} and \ce{C_2H} from Herbig disk population with the T-Tauri population. We take the fluxes for 18 T-Tauri disks 
(AS 209
CI Tau,
DM Tau,
DO Tau,
FP Tau,
GM Aur,
HD 143006,
IM Lup,
J0432+1827,
J1100-7619,
J1545-3417,
J1604-2130,
J1609-1908,
LkCa 15,
PDS 70,
Sz 69,
TW Hya,
V4046 Sgr) 
as collated in \citet{2023ApJ...948...57P} for the \ce{C^{18}O} $J=2-1$, CS $J=5-4$, \ce{H_2CO} $J=3_{0,3}-2_{0,2}$ and \ce{C_2H} $N=3–2, J=\frac{7}{2}–\frac{5}{2}$ transitions. We supplement this sample with CS fluxes from \citet{2016A&A...592A..49T} and Law et al. (in review).  We scale all of the T-Tauri data to our benchmark transitions assuming optically thin gas in LTE with temperatures of 10, 20 and 30~K. These temperatures are motivated by those measured for these species in T-Tauri disks \citep[][Law et al. (in review)]{2020ApJ...890..142P,2021ApJS..257....6G}. From these comparisons there are clear overlaps in the two populations of disks and the general correlations shown in Figure~\ref{fig:scatter_c18o} between these line fluxes and the disk mass hold, although the strength of the linear relationships vary. In particular, the \ce{H_2CO} linear correlation becomes stronger with this expanded sample but the \ce{C_2H} weaker. In addition depending on the temperature assumed the T-Tauri disks are generally brighter in \ce{H_2CO} - as suggested by \citet{2020ApJ...890..142P}.
More uniform line data from ongoing ALMA Large programs 2022.1.00875.L (DECO) and 2024.1.01001.L will allow for some robust comparisons of observable disk chemistry across different spectral type systems. \\

\clearpage
\newpage

\section{Directly imaged exoplanet atmospheric data}

In Table~\ref{table:exoplanets} we list the atmospheric C/O constraints for directly imaged exoplanets around F-, A- and B-type stars as show in Figure~\ref{exoplanets} along with their associated discovery papers as listed in the NASA Exoplanet archive. For completeness we also include systems not considered in our comparison due to their wide $>$200 au orbits, low precision C/O inferences, or lack of atmospheric data.

\begin{table*}[h!]
\caption{List of directly imaged exoplanets around F-, A- and B-type and constraints on their atmospheric C/O. }
\centering
\begin{tabular}{ccccc}
\hline 
Source         & Semi-major axis & Discovery paper & C/O & Atmospheric Reference(s) \\ 
& [au] & & &  \\ \hline \hline
51 Eri b       &     13       &  \citet{2015Sci...350...64M}        &   $0.38\pm0.09$  &  \citet{2023A\string&A...673A..98B} \\
AF Lep b       &      9       & \citet{2023A\string&A...672A..93M}  &   $0.55\pm0.1$   &  \citet{2025A\string&A...696A...6D} \\
Beta pic b     &     10       &  \citet{2010Sci...329...57L}        &   $0.43\pm0.05$  &  \citet{2020A\string&A...633A.110G} \\
HD 206893 b    &     10       &  \citet{2021AJ....161....5W}        &   $0.52\pm0.02$  &  \citet{2025AJ....169..175S}       \\
HIP 65426 b    &     92       &  \citet{2017A\string&A...605L...9C} &   $0.59\pm0.01$  &  \citet{2023AJ....166..257B}  \\  
HIP 99770 b    &     16       &  \citet{2023Sci...380..198C}        &   $0.55\pm0.06$  &  \citet{2024AJ....168..131Z}          \\
HR 8799 b      &     71       &  \citet{2008Sci...322.1348M}        &   $0.73\pm0.02$  &  \citet{2024A\string&A...687A.298N} \\
HR 8799 c      &     40       &  \citet{2008Sci...322.1348M}        &   $0.62\pm0.01$  &  \citet{2024A\string&A...687A.298N} \\
HR 8799 d      &     26       &  \citet{2008Sci...322.1348M}        &   $0.68\pm0.04$  &  \citet{2024A\string&A...687A.298N} \\
HR 8799 e      &     16       &  \citet{2010Natur.468.1080M}        &   $0.83\pm0.02$  &  \citet{2024A\string&A...687A.298N} \\
$\kappa$ And b &     61       &  \citet{2013ApJ...763L..32C}        &   $0.64\pm0.04$  & \citet{2024ApJ...970...71X} \\  
\hline
HD 95086 b     &     56       &  \citet{2013ApJ...772L..15R}        & sub-solar             &  \citet{2024A\string&A...690A.316M} \\ 
HD 106906 b    &     650      & \citet{2014ApJ...780L...4B}         & $0.53^{+0.15}_{0.25}$ &  \citet{2023AJ....166..192A}     \\
HD 206893 c    &     3.5      & \citet{2023A\string&A...671L...5H}  & n/a                   &   -    \\
HIP 21152 b    &     18       & \citet{2022ApJ...934L..18K}         & n/a                   &   -    \\
HIP 39017 b    &     22       & \citet{2024AJ....167..205T}         & 1                     &  \citet{2024AJ....167..205T}      \\ 
HIP 78530 b    &     740      & \citet{2011ApJ...730...42L}         & n/a                   &   -   \\
HIP 79098 AB b &     345      & \citet{2019A\string&A...626A..99J}  & $0.54\pm0.03$         & \citet{2024ApJ...970...71X}      \\  
HR 2562 b      &     27       & \citet{2016ApJ...829L...4K}         & unconstrained         & \citet{2024A\string&A...689A.185G}   \\ 
 \hline
\end{tabular}
\label{table:exoplanets}
\tablecomments{Exoplanets below the line are not included in Figure~\ref{exoplanets}}
\end{table*}

\end{appendix}

\end{document}